\documentclass[twocolumn,trackchanges]{aastex701}

\usepackage{graphicx}
\usepackage{adjustbox}
\usepackage{hyperref}
\usepackage{amsmath}

\usepackage{multirow}

\let\oldmaketitle\maketitle

\renewcommand{\maketitle}{\oldmaketitle\setcounter{footnote}{0}}

\begin{document}

\title{The Solar Neighborhood LVI: The RMSTAR Catalog of the Nearest 3352 M Dwarf Systems and 305 of their Wide Companions}

\author[0009-0009-2960-4020]{Madison R. LeBlanc}
\affiliation{Department of Physics and Astronomy, Georgia State University, Atlanta, GA 30303, USA}
\affiliation{RECONS Institute, Chambersburg, PA 17201, USA}
\affiliation{NSF Graduate Research Fellow}
\email{mleblanc@gsu.edu}  

\author[0009-0006-4398-4654]{Tim M. Johns}
\affiliation{Department of Physics and Astronomy, Georgia State University, Atlanta, GA 30303, USA}
\affiliation{RECONS Institute, Chambersburg, PA 17201, USA}
\email{tjohns6@gsu.edu}  

\author[0000-0002-9061-2865]{Todd J. Henry}
\affiliation{RECONS Institute, Chambersburg, PA 17201, USA}
\email{thenry88@gsu.edu}  

\author[0000-0000-0000-0000]{Kenneth J. Slatten}
\affiliation{RECONS Institute, Chambersburg, PA 17201, USA}
\email{kenslatten@gmail.com}  

\author[0000-0001-6031-9513]{Jennifer G. Winters}
\affiliation{Bridgewater State University, Bridgewater, MA 02324, USA}
\affiliation{RECONS Institute, Chambersburg, PA 17201, USA}  
\affil{Center for Astrophysics $\vert$ Harvard \& Smithsonian, 60 Garden Street, Cambridge, MA 02138, USA}
\email{j3winters@bridgew.edu}

\author[0000-0002-1864-6120]{Wei-Chun Jao}
\affiliation{Department of Physics and Astronomy, Georgia State University, Atlanta, GA 30303, USA}
\affiliation{RECONS Institute, Chambersburg, PA 17201, USA}
\email{wjao@gsu.edu}  

\author[0000-0002-9811-5521]{Aman Kar}
\affiliation{Department of Physics and Astronomy, Georgia State University, Atlanta, GA 30303, USA}
\affiliation{RECONS Institute, Chambersburg, PA 17201, USA}
\email{akar5@gsu.edu}  

\author[0000-0003-2565-7909]{Michele L.~Silverstein}
\affiliation{RECONS Institute, Chambersburg, PA 17201, USA}
\email{mlsilverstein@proton.me}  

\author[0000-0002-1864-6120]{Eliot Halley Vrijmoet}
\affiliation{RECONS Institute, Chambersburg, PA 17201, USA}
\affiliation{Five College Astronomy Department, Smith College, Northampton, MA 01063, USA}
\email{evrijmoet@smith.edu}  

\author[0000-0001-6031-9513]{Casey Rush}
\affiliation{Bridgewater State University, Bridgewater, MA 02324, USA}
\email{C1RUSH@student.bridgew.edu}

\author[0000-0001-9834-5792]{Andrew A. Couperus}
\affiliation{RECONS Institute, Chambersburg, PA 17201, USA}
\affiliation{Five College Astronomy Department, Smith College, Northampton, MA 01063, USA}
\email{acouperus@smith.edu}

\begin{abstract}

We present the RMSTAR (RECONS M STAR) catalog, a 25 pc volume-limited and effectively volume-complete sample of the nearest 3352 M dwarf systems and their 305 wide stellar and 9 brown dwarf companions. RMSTAR has been created using only results from \textit{Gaia} Data Releases 3 (GDR3) and 2 (GDR2), \textit{Hipparcos}, and ground-based discoveries of M dwarf systems that are not available in the space-based results. The wide companions are identified using only results from GDR3 and GDR2, and have separations $\rho\ge0.46\arcsec$. There are 291 primaries with stellar companions, yielding a multiplicity rate of 8.68$\pm$0.49\% for these widely separated systems, with a rate of 6.86$\pm$0.44\% for projected separations $s\geq30$ au, where the sample of stellar companions is complete except perhaps for a few fringe cases. It is shown that the rate of stellar companions increases from the search limit of 30,000 au down to separations of 30 au, with a large set of companions to be characterized at closer separations in future work. We determine luminosity and mass functions for all stellar red dwarfs and their wide secondaries, finding that the luminosity function displays a classic turnover at $M_G\sim11$, whereas the mass function is described by an exponential function that rises from 0.60 M$_{\odot}$ to the end of the stellar main sequence at 0.075 M$_{\odot}$. We assess the large population of close, unresolved companions to M dwarfs by analyzing several \textit{Gaia} parameters, identifying 1089 (30\%) individual sources as having at least one of these elevated unresolved companion indication parameters. 

\end{abstract}


\section{Introduction} \label{sec:intro}

The very characteristics that have historically made M dwarfs difficult to study are those that make them fascinating astrophysical objects. Luminosity and mass --- the two most fundamental properties of stars --- are the main contributors to M dwarfs' complexities. Their low luminosities, and resulting observational faintness, have made it difficult to discover and study them at far distances or in a complete manner. Their small masses result in closer stellar and sub-stellar companion populations than for more massive stars, yet many surveys have revealed secondaries (see references in \cite{Winters_2019} Table 1 and \cite{Cifuentes_2025} Table 1). The low luminosities of M dwarfs that are set by their low hydrogen fusion rates result in long lifetimes and long-term consistent flux outputs. These stars are ubiquitous, comprising 75\% of all stars in the solar neighborhood, and presumably our Milky Way and other galaxies (\citealp{Henry_2006}, \citealp{Henry_2018}, \citealp{Henry_Jao_2024}), and have not changed in any significant way since birth, making them historical touchstones. Given these facts, it is important that we improve our understanding of the formation and evolution (albeit very slow) of M dwarfs, as well as their distribution throughout the Milky Way.

When assessing stellar populations, larger, more complete samples lower statistical uncertainties. With M dwarfs spanning a factor of eight in mass \citep{Benedict_2016}, it is salient to not only have large samples that include all types of M dwarfs, but to also ensure that subsamples of masses, metallicities, multiplicity, etc., are large enough to provide statistical assessments with high degrees of reliability. In the past, volume-limited samples of more than $\sim$1000 M dwarfs have been difficult to create, and volume-complete samples have been nearly impossible due to observational limits. Ground-based observations of the twentieth century gave way to samples of hundreds of M dwarfs with parallaxes accurate to $\sim$1 milliarcsecond (\citealp{Henry_1990}, \citealp{Simons_1996}, \citealp{vanAltena_1995}). The space-based, all-sky survey carried out by \textit{Hipparcos} was revolutionary, observing more than 120,000 stars with milliarcsecond astrometric precision (\citealp{Perryman_1997}, \citealp{vanLee_2007}), but its magnitude limit of $V$$\sim$12 captured relatively few M dwarfs. Until 2020, the combination of ground-based parallax measurements, primarily by the RECONS (REsearch Consortium On Nearby Stars)\footnote{\it{www.recons.org}} team (e.g., \citealp{Jao_2005,Henry_2006,Riedel_2010,Winters_2015,Winters_2017,Henry_2018}), and \textit{Hipparcos} results led to significant improvements in lists of the nearest M dwarfs, as evidenced in the work of \cite{Winters_2019} that included a sample of 1120 M dwarf systems within 25 pc.  Smaller, more targeted samples have been created with the intent of identifying additional close companions to M dwarfs via speckle imaging, including the northern 15 pc POKEMON survey sample \citep{Clark_2024} and the southern 25 pc sample used by \cite{Vrijmoet_2022} to map the orbits of M dwarf companions. In addition, while searching for exoplanets, radial velocity surveys continue to detect stellar companions, such as the work of \citet{Baroch_2021} and J. G. Winters et al.,~(in press). These various surveys have been useful for studying M dwarfs and their companions, but recent {\it Gaia} spacecraft data releases have created a watershed moment for studies of these small stars that dominate the solar neighborhood.

\textit{Gaia} Data Release 3 \citep{Gaia_2023}, hereafter GDR3, has enhanced our ability to create a volume-limited, volume-complete sample of M dwarfs out to the 25 pc horizon we adopt here. The results of GDR3 are unprecedented. Probing magnitudes as faint as $G \approx 21.4$ with sub-milliarcsecond astrometry, the \textit{Gaia} survey has collected photometric data on over 1.8 billion sources and astrometric data on over 1.4 billion sources. Among these are more than 2000 M dwarf systems with parallaxes of at least 40 milliarcseconds (within 25 pc) newly identified via extractions from Gaia for RMSTAR and often included in other Gaia compilations reaching 25 pc or beyond \citep{Smart_2021}. In addition, \textit{Gaia}'s faint magnitude limit and ability to resolve companions at separations of $\rho = 0\farcs4$ or perhaps better in a few cases (although none in this sample) permits the direct identification of a plethora of new multiple systems, and several \textit{Gaia} parameters provide clues to unseen companions. Together, the seen and unseen companions provide crucial information for sample completeness and multiplicity statistics, as evidenced by the keen work of \cite{Cifuentes_2025}. The stellar multiplicity rate of a sample --- how many of the stars in a sample have gravitationally bound stellar companions, defined here as ``MR'' to avoid confusion with the mass function ``MF'' --- provides insight into how stars form, interact, and evolve in multiples versus alone. Multiplicity rates vary from nearly 100\% for the most massive O type stars to 54\% for solar type stars (\citealp{Raghavan_2010}, \citealp{Offner_2023}), and the rate is estimated to be $\sim$27\% for M dwarfs \citep{Janson_2012,Winters_2019}. However, these determinations often rely on approximations for unresolved close companions. While there is no precise definition of the boundary between ``wide'' and ``close'' companions, there are practical resolution limits set by the telescopes and instruments used. It is necessary to consider these limits when assessing the completeness of a sample and the accuracy of a derived multiplicity rate. 

Having a comprehensive grasp of companions to M dwarfs is key to understanding the outcomes of the star formation process. Furthermore, an individual star cannot be completely characterized until all close companions are identified, otherwise there is a chance of contamination that can complicate measurements of its luminosity, abundance, space motion, etc.~\citep{ElBadry_2019}. Without a complete evaluation of single and multiple star systems, and the individual stars within the multiples, accurate determinations of the true luminosity and mass functions remain elusive \citep{Kroupa_1991}. 

Disentangling the multiplicity of all M dwarfs is necessary not only for characterizing stars with stellar companions, but also for creating pure samples of single stars that serve as pristine astrophysical environments. Given the possibility that most M dwarfs are orbited by at least one planet \citep{Ribas_2023, Kaminski_2025} and the fact that most planet searches target single star systems \citep{Bonfils_2013}, it is increasingly important to know which stars have, or do not have, stellar companions for future missions such as the Habitable Worlds Observatory \citep{Tuchow_2025}.

In this paper, we present the 2026 edition of the RECONS M STAR Catalog (hereafter, simply RMSTAR or RMSTAR 2026), a volume-limited catalog of M dwarf systems within 25 pc of the Sun. In $\S$2, we define M dwarfs and outline the construction and limits of the catalog. We identify all systems with M dwarf primaries, excluding any M dwarfs with higher-mass or white dwarf companions, although those are preserved for luminosity and mass function evaluations. 

In $\S$3 we define the constraints for gravitationally bound, lower-mass stellar companions\footnote{A total of nine brown dwarf companions are also identified and included in RMSTAR, but because they comprise only a small fraction of such objects likely to be ultimately found, we do not include them in any statistical calculations.} and add these to the catalog. We identify and characterize wide companions, wherein ``wide'' has two meanings --- separations large enough to be resolved by \textit{Gaia} and \textit{Hipparcos}, and in physical terms the set of (nearly) complete stellar companions spanning s $=$ 30--30000 au. The 30 au benchmark is convenient for this sample as it corresponds to the size of Neptune's orbit in our Solar System, within which companions can only be revealed via high-resolution techniques.

In $\S$4 we discuss the implications of close companions to M dwarfs that are not resolved by \textit{Gaia} but are flagged as possible multiples using \textit{Gaia} indicators. In $\S$5 we use these results to derive luminosity and mass functions for M dwarfs within 25 pc, acknowledging unseen companions and their expected impacts on future assessments. In $\S$6 we describe a few caveats associated with RMSTAR and place the results in context with previous work. In $\S$7 we highlight a few key conclusions of this work. Finally, we include two Appendices with details on some noteworthy systems ($\S$A) and nearby M dwarfs that are not included in RMSTAR ($\S$B).

\section{The RMSTAR Catalog \label{sec: RMSTAR}} 

\subsection{Defining the Sample \label{subsec: Defining Sample}} 

RMSTAR aims to define a volume-limited, volume-complete sample of M dwarf stars within a 25 pc radius of the Sun. The purpose of this catalog is to understand, in a statistically complete context, the contents of the solar neighborhood and to provide target samples for large surveys of stellar, brown dwarf, and planetary companions to M dwarfs. RMSTAR has been created in conjunction with the RECONS K STAR Catalog (hereafter, simply RKSTAR) that includes K dwarf systems within a distance horizon of 50 pc (Johns et al.~in review). Here we outline the steps to construct the initial (2026 edition) of RMSTAR.

RMSTAR consists of M dwarf primary stars and their lower mass companions, often referred to as ``secondaries''.  The primary is defined as the brightest star in the system by absolute \textit{Gaia} magnitude ($M_G$) and companions are relatively fainter stars or brown dwarfs with similar proper motions and parallaxes to the primaries. Thus, this catalog includes M dwarf primaries, M dwarf companions, and brown dwarf companions; this RMSTAR version does not list reported planets. Removing systems with earlier spectral type primaries or white dwarf primaries is essential to the integrity of this sample, as earlier type primaries would bias the conclusions to be drawn about the formation, evolution, and orbital architectures of M dwarf systems. 

We define membership in RMSTAR by the parallax of the primary source and, subsequently, by $M_G$. The primary's parallax must be at least 40 mas and is therefore the original membership source for that star and any companions\footnote{Note that in a few very rare cases, an M dwarf primary may not have a parallax because of a close companion, but a wide M dwarf companion may have a parallax greater than 40 mas. In such a case, the wide component temporarily functions as the ``primary", but we must wait for a reliable parallax for the true primary to establish the full system in RMSTAR.}.  RMSTAR primaries have 8.1 $\le$ $M_G$ $\le$ 17.8; brown dwarfs are fainter than this limit. These magnitude limits for ``red dwarfs'' separate late-K and early-M types and mark the end of the stellar main sequence; both cutoffs are explained in detail by \cite{Henry_Jao_2024}. In short, stars classified as late-K and early-M types in large, reliable, spectroscopic surveys show overlapping magnitudes on the main sequence. An equal split of these overlapping stars is made at $M_G =$ 8.1. The end of the main sequence is based on the hydrogen burning limit identified by \cite{Dieterich_2014}, who used astrometric and photometric measurements to measure the luminosities and temperatures of objects corresponding to the radius inflection point at L2.5V for the smallest objects where the transition from fusion to electron degeneracy support occurs. This inflection point also corresponds to the gap in the smooth distribution of main sequence stars at $M_G$ = 17.8 \citep{Henry_Jao_2024}. While not formally called M dwarfs, stars with spectral types L0.0V to L2.5V are still on the main sequence above the hydrogen burning limit, thereby representing the lowest mass red dwarfs; thus, the ``M dwarfs'' listed in RMSTAR include all red dwarfs with spectral types M0.0V through L2.5V. A very small number of young M-type and L-type objects are in fact young brown dwarfs. These will be included in RMSTAR and noted as potential brown dwarfs should data support such a designation. $M_G$ is a necessary, but less reliable, discriminant for membership than parallax because a star can appear brighter than it actually is due to an unresolved companion, thereby moving it to brighter than the adopted upper limit $M_G$ cutoff, and lost to the sample. Alternately, a pair of unresolved brown dwarfs may together have an $M_G$ value brighter than the adopted lower limit $M_G$ cutoff, although the objects should not be in RMSTAR. 

\subsection{Gathering Primaries for Catalog Construction \label{subsec: Gathering Prims}} 
\subsubsection{Primaries from \textit{Gaia} DR3}\label{subsubsec:GDR3}

GDR3 includes photometric data for over 1.8 billion sources as well as astrometric and spectroscopic data for most of these sources. For M dwarfs within 25 pc, GDR3 provides accurate photometry and parallaxes that allow a rich sample of M dwarfs to be created. The detection limit of a source by \textit{Gaia} is $G$ $\simeq$ 21. At 25 pc this corresponds to $M_G$ $\simeq$ 19, more than a magnitude fainter than the $M_G$ = 17.8 cutoff for RMSTAR M dwarfs. Thus, RMSTAR uses the detection limit of GDR3 to reach a horizon of 25 pc for the full range of M dwarfs, as defined in $\S$\ref{subsec: Defining Sample}, and also picks up some brown dwarfs as companions to the M dwarfs.

\begin{figure*}
    \centering
    \includegraphics[scale = .47]{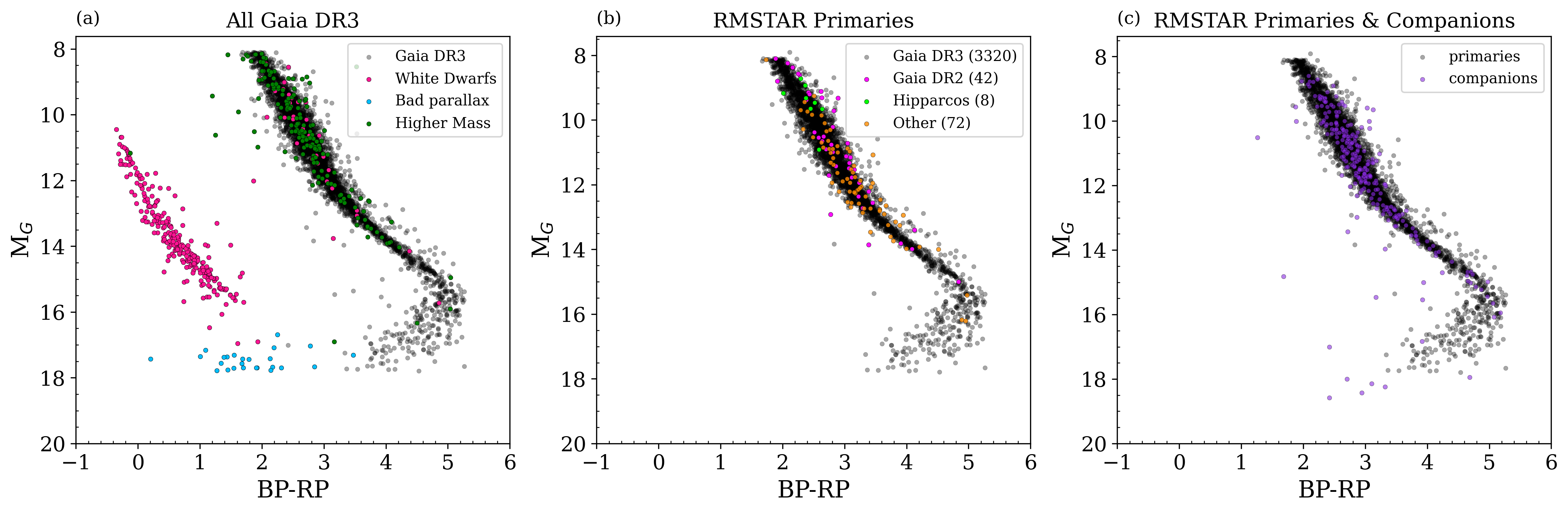}
    \caption{Color-magnitude diagrams describing entries in RMSTAR. Panel $a$ shows all sources in \textit{Gaia} DR3 within 25 pc and with absolute \textit{Gaia} magnitude $M_G \geq 8.1$. Grey points represent M dwarfs from \textit{Gaia} DR3, pink points represent white dwarfs or M dwarfs with white dwarf companions, blue points represent sources that were removed because of poor parallax measurements, and green points represent sources that were removed because they have higher mass stellar companions, as discussed in $\S$\ref{subsec:nonprims}. Panel $b$ shows all primary stars in the RMSTAR catalog colored by the source of each stars' astrometric data, including those from GDR3 (grey), GDR2 (magenta), \textit{Hipparcos} (green), and RECONS/other literature sources (orange). Panel $c$ shows all primaries (grey) and wide companions (purple) in the RMSTAR 2026 catalog.
    }
    \label{fig:HRDs}
\end{figure*}

RMSTAR primaries were selected and vetted through a process beginning with querying GDR3 for all sources within 25 pc (parallax $\leq$ 40 mas). An absolute magnitude cut was then made to keep stars only within the brightness range 8.1 $\leq$ $M_G$ $\leq$ 17.8; the single point brighter than $M_G=8.1$ is the known young star AU Mic, that is the primary in a triple M dwarf system. This initial selection of stars includes the majority of M dwarfs within 25 pc along with some white dwarfs and a few objects with large errors in parallax, all of which are categorized and highlighted in Figure \ref{fig:HRDs}a\footnote{A few sources have no $BP$ and/or $RP$ magnitudes and are not shown in Figure \ref{fig:HRDs}a even though they are included in RMSTAR.}. Within this initial draft selection are M dwarfs that are wide secondaries to more massive or white dwarf primaries. These are removed from RMSTAR as described in $\S$\ref{subsec:nonprims} and are kept in a separate list to utilize for the luminosity and mass function analysis in $\S$\ref{subsubsec:LF} and $\S$\ref{subsubsec:MF}.

A total of 3534 M dwarfs were identified for RMSTAR using GDR3, which were then separated into 3230 primaries and 276 companions. These are shown with black points in the color-magnitude diagram in Figure \ref{fig:HRDs}b. RMSTAR includes identifiers and basic system information, plus astrometric, photometric, and spectroscopic parameters referenced in the RMSTAR table as ``Gai22'' (see entries in Table \ref{tab:RMSTAR}). The sources for data in RMSTAR columns are listed in Table \ref{tab:params}.

\begin{deluxetable*}{ll|ll|ll}[h]
\tablewidth{0pt}
\tablecaption{Table of observational data included in RMSTAR 2026. } 
\label{tab:params}
\tablehead{\multicolumn{2}{c}{\textbf{Astrometry}} & \multicolumn{2}{c}{\textbf{Photometry}} & \multicolumn{2}{c}{\textbf{Spectroscopy}} \\ \colhead{Column}  & \colhead{Source} & \colhead{Column} & \colhead{Source} & \colhead{Column} & \colhead{Source}}
\startdata
RA J2000.0 & Gai22 & $G$ & Gai22 & RV & Gai22 \\
Dec J2000.0 & Gai22 & $V$ & Lit./estimation & RV error & Gai22 \\
Parallax & Gai22/Gai18/vLe07/Lit. & $BP$ & Gai22 &  & \\
Parallax error & Gai22/Gai18/vLe07/Lit. & $RP$ & Gai22 & & \\
Proper Motion (RA) & Gai22/Gai18/vLe07/Lit. & IPDfmp & Gai22 &  & \\
Proper Motion (Dec) & Gai22/Gai18/vLe07/Lit. & Separation & Gai22 &  &  \\
RUWE & Gai22 & Position Angle & Gai22  &  &  \\ \hline
\enddata
\tablecomments{The data are primarily from GDR3 (``Gai22''), GDR2 (``Gai18''), and Hipparcos (``vLe07''), with additional contributions from a variety of literature sources that are all listed and referenced in Table \ref{tab:refs} in Appendix \ref{references}.}
%\tablecomments{}
\end{deluxetable*}

We find 132 M dwarfs in GDR3 that lack reliable five-parameter solutions, typically because they are close binaries with not-yet-solved orbits. To ensure that as many of these stars as possible are included in RMSTAR, additional catalog and literature searches were done, as described next in $\S$\ref{subsubsec:GDR2}, $\S$\ref{subsubsec:Hipparcos}, and $\S$\ref{subsubsec:RECONS}. 

\subsubsection{Primaries from \textit{Gaia} DR2}\label{subsubsec:GDR2}

\textit{Gaia} Data Release 2 (GDR2) \citep{GaiaDR2_2016,GaiaDR2_2018b} included 22 months of data, compared to the 34 months used in GDR3. While GDR3 data yielded improved precision for parallaxes and other results, some stars ``dropped out'' because the additional 12 months of data revealed orbital motion that differed significantly from \textit{Gaia}'s single-star astrometric model. This consequently caused a small number of stars to be excluded from GDR3 when they had previously been in GDR2. 

To recover these stars for RMSTAR, we conducted a thorough search of GDR2. The procedure to identify these stars is similar to that described in $\S$\ref{subsubsec:GDR3}. All sources with parallaxes of 40 mas or larger were extracted from GDR2, $M_G$ was calculated for each source, and only those with $M_G$ $\ge$ 8.1 were kept. Next, a cross-match between the GDR2 sources and the sources found in GDR3 ($\S$\ref{subsubsec:GDR3}) was performed. Sources with astrometry in both data releases were removed, as these have already been included or excluded from the RMSTAR catalog based on their superior GDR3 parallaxes. Sources with astrometric solutions in GDR2 but $not$ in GDR3 were investigated to determine whether or not they should be included in the catalog. We validated each source by checking it against its GDR3 values: each source should have an entry in GDR3 with approximately the same brightness in the $G$ band, but no parallax. It is important that each of these sources is checked carefully, as some sources were within 25 pc in GDR2, but then pushed beyond the distance limit in GDR3.

M dwarfs in GDR2 that were found to fit RMSTAR criteria were then added to the catalog. The included parameters are a combination from GDR2 (astrometry) and GDR3 (all other data). In these cases, the references on the astrometric data are ``Gai18'' for GDR2 rather than ``Gai22'' for GDR3. A total of 42 M dwarf systems were added to RMSTAR from GDR2. These are shown with green points in the color-magnitude diagram in Figure \ref{fig:HRDs}b.

\subsubsection{Primaries from \textit{Hipparcos}}\label{subsubsec:Hipparcos}

A search of \textit{Hipparcos} results \citep{vanLee_2007} was also conducted to find additional M dwarfs that were not included in GDR2 or GDR3. Sources in the \textit{Hipparcos} catalog generally have larger astrometric parameter errors than from \textit{Gaia}, and it is likely that these stars have unresolved companions confounding astrometric solutions that led to GDR2 and GDR3 omissions. Nevertheless, until better astrometric solutions for these stars are available, a few are currently considered to be within 25 pc and thus are included in RMSTAR.

\begin{figure} 
    \centering
    \includegraphics[scale = .65]{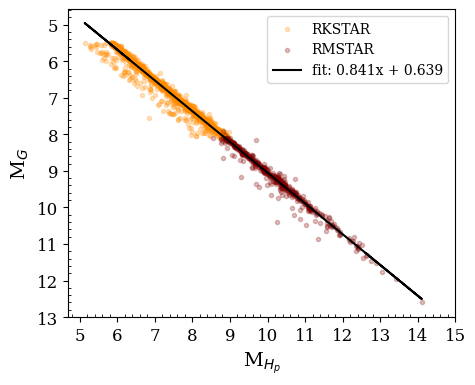}
    \caption{$M_G$ vs.~$M_{H_P}$ for K dwarfs in RKSTAR (orange) and M dwarfs in RMSTAR (dark red). A linear fit, shown in black, is used to convert $M_{Hp}$ to $M_G$ to define \textit{Hipparcos} boundaries for K and M dwarfs.}
    \label{fig:HIPrel}
\end{figure}

Magnitude limits for RMSTAR membership were defined using $M_G$, so the magnitudes deriveable from \textit{Hipparcos}, $M_{Hp}$, needed to be transformed to $M_G$. This was straightforward because GDR3 includes \textit{Hipparcos} identifiers that were used to extract $Hp$ values that could be combined with GDR3 parallaxes to calculate $M_{Hp}$. Figure \ref{fig:HIPrel} shows $M_G$ versus $M_{Hp}$ for presumed single stars in both RKSTAR (orange points) and RMSTAR (dark red points) having magnitudes in both \textit{Hipparcos} and \textit{Gaia}, with a linear fit that is used to convert $M_{Hp}$ to $M_G$. Because \textit{Hipparcos'} faint limit of $Hp$ $\sim$ 12.4 \citep{Perryman_1997} is much brighter than \textit{Gaia's} corresponding limit, there are not many late M dwarfs available for the relation.  So, we fit both the RKSTAR and RMSTAR members simultaneously, enhancing the quality of the fit. Using the conversion 

\begin{equation}\label{MG_MH}
    M_G = 0.841 M_{Hp} + 0.639
\end{equation}

\noindent we find that the bright limit for M dwarfs in RMSTAR corresponds to $M_{Hp}$ = 8.88. Note that the spray of points below the fit are multiples that were unresolved by \textit{Hipparcos} but have been resolved by \textit{Gaia}.

With this criterion in hand, we then queried the \textit{Hipparcos} results for sources within 25 pc and calculated $M_{Hp}$ for each. As with GDR2 sources, we cross-matched potential additions to RMSTAR entries to create a short list of stars with \textit{Hipparcos} entries missing from both GDR3 and GDR2. The result is that eight stars were added to RMSTAR from \textit{Hipparcos}, for which the astrometry is from \textit{Hipparcos} (referenced as ``vLe07'') and other data are from GDR3, if available. These are shown with cyan points in the color-magnitude diagram in Figure \ref{fig:HRDs}b.

\subsubsection{Primaries from Other Sources}\label{subsubsec:RECONS}

There are a number of additional M dwarfs that do not have astrometry in \textit{Gaia}  or \textit{Hipparcos}. These entries are usually multiple star systems, complicating their motions and making it difficult to solve for their astrometry, unless long-term datasets are available that can yield solutions including orbital motion. We cross-matched RMSTAR with the RECONS database and the Yale Parallax Catalog \citep{vanAltena_1995}, and searched the literature for additional M dwarf systems (e.g., \citealp{Dittmann_2014,Finch_2018,Winters_2017}). We find 72 M dwarf systems with parallaxes from other sources that meet the $M_G$ limits for M dwarfs. Many of these parallax measurements are comparable to \textit{Hipparcos} results, with errors of $\sim$1 mas, but some are significantly larger. For these systems, we include the systems' J2000.0 coordinates, parallaxes, proper motions, and errors in RMSTAR with appropriate references, using the first three letters of the first author's last name and the last two digits of the year published, e.g., \cite{Benedict_2016} $=$ Ben16. All other available data included in RMSTAR come from GDR3.

\begin{figure}
    \centering
    \includegraphics[scale = 0.65]{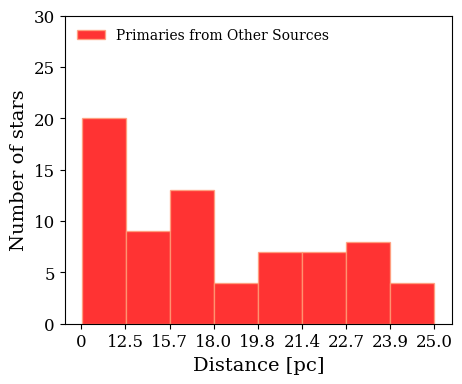}
    \caption{Histogram of 72 primary M dwarfs with astrometry from ground based efforts by the RECONS team, the Yale Parallax Catalog, or from other sources in the literature, plotted in equal-volume shells with increasing distance from the Sun. Most of these M dwarf systems are well within 25 pc and likely to remain in RMSTAR after more precise parallaxes are determined in future \textit{Gaia} data releases.}
     \label{fig:other prims}
\end{figure}

As shown in Figure \ref{fig:other prims}, which plots these additions in eight equal-volume shells, most of these systems are closer than 18 pc.  This is an observational bias towards the closest stars primarily due to the RECONS team's effort to identify all stars within 10 pc \citep{Henry_2004}, as well as other efforts to identify nearby stars (\citealp{Lepine_2005,Scholz_2005}). Most of these added systems were selected photometrically and found to be unresolved multiples made up of M dwarfs of appropriate colors, but were overluminous and therefore more distant than 10 pc. These are precisely the systems for which parallaxes are not yet available in \textit{Gaia} results because they are multiples, nor in \textit{Hipparcos} results because they are too faint. In most cases, these additions are much closer than 25 pc and will remain in RMSTAR when revised parallaxes are available, but a few near 25 pc may be dropped in future RMSTAR editions.

\subsection{Weeding Out M Dwarfs That Are Not Primaries} \label{subsec:nonprims}

There are three broad categories of companions to the M dwarf primaries in the RMSTAR catalog. First are the higher mass companions that are, of course, the true primaries in these systems. Second are the white dwarf companions that were formerly more massive than the M dwarfs. Systems with either more massive or white dwarf primaries are not included in RMSTAR, but are removed and set aside into a second list used to calculate accurate luminosity and mass functions for M dwarfs (see $\S$\ref{subsubsec:LF} and $\S$\ref{subsubsec:MF}). The third category of companions includes those with masses lower than the M dwarf primaries --- these are the subject of the companion search in this paper and will be discussed in detail by $\S$\ref{sec:Companions}, after removals are accomplished to set the list of systems with M dwarfs as primaries.

\subsubsection{Removing M Dwarfs with Higher Mass Primaries} \label{subsubsec:highermass}

A portion of the M dwarfs included in the initial RMSTAR search have higher mass primaries that are OBAFGK dwarfs and a few are companions to more evolved stars, e.g., giants. Higher mass primaries influence both the formation processes and the orbital architectures of the systems, thus affecting statistical analyses of M dwarf samples. Because our data collection included an absolute magnitude cut, the higher mass primaries would have been removed from the catalog while their M dwarf companions remained. So, we identified higher mass primaries through a series of checks (using $M_G$ as a proxy for mass), and the 201 M dwarfs that were found to have higher mass primaries were removed from the RMSTAR catalog. These are added to a separate removals list captured in Table \ref{tab:removals} in Appendix B, where the reasons for removal are listed.

We begin this search by cross-matching the RMSTAR catalog with GDR3 using a projected 30,000 au search radius. \cite{ElBadry_2021} found that at separations s $\gtrsim$ 30,000 au, ``binary'' populations are dominated by non-physical chance alignments of stellar pairs. For RMSTAR M dwarfs, when a second star is found within 30,000 au, it is checked for whether or not the parallaxes, proper motions in right ascension and declination, and, when available, radial velocities, match. The criteria used to confirm or refute matches in these parameters are explained in $\S$\ref{sec:Companions}. If the values agree, and the primary star is brighter than the M dwarf cutoff at $M_G$ = 8.1, the system was removed from RMSTAR.

Next, RMSTAR entries were cross-matched against both GDR2 and \textit{Hipparcos} results using the same matching criteria, without the radial velocity criterion for the \textit{Hipparcos} query. The search for higher mass companions in GDR2 was completed in the exact same manner as that of GDR3. For \textit{Hipparcos}, the upper brightness limit of $M_{Hp}$ = 8.88 was used in place of the \textit{Gaia} brightness limit of $M_G$ = 8.1. As with the GDR3 match, M dwarfs with higher mass primaries were removed from RMSTAR.

\subsubsection{Removing M Dwarfs with White Dwarf Primaries} \label{subsubsec:wds}

White dwarfs were formerly higher mass stars --- i.e., they are the primaries in systems with M dwarfs despite sometimes \textit{currently} being fainter in $M_G$ --- and such systems, similarly to higher mass stars, have different formation processes and orbital architectures than M dwarfs and their lower mass companions \citep{Kippenhahn_1990}. As can be seen in Figure 1a, white dwarfs fall within the $M_G$ range of M dwarfs, but are much bluer in $BP-RP$ color. We identify and remove individual white dwarfs and their companions from RMSTAR, including M dwarfs with white dwarf primaries. 

Sources with blue colors, defined to be $BP-RP$ $\le$ 1.8, and absolute magnitudes $10.0 \le M_G \le 17.0$, were flagged as white dwarfs in a by-eye analysis of the color-magnitude diagram shown in Figure \ref{fig:HRDs}a. All white dwarfs were checked for common right ascensions and declinations, parallaxes, proper motions, and radial velocities to determine whether or not they were companions to RMSTAR M dwarfs with the same process used to check for companions detailed in $\S$\ref{sec:Companions}. These checks allowed single white dwarfs to be removed from RMSTAR as well as those with M dwarf companions. The samples before and after white dwarf cuts are shown in the color-magnitude diagrams of Figure \ref{fig:HRDs}. 304 removed sources, both white dwarfs and M dwarfs with white dwarf companions, are captured in Table \ref{tab:removals} with the removal reasons listed.

\subsubsection{Removing M Dwarfs via Special Considerations} \label{subsubsec:special}

\textbf{Bad Parallaxes:} An additional subset of 32 sources scrubbed from RMSTAR are those with poor quality parallax measurements in GDR3. These sources fall within the brightness range of the sample, but their faintness and/or location in crowded fields, e.g., toward the direction of the Galactic bulge, results in large parallax errors ($\gtrsim$ 0.85 mas) and consequently more uncertain absolute magnitudes. These sources are shown with red points in the color-magnitude diagram in Figure \ref{fig:HRDs}a with $16.0 < $ M$_G < 17.8$ and typically have $BP-RP$ $< 3.0.$ We are unable to classify these sources reliably as M dwarfs, so we removed them from RMSTAR and listed them in Table \ref{tab:removals} with the removal reason given as ``bad parallax.'' This parallax error cut was made only for primary stars in GDR3, as GDR2, \textit{Hipparcos}, and ground based parallax sources tend to carry larger errors. 

There are two companions in RMSTAR that were resolved in GDR3 with parallax errors greater than 0.85 mas --- RMS 0639-2101B (LP 780-032B) and RMS 1214-2345B (RX J1214.1-2345B). After careful examination of their astrometric data and that of their primaries, we determined that these are true companions. RMS 0639-2101B is only 0\farcs56 from its primary and RMS 1214-2345B is a very faint brown dwarf with $M_G$ = 20.4.

\textbf{Unresolved White/Red Dwarf Pair:} In a single case, G 203-047AB was removed from RMSTAR because it appears that an unresolved companion to the observed M dwarf is a white dwarf (\citealp{Delfosse_1999}, \citealp{Hollands_2018}). The smattering of a few similar points on the color-magnitude diagram (Figure \ref{fig:HRDs}a) indicates that this may not be the only such system that will merit removal from RMSTAR, but until other compelling data indicate that these have unresolved white dwarf companions, they remain in RMSTAR.

\subsection{RMSTAR 2026 Primaries}\label{subsec:finalprims}

After refining the sample, we find 3352 systems with M dwarf primaries in RMSTAR. 

All columns of the RMSTAR catalog are outlined in Table \ref{tab:RMSTAR}, along with two example systems, including a single and a binary. The column titles have been arranged into a vertical list on the left to fit on the page. Descriptions of column contents will be given as needed in the sections below. We start here with the first entry ``char'' that characterizes each object, where ``p'' indicates the primary M dwarf in each system, and ``g'' indicates a companion from GDR3 or GDR2. There are no companions from \textit{Hipparcos} or other sources in this version of RMSTAR. The resulting primaries are represented with grey points in the rightmost color-magnitude diagram in Figure \ref{fig:HRDs}, with wide companions shown in purple.

%\documentclass[twocolumn]{aastex701}
%%\documentclass[twocolumn,linenumbers]
%%\usepackage[latin1]{inputenc}
%\let\tablenum\relax 
%%\usepackage{savesym}
%\usepackage{array,booktabs,siunitx}
%%\newcolumntype{T}[1]{S[table-format=#1]}
%%\usepackage[skip=0.333\baselineskip]{caption}
%\usepackage{longtable}
%\usepackage{placeins}
%
%\UseRawInputEncoding
%\usepackage{float}
%\usepackage{color}
%\newcommand{\vdag}{(v)^\dagger}
%\newcommand\aastex{AAS\TeX}
%\newcommand\latex{La\TeX}
%
%\def\pers{\hbox{s$^{-1}$}}
%\providecommand{\msun}{\ensuremath{\,M_\Sun}}
%\providecommand{\rsun}{\ensuremath{\,R_\Sun}}
%\def\coreno{\hbox{$413$}}
%\providecommand{\vsini}{$v \sin i$}
%
%
%
%%\submitjournal{AJ in the future}
%
%\shorttitle{RMSTAR 2026} 
%\shortauthors{LeBlanc et al.}
%
%\begin{document}
%
%\title{Radial and Rotational Velocities of a Volume-Complete Sample of M Dwarfs with Masses $0.1-0.3$~M$_{\odot}$ within 15 parsecs}
%\newpage

\startlongtable
\begin{deluxetable*}{llllrrr}
%\centering
%\setlength{\tabcolsep}{0.03in}
%\tablewidth{0pt}
\tabletypesize{\scriptsize}
\tablecaption{List of columns in RMSTAR 2026, outlining parameters included in the catalog with headers listed vertically to fit on the page. One example of a single star and one example of a binary with a star and a brown dwarf are shown. \label{tab:RMSTAR}}
\tablehead{\colhead{Index}          &
	   \colhead{Units}               &
	   \colhead{Column Name}   &
    \colhead{Description}
	   }
\startdata
1  &          & \textbf{char}      & Primary $=$ p, Gaia Companion $=$ g                      & p            & p            & g            \\
2  &          & \textbf{RMS}       & RMSTAR Name                                              & RMS0001+0659 & RMS0004-4044 & RMS0004-4044 \\
3  &          & \textbf{cmp}       & Component                                                &              & A            & bd           \\
4  & mag      & \textbf{MG}        & Absolute Gaia Magnitude                                  & 12.824       & 11.059       & 17.951       \\
5  &          & \textbf{num\_st}   & Number of Stars in the System                            & 1            & 1            &              \\
6  &          & \textbf{f\_RUWE}   & RUWE Flag                                                & FALSE        & FALSE        & FALSE        \\
7  &          & \textbf{f\_IPDfmp} & IPDfmp Flag                                              & FALSE        & FALSE        & FALSE        \\
8  &          & \textbf{f\_e\_RV}  & RV error Flag                                            & FALSE        & FALSE        &              \\
9  &          & \textbf{f\_NSS}    & NSS Flag                                                 & FALSE        & FALSE        & FALSE        \\
10 & hh mm ss & \textbf{RAJ2000}   & Right Ascension (J2000.0)                             & 00 01 15.82   & 00 04 36.43   & 00 04 34.86   \\
11 & dd mm ss & \textbf{DecJ2000}  & Declination (J2000.0)                                  & $+$06 59 35.5&$-$40 44 02.8 &$-$40 44 06.4 \\
12 & mas\,yr$^{-1}$   & \textbf{muRA}      & Proper Motion in RA                                      &$-$436.302    & 677.675      & 668.888      \\
13 & mas\,yr$^{-1}$   & \textbf{e\_muRA}   & Uncertainty in Proper Motion in RA                       & 0.044        & 0.029        & 0.232        \\
14 & mas\,yr$^{-1}$   & \textbf{muDec}     & Proper Motion in Dec                                     &$-$83.364     & $-$1505.616  & $-$1498.236  \\
15 & mas\,yr$^{-1}$   & \textbf{e\_muDec}  & Uncertainty in Proper Motion in Dec                      & 0.022        & 0.026        & 0.199        \\
16 & mas      & \textbf{pi}        & Parallax                                                 & 42.7802      & 81.2226      & 82.3466      \\
17 & mas      & \textbf{e\_pi}     & Uncertainty in Parallax                                  & 0.0365       & 0.0344       & 0.2554       \\
18 &          & \textbf{r\_ast}    & Reference for Astrometry                                 & Gai22        & Gai22        & Gai22       \\
19 &          & \textbf{map}       & Configuration of Multiple Components                     &              & Abd          & Abd          \\
20 & arcsec (\arcsec)   & \textbf{sep}       & Separation of Multiple Components                        &              &              & 18.32        \\
21 & deg ($^\circ$)     & \textbf{PA}        & Position Angle of Multiple Components                    &              &              & 259.1        \\
22 & 0/100    & \textbf{IPDfmp}    & IPD fraction of multiple peaks                           & 1            & 2            & 0            \\
23 &          & \textbf{RUWE}      & Renormalized Unit Weight Error                           & 1.143        & 1.404        & 1.294        \\
24 & km/s     & \textbf{RV}        & Radial Velocity                                          &$-$2.42       & 32.47        &              \\
25 & km/s     & \textbf{e\_RV}     & Uncertainty in Radial Velocity                           & 1.51         & 0.19         &              \\
26 &          & \textbf{NSS}       & Gaia Non-Single Star tables flag                         & 0            & 0            & 0            \\
27 & mag      & \textbf{G}         & G Band Magnitude                                         & 14.667       & 11.510       & 18.372       \\
28 & mag      & \textbf{BP}        & BP Magnitude                                             & 16.769       & 13.081       & 21.391       \\
29 & mag      & \textbf{RP}        & RP Magnitude                                             & 13.329       & 10.287       & 16.704       \\
30 & mag      & \textbf{BP-RP}     & BP-RP Color                                              & 3.439        &  2.793       & 4.686        \\
31 & mag      & \textbf{V}         & V Band Magnitude                                         & 16.58        & 12.83        & 22.77        \\
32 & mag      & \textbf{r\_V}      & Reference for V Band Magnitude                            & Sil19        & Win15        & Die14        \\
33 &
   &
  \textbf{GDR3\_ID} &
  Gaia DR3 Source Identifier &
  \multicolumn{1}{r}{2745860763717896448} &
  \multicolumn{1}{r}{4996141155411983744} &
  \multicolumn{1}{r}{4996141155411984128} \\
34 &          & \textbf{HIP\_ID}   & Hipparcos Source Identifier                              &              &              &          
\enddata
\tablecomments{This table is available in its entirety in machine-readable form.}
\tablecomments{All references are listed in Appendix \ref{references}.}
\end{deluxetable*}

% \end{document}

\begin{figure}[h!] 
    \centering
    \includegraphics[scale = 0.65]{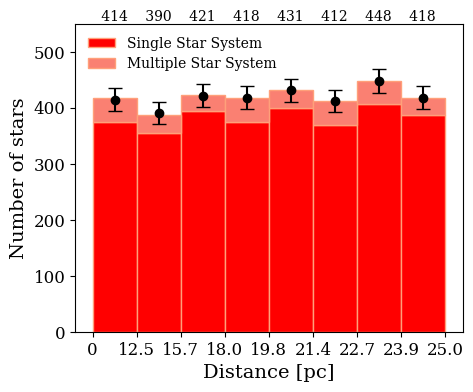}
    \caption{A histogram of the numbers of M dwarf primaries, and hence, systems, in RMSTAR through eight equal-volume shells of 8181 pc$^3$ from 0 to 25 pc. Dark red bars show star systems listed as single in RMSTAR and lighter red bars show multiple star systems in which at least one wide companion is included. Errorbars shown in black represent the square-root of the total number of systems in each volume bin, corresponding to Poisson counting errors. Numbers across the top x-axis indicate the total number of systems (single plus multiple) in each volume bin.}
    \label{fig:completeness}
\end{figure}

Given the importance of creating a complete list of the nearest M dwarfs for statistical investigations, it is important to assess the space densities of RMSTAR systems. Figure \ref{fig:completeness} maps the numbers of primaries in the sample per equal volume bin, where the 25 pc volume has been divided into eight shells with radii marked along the x-axis. Only M dwarf primaries were used to create this histogram, with these subdivided into single and multiple systems. The set of 305 stellar companions is described in $\S$\ref{sec:Companions}. It is clear that all of the volume bins have similar numbers of M dwarf systems, and thus the number of primary stars in the catalog is effectively complete out to 25 pc. In fact, the first bin from 0--12.5 pc is likely to be missing very few systems, if any, as described in some detail for the 10 pc sample by \cite{Henry_Jao_2024,Gonzalez-Payo_2026}. In contrast, the numbers of multiple systems are less consistent through the shells, with smaller populations in the more distant bins. As described below, this is because there are more companions at closer separations, and more distant stars will have fewer of these closer companions resolved by the spacecraft used to identify wide companions here.

\section{Confirmed Resolved Companions in \textit{Gaia} and \textit{\textit{Hipparcos}}} \label{sec:Companions}

\subsection{Identifying Lower Mass Companions within 30,000 AU of the M Dwarfs} \label{subsec:lowermass}

The key research effort of this paper is to identify the population of wide companions to the M dwarf primaries in RMSTAR. To identify resolved, gravitationally bound companions, we searched regions around each of the 3352 primaries in GDR3, GDR2, and \textit{Hipparcos}. Thus, we define ``wide'' companions as those resolved by, and that have astrometric solutions from, \textit{Gaia} or \textit{Hipparcos}. We note that although the \textit{Hipparcos} catalog was searched for wide companions, none were recovered that were not also in GDR3 or GDR2.

To reveal wide companions, we performed a radial search of each M dwarf primary out to a projected separation of 30,000 au, as described in $\S$\ref{subsubsec:highermass}. Some companions do not have complete astrometric solutions in GDR3, so we also searched RMSTAR primaries out to 30,000 au for companions in GDR2 and \textit{Hipparcos}, with no restrictions on parallax because some may have astrometric solutions placing them beyond 25 pc. If the primary M dwarf is within 25 pc but its companion is verified to be physical, even if its parallax places it beyond 25 pc, the entire system is still included in RMSTAR. This is the case for only three companions --- RMS 0621+1554B, RMS 1719-2949B, and RMS 1047+4026B.

\begin{figure*}[t]
    \centering 
    \includegraphics[scale = 0.38]{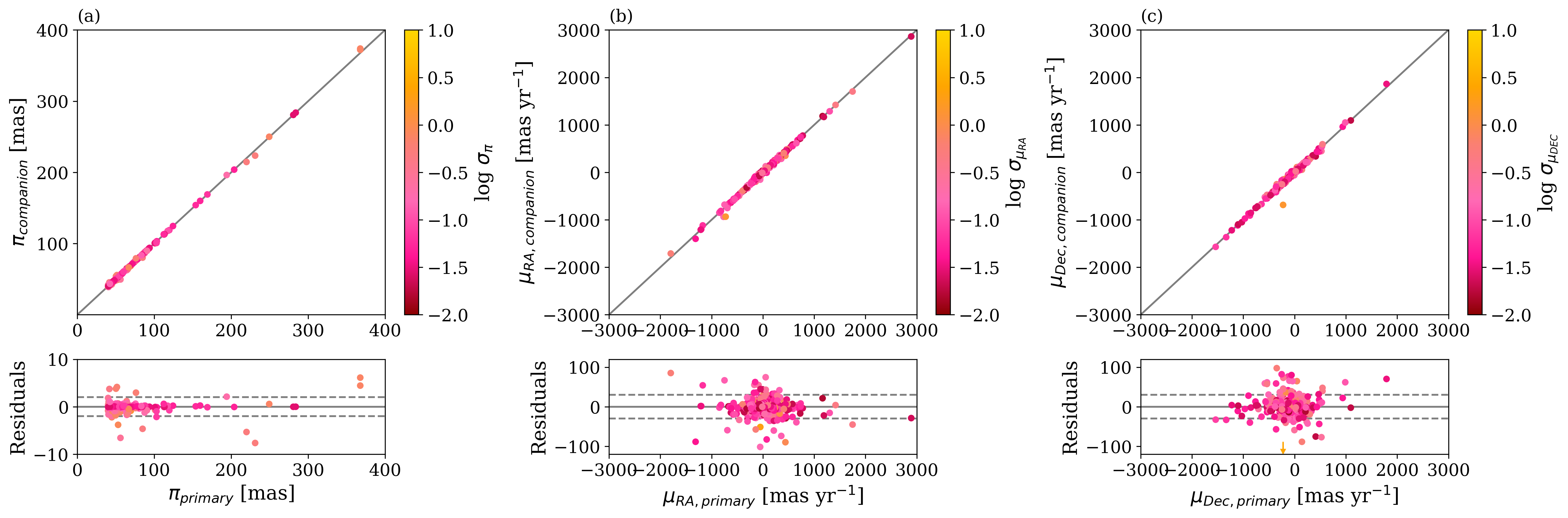}
    \caption{Comparison plots of astrometric parameters for each multiple in RMSTAR. Parallax, proper motion in right ascension, and proper motion in declination, respectively, are plotted for each system, with primary stars' values on the x-axis and the companions' values on the y-axis. For stars with more than one companion, a primary star's parameters are plotted against each companion. Points are colored by the larger error of each pair for each parameter. Solid black one-to-one lines are drawn on each plot. The dashed lines in the residual plots are drawn at $\pm$2 mas for parallax (panel a) and $\pm$ 30 mas\,yr$^{-1}$ for proper motions (panels b and c); these are thresholds for automatic inclusion of companions to RMSTAR. Companions beyond these cutoffs have been verified as physical using additional information (see $\S$\ref{subsec:widecomps} for more details).}
    \label{fig:onetoone}
\end{figure*}

Companions were determined to be gravitationally bound using match requirement thresholds of $\pm$2 mas in parallax and $\pm$30 mas\,yr$^{-1}$ in both right ascension and declination proper motions. The three panels of Figure \ref{fig:onetoone} illustrate these astrometric measurements for companions vs.~primaries, color-coded by the larger errors of the two objects' values and with residuals shown. Primary-companion pairs with offsets larger than these values are those that typically have one or both components that are multiples themselves, as confirmed by checking \textit{Gaia} multiplicity indicators (described in $\S$\ref{sec:unresolved}). These were checked individually and those determined to be true physical systems, based on locations in the color-magnitude diagram and/or other ancillary data (e.g., the Washington Double Star Catalog\footnote{The Washington Double Star Catalog (WDS) \citep{Mason_2001} is a key resource with its records of known close and wide companions to all types of stars, although it also contains non-gravitationally bound background sources that must be removed when using it to derive lists of true multiples.  The WDS is also a compilation of data from published and unpublished sources, and it is neither volume-limited nor volume-complete.}), were considered verified and are included in RMSTAR.  Such offsets typically happened if the primary or companion has a multiplicity flag in GDR3 or if the secondary is very faint. The single most extreme case is RMS 2227+5741AB, for which the offset in $\mu$$_{DEC}$ is 462.783 mas\,yr$^{-1}$; details are given in Appendix \ref{sec:worthyofnote}.

For multiple systems, components ``A'', ``B'', ``C'', etc., are assigned in rank-order of $M_G$ values. These letters are given in the RMSTAR ``cmp'' column; this column is left blank for single systems. Nine brown dwarf companions are noted as ``bd'' components rather than with a capital letter and are {\it not} included in the counts of stars in those systems in the {``num\_st''} column on the primary star line. The nine brown dwarf companions in RMSTAR have $M_G$ $>$ 17.8 and six of them can be seen in the lower portion of Figure \ref{fig:HRDs}c --- the other three do not have $BP$ values; these are labeled ``bd'' in the ``cmp'' column. We note that more are known to exist that are not included in this version of RMSTAR because they are fainter than the \textit{Gaia} magnitude limit. It should be noted that secondaries (and of course primaries) without astrometric data are not yet included in RMSTAR. Thus, even if an object is a likely companion, without any parallax in GDR3, GDR2, or \textit{Hipparcos}, membership and companionship cannot be confirmed.

\subsection{Characterizing Wide Companions to M Dwarfs}\label{subsec:widecomps}

\textbf{Among the 3352 M dwarf systems in the current version of RMSTAR, 3061 are singles, 277 are binaries, and 14 are triple systems, when considering only wide companions. The suite of companions to RMSTAR primaries contains 305 M dwarf companions and nine brown dwarf companions.}

All projected separations and position angles for the wide multiples included in RMSTAR are from GDR3, regardless of the sources of their parallaxes. Hence, there must be discrete entries for both objects in a pair in GDR3. If the parallaxes and proper motions for a primary and/or its companion are from a source other than GDR3, we find the matching source in GDR3 by right ascension, declination, and $G$ magnitude. In the RMSTAR systems table (Table \ref{tab:RMSTAR}), separations are given in arcseconds in column ``sep'' and position angles are given in degrees using north as 0$^{\circ}$ through east at 90$^{\circ}$ in column ``PA.'' All separations are listed as the angular separation ($\rho$) between A and a given companion, this information can be found in column ``map.'' The smallest angular separations reached are dictated by the angular resolution limit of the \textit{Gaia} telescope, whereas the largest separations correspond to our 30,000 au cutoff for the companion search. In this version of RMSTAR, separations range from 0\farcs46 (RMS 1937+3147AB) to 1127\farcs34 (RMS 0534+5112AB).

\begin{figure*}
    \centering
    \includegraphics[scale = .45]{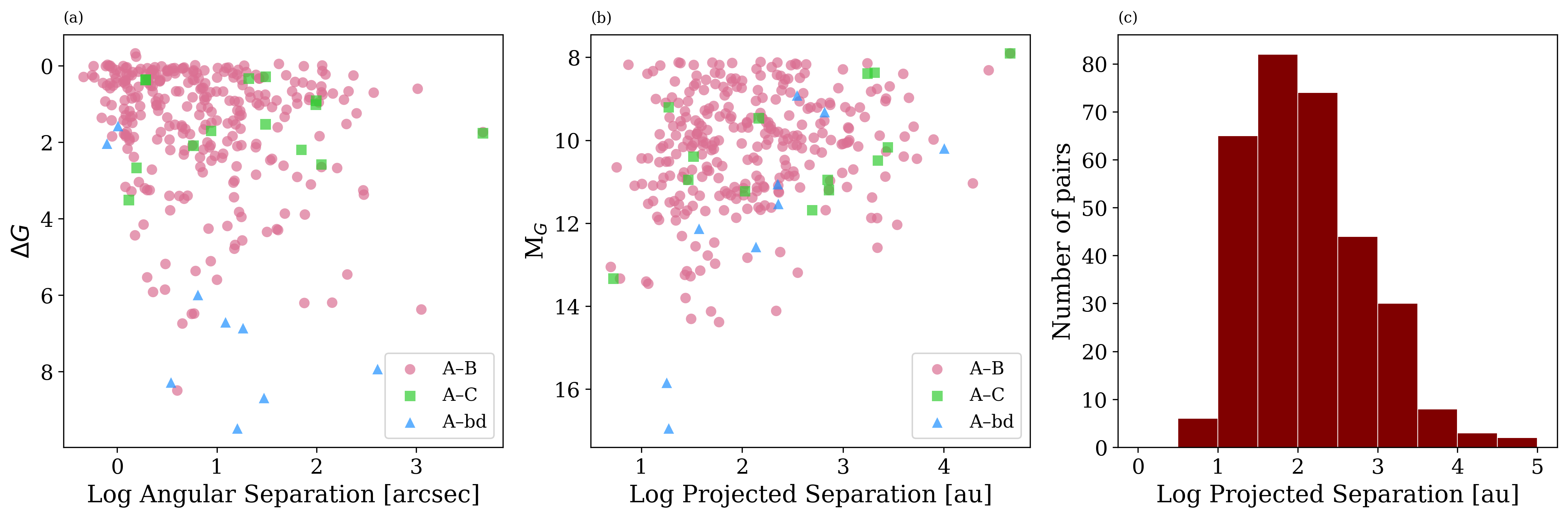}
    \caption{Three ways of evaluating the wide companion population of M dwarfs in RMSTAR are shown. Points indicate data for pairs using the absolute magnitudes and separations between the primary and each companion, with separations between A and B shown as pink circles, separations between A and C shown as green squares, and separations between A and a brown dwarf companion shown as blue triangles. (a) $\Delta$$G$ vs.~log projected separation (arcseconds) is a purely observational schematic that illustrates the observational biases. For each system, $\Delta$$G$ is calculated between the primary star and each companion. (b) Primary star $M_G$ vs.~log projected separation in astronomical units (au) places the systems into physical context. (c) Histogram of projected separations in log au reveals a peak in the distribution at $\sim$30 au. The decrease in the number of systems toward smaller separations is due to observational biases, whereas the decrease in the number of systems toward larger separations is astrophysical --- few companions are found at large separations.}
    \label{fig:logsepHist}
\end{figure*}

The three panels of Figure \ref{fig:logsepHist} provide various analyses of the wide companion populations, where the x axes are in log space to span large ranges in separations.  

Panel 6a is purely observational, showing the magnitude differences in the \textit{Gaia} $G$ band vs.~separations in arcseconds. Many companions have $\Delta$$G$ $<$ 2, with a smattering of companions at larger magnitude differences up to $\Delta G\approx$9.5. The classic trend of higher sensitivity at larger separations is evident, with a thin triangular gap on the left side of the plot where high-contrast, close companions are not detected. Separations peak near 1\arcsec~and there is a steep decrease in the number of companions resolved at smaller separations --- this is an observational bias due to the resolution limits of \textit{Gaia} and \textit{Hipparcos}, resulting in many companions with $\rho \lesssim 1$\arcsec~not being detected. Although many RMSTAR primaries have multiplicity flags in GDR3 indicating unresolved companions, these unresolved close companions are not included in this version of RMSTAR.

Panel 6b compares physical attributes of the systems, using $M_G$ vs.~separations in au, where the projected separations are simply calculated by multiplying the angular separations by the inverse of the primary stars' parallaxes. Several important trends are evident. First, companions with $M_G$ = 8.0--10.0 fill the separation range from 20 au--5000 au. Second, intrinsically fainter companions are found at generally smaller separations, as evidenced by the empty lower right corner of the plot. This is an astrophysical, not observational, effect, in which separations narrow for lower mass, fainter companions. Third, there are very few companions detected at separations $<$ 10 au, reflecting the limits of the spacecraft. Fourth, there are only three companions at separations larger than 10,000 au (RMS 0534$+$5112, RMS 2045$-$3120, and RMS 2100$-$4131), indicating that such wide binaries rarely form or are destroyed over time, as previously found by \citet{Weinberg_1987}, \citet{Binney_2008}, and \citet{Jiang_2010}. Fifth, the 23 systems with primaries having $M_G$ $>$ 12.0 is a modest number because only a few low mass M dwarfs have wide companions; note that there are only two such systems with primaries fainter than $M_G$ = 15.0 (RMS 1520$-$4422 and RMS 2045$-6$332) and both of those companions are brown dwarfs. Overall, panels (a) and (b) indicate that at close separations, primaries' fluxes overpower fainter companions that are consequently missing from RMSTAR, while there are very few companions at the widest separations.

Panel 6c is a histogram of panel 6b, showing the separation distribution between primaries and companions. The histogram peak at $\sim$30 au (log separation 1.5) is due to observational bias because of \textit{Gaia's} resolution limits, wherein companions with separations closer than the distance of the Sun to Neptune in our Solar System are not detected. We have speckle imaging and radial velocity surveys underway to recover these close companions. Nonetheless, a key result is that the number of companions decreases steadily from 30 au to 30,000 au. In fact, the M dwarfs with the widest companions are usually in triple systems, with dynamics and/or extra mass allowing them to hold onto companions at  larger separations, as described by \citet{Law_2010}, \citet{Reipurth_2012}, and \citet{Tokovinin_2017}.

%--------------------------------------------------------------------------------

\section{Candidate Unresolved Companions}\label{sec:unresolved}

While RMSTAR 2026 contains only wide companions at separations $>$ 0\farcs46 found in GDR3, GDR2, and \textit{Hipparcos}, a plethora of companions exist at closer separations than resolved by these space-based platforms. High-resolution work such as speckle imaging surveys and radial velocity efforts have revealed many companions, and these are important to consider in the overall assessment of M dwarf systems, particularly because they provide insight to crucial orbital regions on the scale of our Solar System.  

While an exhaustive assessment of these close companions is planned for a future RMSTAR version, in large part because of our own ongoing surveys, here we take advantage of multiplicity flags in GDR3 and already-published speckle work from our group to assess what is yet to be revealed. We cross-matched RMSTAR primaries with the speckle imaging survey of nearby M dwarfs conducted by \cite{Vrijmoet_2022} which detected companions in the separation range 0.\arcsec024--2.\arcsec0, finding 66 stars with companions. The purpose of this cross-match was to obtain separations of M dwarf systems both resolved and unresolved by \textit{Gaia} or \textit{Hipparcos} to map out the detection parameter spaces for {\it Gaia's} RUWE and IPDfmp multiplicity parameters. 

We also cross-matched the RMSTAR catalog with The Ninth Catalog of Spectroscopic Binary Orbits (SB9) \citep{SB9}, finding 14 stars with companions. These are used to evaluate RV errors across the observational baseline of GDR3 ($\lesssim$ 1000 days) to reveal potential systems that were not resolved by \textit{Gaia} or \textit{Hipparcos}. The unresolved companions at the separations and orbital periods sampled by these efforts are not included in this version of RMSTAR, but are used as examples to map the parameter space yet to be explored, with an eye on the discovery of new, close secondaries.

\subsection{\textit{Gaia} RUWE}\label{subsec:RUWE}

Re-normalized unit weight error --- \texttt{ruwe} --- henceforth RUWE, is a parameter released in both GDR2 and GDR3. RUWE serves as a goodness-of-fit statistic and is formally given as an indicator of the reliability of an astrometric fit to a single source \citep{LL:LL-124} for the five- and six-parameter astrometry published in GDR2 and GDR3 \citep{Gaia_2023}. The unit weight error (UWE) is calculated by dividing the chi-squared of the astrometric fit by the number of good observations of the source minus the number of astrometric parameters, either 5 or 6, and then taking the square root of that value. Both UWE and RUWE serve as indicators of the quality of astrometric fits, but UWE has a strong dependence on color and magnitude. Thus, UWE is re-normalized to account for these dependencies to generate the more commonly used RUWE.

For a single star, a high quality astrometric solution will yield RUWE$\sim$1.0. Because astrometry involves precisely measuring the position of a source's photocenter, many external factors can affect the quality of the published astrometry, and consequently, increase RUWE. These factors include faint source magnitudes, crowded fields, limited numbers of observations, and unseen companions \citep{LL:LL-124}. RUWE is particularly useful in identifying unresolved multiple star systems, specifically in the case where an unresolved companion is within $\sim$1\arcsec~of the primary --- a single photocenter is poorly fit because there is an orbiting companion causing the photocenter to ``wobble'' about the system's center of mass, increasing RUWE. \cite{LL:LL-124} describes a ``natural breakpoint'' at RUWE = 1.4 for reliable astrometric solutions, above which it is commonly considered to be a sign of multiplicity warranting follow-up observations (\citealp{Belokurov_2020}, \citealp{Vrijmoet_2020}, \citealp{Ziegler_2021}), although there are exceptions of close multiple systems with RUWE $<$ 1.4 that have been resolved with ground-based observations \citep{Cifuentes_2025}.

\begin{figure*}[]
    \centering
    \includegraphics[scale = 0.43]{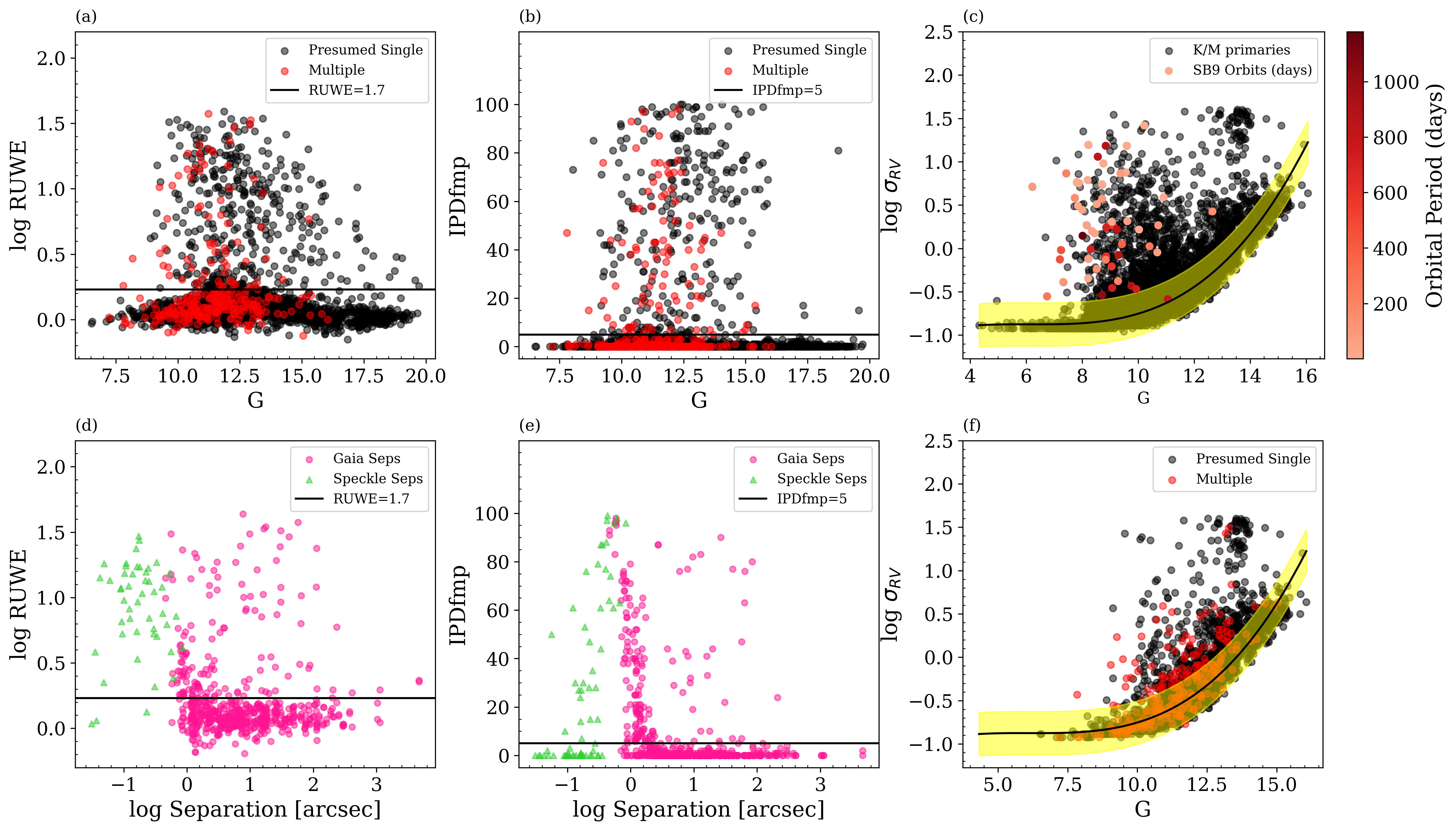}
    \caption{\textit{Gaia} companion indicators for stars in the RMSTAR catalog are shown. In each case, see the text in $\S\ref{sec:unresolved}$ for rationales for the black line cutoffs adopted.
    (a) shows log RUWE vs.~$G$ magnitude for all RMSTAR primaries, with presumed singles in black and known multiples in red. A horizontal black line is drawn at RUWE = 1.7.
    (b) shows IPDfmp vs.~$G$ magnitude for all RMSTAR primaries, with presumed singles in black and known multiples in red. A horizontal black line is drawn at IPDfmp = 5.
    (c) shows log RV error vs.~$G$ magnitude for all RMSTAR and RKSTAR primaries, with presumed singles in black and known multiples in SB9 \citep{SB9} in red, shaded by orbital period. A third-degree polynomial fit (Equation \ref{eq:RVerror}) is shown in black with a $\pm0.25$ log $\sigma$$_{RV}$ envelope in yellow around the best fit line.
    (d) shows log RUWE vs.~log separation in arcseconds for RMSTAR systems with companions detected by \textit{Gaia} or \textit{Hipparcos} in pink and companions detected by speckle imaging reported by \citet{Vrijmoet_2022} in green. A horizontal black line is drawn at RUWE = 1.7. 
    (e) shows IPDfmp vs.~log separation in arcseconds for RMSTAR systems with companions detected by \textit{Gaia} or \textit{Hipparcos} in pink and companions detected by speckle imaging reported by \citet{Vrijmoet_2022} in green. A horizontal black line is drawn at IPDfmp = 5.
    (f) shows log RV error vs.~$G$ magnitude for all RMSTAR primaries, with presumed singles in black and known multiples in red. The polynomial fit is the same as in (c).}
    \label{fig:6panel}
\end{figure*}

For the RMSTAR sample, we have defined a conservative cutoff of RUWE = 1.7 to separate singles and likely multiples. This cutoff is based on the distribution of points in Figure \ref{fig:6panel}a, which shows log RUWE vs.~$G$ magnitude values for all RMSTAR primaries, and is supported by the data shown in \ref{fig:6panel}d. There is a large mass of likely singles below RUWE = 1.7, whereas the multiples below this value have separations larger than a few arcseconds, beyond the limit of where an astrometric perturbation would be evident in the 34 months of data used for GDR3. 

We explore the behavior of RUWE values as a function of separation in Figure \ref{fig:6panel}d, where pink points are the wide binaries from \textit{Gaia} and \textit{Hipparcos} with separations from $\sim$0\farcs4 to more than 1000\arcsec, and green points represent multiples reported by \cite{Vrijmoet_2022}, who used speckle imaging to reveal companion at separations $\rho\lesssim 1$\arcsec. Note that RUWE values quickly increase near the convenient value of $\rho = 1$\arcsec. The RUWE = 1.7 cutoff is supported by these data because there are only three companions found with RUWE $<$ 1.7 --- two are at {\it very} small separations for which the astrometric perturbation is minimal and the third has components of similar brightness that also results in a minimal astrometric perturbation. Large RUWE values for wide binaries are caused by a third component within $\sim$1\arcsec~of one of the two stars, hence the pink points in the upper right of Figure \ref{fig:6panel}d. The 382 black points above the line at RUWE = 1.7 in Figure \ref{fig:6panel}a are RMSTAR primaries unresolved by \textit{Gaia} or \textit{Hipparcos} that we identify to be likely multiples. Some examples of stars that are flagged as having possible companions in RMSTAR are shown in Table \ref{tab:stars}, where TRUE indicates the M dwarf has a potential unresolved companion because the value is above the given adopted cutoff line in the corresponding panel of Figure \ref{fig:6panel}.

\begin{deluxetable}{lcccc}[h]\label{tab:stars}
%\tablecolumns{2}
\tablewidth{0pt}
\tablecaption{Examples of five stars in RMSTAR that show elevated RUWE, IPDfmp, RV error, and/or are in the NSS list.}
\tablehead{\colhead{RMS Name}  & \colhead{\texttt{f\_RUWE}} & \colhead{\texttt{f\_IPDfmp}} & \colhead{\texttt{f\_e\_RV}} & \colhead{\texttt{f\_NSS}}}
\startdata
RMS 0021$-$4605&  TRUE&FALSE&TRUE&FALSE\\
RMS 0058$-$2751&  TRUE&FALSE&TRUE&FALSE\\ 
RMS 0105$+$2829&  TRUE&FALSE&    &TRUE\\ 
RMS 0617$+$8353&  TRUE&TRUE&TRUE&FALSE\\ 
RMS 0707$+$6712&  TRUE&TRUE&FALSE&FALSE\\ 
\enddata
%\vspace{-0.8cm}
\tablecomments{In total, RMSTAR includes 382 primaries with high RUWE, 277 with high IPDfmp, 430 with high RV error, and 57 in the NSS list. Overall, 1089 unique sources have at least one of these flags.}
\end{deluxetable}

\subsection{\textit{Gaia} IPDfmp}\label{subsec:IPDfmp}

A second parameter from GDR3 that contains useful information on multiplicity is the Image Parameter Determination fraction of multiple peaks --- \texttt{ipd\_frac\_multi\_peak} --- henceforth IPDfmp. IPDfmp represents the percentage of \textit{Gaia} observations where more than one peak in the point spread function of the target was detected \citep{GaiaEDR3}. Multiple source detection depends on the spacecraft scan direction, and therefore is able to resolve multiple star systems during some scans but not others.

Generally, IPDfmp $>$ 2\% has been used to indicate a possible companion to the source (\citealp{Tokovinin_2023}, \citealp{Clark_2024}). For the RMSTAR sample we choose a more conservative cutoff of IPDfmp $>$ 5\% based on Figure \ref{fig:6panel}b, which shows the distribution of RMSTAR primaries' IPDfmp values vs.~$G$ magnitude, where black points represent presumed singles and red points are the wide multiples listed in RMSTAR. The large collection of points below this cutoff are likely singles (or very wide multiples), whereas the points above the line are multiples separated by $\sim$0\farcs4--2\arcsec\ that were resolved by \textit{Gaia} or \textit{Hipparcos} (red points) or are likely additional multiples not reported as separate sources in GDR3 (black points). The complementary plot in Figure \ref{fig:6panel}e explores IPDfmp vs.~separation, where pink points represent systems detected by \textit{Gaia} or \textit{Hipparcos} and green points are the speckle binaries reported by \citet{Vrijmoet_2022}. From right to left, IPDfmp values increase quickly at separations of a few arcseconds, then fall when companions are separated by only a few tenths of an arcsecond, where the point spread functions of the two stars blend together. Similarly to RUWE, some pairs at large separations still have IPDfmp $>$ 5\%; these have third components close to one of the stars in the wide pair. We identify 277 primaries (black points) in Figure \ref{fig:6panel}b having IPDfmp $>$ 5\% without companions resolved by \textit{Gaia} or \textit{Hipparcos}. Some examples of stars that are flagged as ``TRUE'' for having IPDfmp $>$ 5\% are listed in Table \ref{tab:stars}.

\subsection{\textit{Gaia} Radial Velocity Error}\label{subsec:RVerr}

Along with astrometric solutions and image evaluation, GDR3 also provides RV measurements and associated errors --- \texttt{radial\_velocity} and \texttt{radial\_velocity\_error}, respectively --- from spectroscopic observations for some sources. RVs are useful because they provide the third component of space motion that cannot be acquired with astrometry, and in the current context can be used to detect unseen companions. As it orbits, a secondary's gravitational pull causes changes in \textit{Gaia's} RV measurement over time, resulting in large errors reported in GDR3 because the values given are simply averages of individual sets of available measurements with standard deviation errors.

GDR3 only provides RV measurements for stars brighter than $G \approx$ 16, meaning that not all stars in the RMSTAR catalog have RV measurements. In order to improve our understanding of the impacts of companions on RV errors, we augment RMSTAR primaries with K dwarf primaries within 50 pc in RKSTAR (Johns et al.~submitted). This more than doubles the number of stars with RV errors that can be analyzed for trends.

As can be seen in Figure \ref{fig:6panel}c RV errors in GDR3 are strongly dependent on the apparent magnitudes of RMSTAR and RKSTAR primaries with RV measurements. Because fainter stars are more difficult to measure and have larger RV errors, we are unable to simply define a certain RV error that would indicate a close companion, as we did for RUWE and IPDfmp. Instead, we derive the third-order, least-squares, polynomial fit:

\begin{equation}\label{eq:RVerror}
\begin{split}
    log( \sigma_{RV}) = 0.00214 \cdot G^3 - 0.03920 \cdot G^2 \\
    + 0.23857 \cdot G - 1.35753
    \end{split}
\end{equation}

\noindent shown with a black curve in Figure \ref{fig:6panel}c as a cutoff. This curve was found by recursively fitting the RV errors for stars with given $G$ magnitudes, utilizing three iterations of 50$\sigma$ sigma-clipping (50$\sigma$ was chosen due to the large number of data points making 1$\sigma$ very small), and removing outliers to better fit the distribution of ``good'' RV errors. The stars below the best fit line have small RV errors and therefore good RV measurements, so we use the distribution below the fit to determine how high above the fit should be considered a poor quality RV measurement. In a by-eye analysis, we determine that stars with RV errors within an envelope of $\pm$0.25 in log $\sigma$$_{RV}$ of the fit have acceptable RV errors, while those above the upper margin are considered to have large RV errors for their $G$ magnitudes. To support the margins used on the line of best fit, we utilized orbital parameters for binaries in SB9 \citep{SB9}, selecting those that have orbital periods $\leq$1200 days, matching the 34-month observational baseline used to generate GDR3 results. Stars with companions are over-plotted in red in Figure \ref{fig:6panel}c, and shaded by orbital period. Those with companions have large RV errors and consistently lie above the outlined yellow envelope. This is the region used to select M dwarfs in RMSTAR that may have unresolved companions based on their RV errors.

Figure \ref{fig:6panel}f shows the RV errors for all RMSTAR primaries with GDR3 RVs. There are 430 nearby M dwarfs without resolved companions in \textit{Gaia} and \textit{Hipparcos} found above the yellow envelope --- these potentially harbor hidden secondaries. Some examples of stars that are flagged as ``TRUE'' for having RV error above the yellow envelope are listed in Table \ref{tab:stars}. However, there are other causes of elevated RV errors, such as the activity level of the targeted star. Follow-up observations are needed to determine if there is a companion or if the large RV errors have other origins. We also acknowledge that 659 stars in RMSTAR that are bright enough to have RV measurements ($G\leq16$) in GDR3 do not have RV data available. This is likely due to the RV error being so large that the measurement was thrown out altogether. It is recommended that these stars with missing RVs also be investigated in follow-up observations. 

\subsection{\textit{Gaia} Non-Single Stars}\label{subsec:NSS}

Finally, there is a fourth source in GDR3 to consider for evidence of unresolved companions --- \textit{Gaia's} list of Non-Single Stars (NSS). The four NSS tables in GDR3 contain orbital models and other information for some astrometric, spectroscopic, and eclipsing multiple star systems \citep{Babusiaux2023}. If a RMSTAR primary is in NSS, the NSS flag is set to ``TRUE'' based on the NSS value being non-zero, and is typically accompanied by other flags for the three other multiplicity parameters discussed above. Additionally, a ``False'' flag does not mean that no companion is present, but that there is no orbital solution in the NSS. There could, in fact, be a wide, faint companion beyond the \textit{Gaia} detection limit or temporal coverage \citep{Tokovinin_2023, Cifuentes_2025}. Some examples of these flags are given in Table \ref{tab:stars}. RMSTAR primaries flagged as multiple stars in the NSS have {\it not} been thoroughly analyzed for this version of RMSTAR, but are high priority targets in our follow-up surveys. 

\subsection{Summary of \textit{Gaia} Indicators}\label{subsec:gaiaindsumm}

The four parameters discussed probe different ranges of separations between primaries and companions, with some overlap. In general, these indicators can be used as flags to create subsamples of RMSTAR M dwarfs that are high priority for follow-up observations. The main RMSTAR Table includes information that can be used to pinpoint likely unresolved multiples via each of the four GDR3 indicators. If RUWE $\ge$ 1.7 in the ``RUWE'' column or IPDfmp $>$ 5 in the ``IPDfmp'' column, the star is flagged with ``TRUE'' in the \texttt{f\_RUWE} and/or \texttt{f\_IPDfmp} columns, respectively. For RV error, potential multiples are flagged with ``TRUE'' in the \texttt{f$\_$e$\_$RV} column if they fall above the yellow envelope in Figure \ref{fig:6panel}f. For NSS, we simply list ``TRUE'' in the \texttt{f\_NSS} column if the star is present in any of the four NSS tables in GDR3 and ``FALSE'' if it is not. If one of these parameters is not available in GDR3, there is no entry for that flag (i.e. blank). Some examples of these flags are given in Table \ref{tab:stars}. Overall, 71.8\% of RMSTAR primaries have no companion indicator flags, while the remaining 28.2\% have one or more
indicators.

\begin{figure}[h]
    \centering
    \includegraphics[width=1\linewidth]{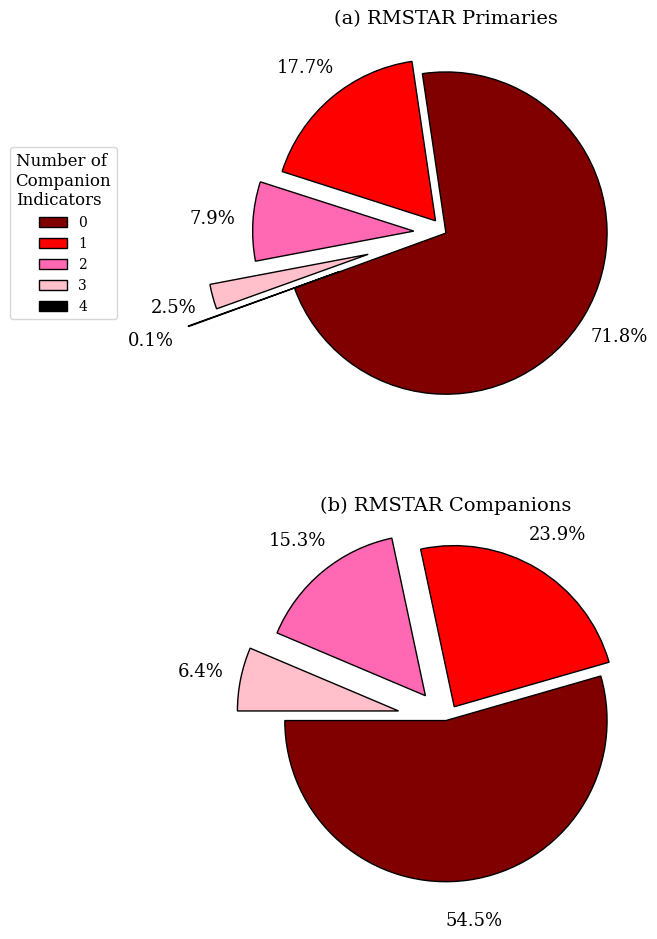}
    \caption{Pie charts summarizing the (a) 3352 RMSTAR M dwarf primaries and (b) 305 RMSTAR M dwarf companions, colored by the number of GDR3 parameters pointing to hidden multiplicity --- RUWE, IPDfmp, RV error, and NSS. 71.8\% of RMSTAR primaries have no companion indicator flags, while the remaining 28.2\% have one or more indicators. Two stars, RMS 1519-1245 and RMS 0754-2920, have all four. Of the known wide companions, 54.5\% have no companion indicator, while the remaining 45.5\% have one or more indicators.}
    \label{fig:piechart}
\end{figure}

Figure \ref{fig:piechart} summarizes the portions of RMSTAR (a) primaries and (b) companions with one or more indicators pointing to unresolved companions. We do not exclude any stars in this pie chart, even if a wide companion is known, as there may also be unresolved close companions that have yet to be discovered. For this reason we also include the secondary pie chart representing the companion indicators of RMSTAR companions, as these stars may also have unresolved secondaries.

These indicators, while useful, are not the final determinants of whether or not a star has a companion, and some should be used with particular caution. For example, primaries with very wide companions will not have indicator flags, and some stars with indicator flags may not have companions, e.g., active or fast-rotating single M dwarfs with high RV errors. There are also stars entirely missing values in GDR3, e.g., stars without astrometric solutions do not have RUWE values, and many faint M dwarfs do not have RV measurements. Overall, 28\% of the presumed single (that is, with no resolved companions present in \textit{Gaia} or \textit{Hipparcos}) M dwarf primaries in this version of RMSTAR have at least one multiplicity indicator. Among companions, nearly half (45\%) may have unresolved companions, pointing to a large population of hierarchical triples.

\section{Results}\label{sec:results}

\subsection{Key Attributes of the Nearby M Dwarf Population}\label{subsec:MR}

The population of the nearest M dwarfs can be characterized in several ways, and given that M dwarfs comprise three of every four stars in the solar neighborhood, the first focus of this paper is to build as complete a sample as possible for statistical results. The second focus of this paper has been to determine the wide stellar companion population to M dwarfs, a specific multiplicity rate that provides important clues about the formation of wide stellar binaries. 

The luminosity (LF) and mass (MF) functions for nearby stars serve as the building blocks for stellar populations throughout the Milky Way and other galaxies. Thus, a key RMSTAR effort is to provide accurate LF and MF determinations for a volume-limited sample, which can now be created because of \textit{Gaia's} astrometry and photometry that reaches all the way to the end of the stellar main sequence at 25 pc. Although RMSTAR 2026 does not include close companions, it is worth providing a first set of LF and MF for the nearest M dwarfs with likely multiples identified, and then build upon these results in future RMSTAR editions. Determinations of the wide companion multiplicity rate, LF, and MF for M dwarfs are given in the next three sections.

\subsection{Multiplicity Rate for Wide Companions}\label{subsec:MR}

To solve for the multiplicity rate (MR) of wide companions in RMSTAR systems, we simply divide the total number of 291 multiple star systems (277 binaries, 14 triples) by the total number of systems, 3352, and find a rate of 8.68$\pm$0.49\%, where the error represents the binomial distribution statistical error. We anticipate that this MR for wide companions is accurate because very few primaries are missing from the sample (see $\S$\ref{subsec:finalprims}), nor are there many wide (sep $>$ 30 au) stellar companions missing.

\begin{figure}
    \centering
    \includegraphics[width=0.99\linewidth]{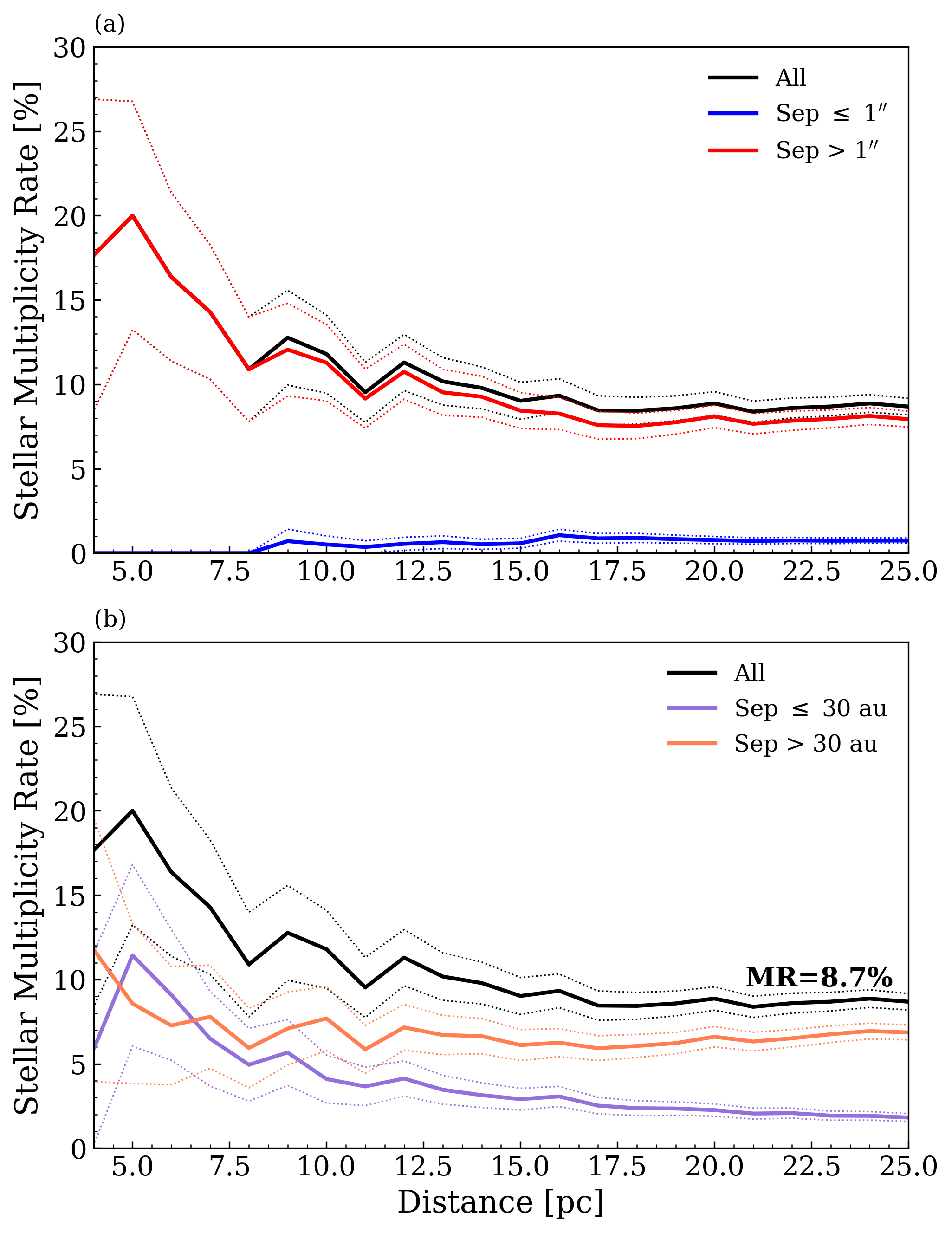}
    \caption{Cumulative stellar multiplicity rates (MRs) as a function of distance (in 1 pc bins) for 291 wide multiples among the total of 3352 RMSTAR systems. (a) shows the MR trends for different angular separations, with a solid black line for all systems, a solid blue line for systems with separations $\le$ 1\arcsec, and a solid red line for systems with separations $>$ 1\arcsec. (b) shows the MR trends for different projected separations, with a solid black line for all systems, a solid purple line for systems with separations $\le$ 30 au, and a solid orange line for systems with separations $>$ 30 au. Dotted colored lines represent binomial errors matched to the heavier MR lines. }
    \label{fig:MultRate}
\end{figure}

Figure \ref{fig:MultRate} shows the stellar MR for wide companions as a function of distance to the RMSTAR primaries for different projected angular (top, in arcseconds) and physical (bottom, in au) separations. At distances farther than $\sim$10 pc, where results are no longer dependent on small number statistics, the MR for systems with projected separations $>$ 30 au (red line, the distance from the Sun to Neptune in our Solar System) is flat out to 25 pc. Alternately, at projected separations $\leq$ 30 au, we find that the multiplicity rate steadily decreases to 25 pc, pointing to the discussed incompleteness for this edition of RMSTAR at separations less than $\sim$1\arcsec.

As shown in Figure \ref{fig:logsepHist}c, the companion separations of 0\farcs46 to more than 1000\arcsec~sampled in RMSTAR translate into physical units corresponding to a few au to more than 10,000 au (reminder: our wide companion cutoff for searches was 30,000 au). We expect that effectively all stellar companions with separations of 30--30,000 au have been found, given that the horizon of 25 pc places a companion at 30 au at 1.2$\arcsec$ is easily resolved with \textit{Gaia} unless the secondary is much fainter than the primary\footnote{The detection limit of a source by \textit{Gaia} is $G\simeq21$ corresponding to $M_G\simeq19$ at 25 pc, which captures the entire RMSTAR catalog and stars more than a magnitude fainter than the $M_G=17.8$ cutoff.}. Of course, these are {\it projected} separations, and possible adjustments include (1) adding a fraction to each separation measured to account for three-dimensional separations being compressed into two-dimensional measurements, and (2) adjusting for orbital mechanics given that companions spend most of their time at large separations where they move slowly in their orbits. However, given that previous work rarely makes these corrections, for direct comparisons we provide multiplicity rates as measured for the bins we consider complete: 4.53$\pm$0.36\% for 30--300 au, 2.00$\pm$0.24\% for 300--3000 au, and 0.33$\pm$0.10\% for 3000--30,000 au. The total MR for separations from 30--30,000 au of 6.86$\pm$0.44\%, corresponds to 1 in 15 M dwarfs having a stellar companion at these separations. The final 1.82\% of multiples represents the 61 systems with companions closer than 30 au.

Additional key facts gleaned from this work relating to the types of binaries seen within this MR include (1) M dwarf wide multiples trend toward equal brightness components ($\Delta$$G$$\sim$0, Figure \ref{fig:logsepHist}a), (2) more massive, intrinsically brighter, M dwarfs ($M_G\lesssim$12 mag) are far more likely to have companions than less massive M dwarfs (Figure \ref{fig:logsepHist}b), and (3) the companion population increases from 30,000 down to 30 au (Figure \ref{fig:logsepHist}c). Given that the companion population decreases for separations $\leq$ 30 au due to observational biases caused by the angular resolution limits of the spacecraft used to build the companion sample, we expect the peak in projected separations to shift to smaller separations when close companions are added to the sample, as illustrated by \cite{Winters_2019}. The additions to RMSTAR from speckle imaging and radial velocity surveys that explore the closer environs of M dwarfs will be companions at separations $\lesssim$30 au, and of course will yield a higher total MR.

\subsection{Luminosity Function Using $M_G$} \label{subsubsec:LF}

In order to create an accurate LF, it is necessary to have photometric magnitudes in the same band for each star in the sample. A natural photometric band in which to produce the LF is $M_G$, given its wide availability via \textit{Gaia} results. With the exceptions of one primary star and one companion star that are explained in $\S$\ref{sec:worthyofnote}, every star in the RMSTAR catalog has both a parallax and apparent \textit{Gaia G} magnitude that together yield $M_G$. The parallax measurements come from GDR3, GDR2, \textit{Hipparcos}, or ground-based astrometry from the literature, while all \textit{Gaia G} magnitudes are from GDR3.

The resulting LF is illustrated with the histogram in Figure \ref{fig:LFMF} (left). The histogram is stacked by subsample to show how the specific groups of M dwarfs contribute to the distribution, and the widths and heights of each of the LF bins are presented in Table \ref{tab:heights}. Eleven bins are used to match the 11 bins mapped for the MF, discussed below. RMSTAR primaries are shown in red and the wide resolved companions in blue. Also included in white are systems with joint photometry, in which an additional stellar component may be present. These joint entries are assigned based on our assessments of the \textit{Gaia} parameters discussed in $\S$\ref{sec:unresolved} --- any primary found to have at least one flag is plotted in white, although some stars flagged may not be multiples, e.g., fast rotators with high RV errors. Of course, additional multiple systems with companions at close separations not flagged by \textit{Gaia} parameters may also be included in the red and blue portions of the histogram. Finally, the grey portions outline additional M dwarfs removed from RMSTAR because they are not primaries, e.g., Proxima Cen. These are important to recapture for the LF to provide a complete picture of the M dwarf population in the solar neighborhood. The close multiples that may be found in any of the four groups represented in the LF would cause an individual entry to be brighter than either individual component. Such entries will split into (at least) two M dwarfs at fainter $M_G$ values and as future work characterizes these systems, the LF shown will shift to fainter magnitudes.

\begin{figure*} \label{fig:LFMF}
   \centering
   \includegraphics[scale = 0.4]{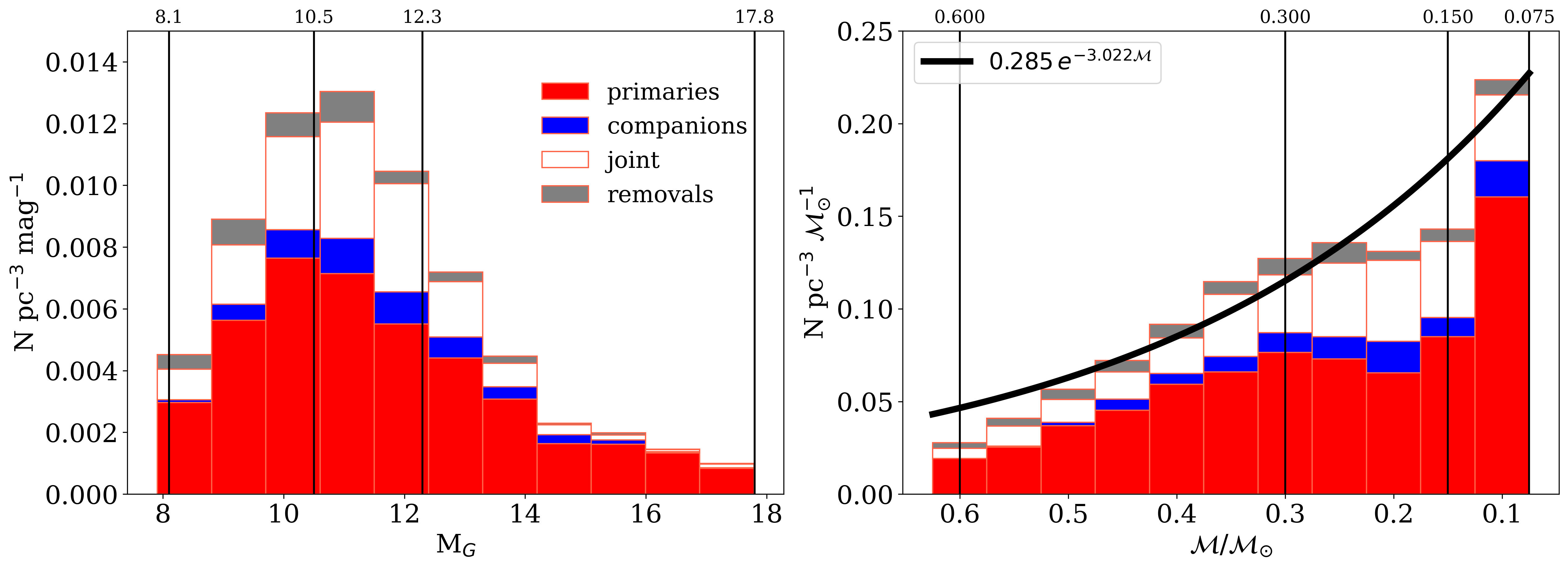}
   \caption{Luminosity function (LF, left) and mass function (MF, right) for a total of 3896 M dwarfs within 25 pc, including the 3352 primary M dwarfs in RMSTAR, their 305 wide stellar companions, plus 239 recovered from systems with higher mass primaries. Masses were calculated using the MLR at $M_V$ of T.~Johns et al.~(submitted); see $\S\ref{subsubsec:MF}$ for details on the methodology. Primaries are shown in red, resolved wide companions in blue, suspected unresolved multiples are in white, and M dwarfs with higher mass companions are in grey. Vertical black lines are drawn at $M_G$ values corresponding to factors of two in mass in both panels, from left to right respectively at $\mathcal{M}$ = 0.60, 0.30, 0.15, and 0.075 $\mathcal{M_{\odot}}$. The derived MF for M dwarfs in the solar neighborhood is shown with a bold black curve.}
   \label{fig:LFMF}
\end{figure*}

We find a peak in the luminosity function at $M_G = 10.8$. The vertical black lines in the LF mark factors of two in the wide range of masses spanned by M dwarfs.  As can be seen by comparing the locations of the lines in the LF to the corresponding factor-of-two lines in the MF in the right panel of Figure \ref{fig:LFMF}, the smallest range of masses from 0.075--0.15 $M_{\odot}$ corresponds to the largest range of brightnesses $M_G$ = 12.3--17.8, explaining the spread and consequent drop-off towards lower luminosities in the LF.

\subsection{Mass Function} \label{subsubsec:MF}

\subsubsection{Estimating Masses Using $M_V$} \label{subsubsec:Vmag}

To estimate the masses of the M dwarfs in RMSTAR, we use the mass-luminosity relation (MLR) in the $V$ band from T.~Johns et al.~(submitted), which is derived using dynamical K dwarf masses from various literature sources and dynamical M dwarf masses from \citet{Benedict_2016}. We use a combination of direct $V$ measurements and provide a conversion from \textit{Gaia} $M_G$ to $M_V$ in the next section for M dwarfs without direct $V$ measurements. Fortunately, more than 1,900 M dwarfs in RMSTAR have $V$ from RECONS efforts to gather $VRI$ photometry for red dwarfs within 25 pc using the ARCSAT 0.5\,m at Apache Point Observatory in the north and the SMARTS 0.9m at Cerro Tololo Inter-American Observatory in the south, in large part because of a focused effort described by \citet{Silverstein_2019}, who provided details about the photometry program and reduction techniques. The relevant $V$ photometry values for RMSTAR members secured in that effort, supplemented with other sources also described by \citet{Silverstein_2019}, are given in the RMSTAR master Table in columns labeled ``V'' for the value and ``ref'' for the photometry source. Overall, there are 1924 RMSTAR M dwarfs for which $V$ magnitudes are available, 1168 of which are used to derive $M_V$ values that can be converted to masses for the MF as described in $\S$\ref{subsubsec:Vmag_calc}.

\subsubsection{Conversions of $M_G$ to $M_V$ for K and M Dwarfs} \label{subsubsec:Vmag_calc}

For stars that do not have measured $V$ magnitudes, we create a relationship to convert from \textit{Gaia} $M_G$ to $M_V$ by utilizing 1323 stars with both $G$ and $V$ magnitudes and their corresponding parallaxes. To support the conversion fit at the bright magnitude limit for M dwarfs at $M_G$ = 8.1, and to provide a uniform conversion for both K and M dwarfs, we augment the 1168 M dwarfs from RMSTAR with 155 K dwarfs from RKSTAR as bright as $M_G$ = 5.5. To make sure the conversion is as accurate as possible, we use only presumed individual K and M dwarfs with no companion indicators, as the presence of a close companion could contaminate the primary star photometry. Additionally, we perform a preliminary fit and use one iteration of 30$\sigma$ sigma-clipping to remove any stars that have poor or blended photometry from background sources.

\begin{figure} 
    \centering
    \includegraphics[scale = .65]{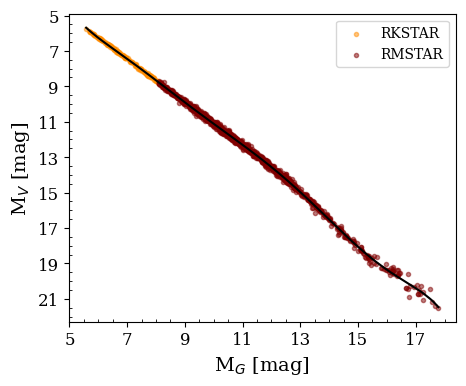}
    \caption{$M_V$ vs.~$M_G$ relation for 1168 primary M dwarfs from RMSTAR (red) and 155 primary K dwarfs from RKSTAR (orange) that have both $V$ and $G$ measurements and parallaxes. The black line shows the selected 7$^{th}$-degree least-squares polynomial fit to the data.}
    \label{fig:MVvsMG}
\end{figure}

Figure \ref{fig:MVvsMG} shows the resulting relation that spans 5.5$\leq M_G \leq$ 17.8 and 5.7$\leq M_V \leq$21.5, with a 7$^{th}$-degree least squares polynomial fit having the form of Equation \ref{eq1}; coefficients for the fit are given in Table \ref{tab:coeffs}. Polynomials were fit with degrees of 1 through 10, and the 7$^{th}$-degree polynomial was selected because the average deviation of the best fit line from the dataset did not substantially improve for higher orders.

\begin{equation}     \label{eq1}
\begin{split}
    M_V = C_7 \ast M_G^7
    + C_6 \ast M_G^6
    + C_5 \ast M_G^5 
    + C_4 \ast M_G^4 \\
    + C_3 \ast M_G^3
    + C_2 \ast M_G^2 
    + C_1 \ast M_G 
    + C_0
    \end{split}
\end{equation}

\begin{deluxetable}{cc}[h]
%\tablecolumns{2}
\tablewidth{0pt}
\tablecaption{Coefficients in the Equation \ref{eq1} polynomial to estimate $M_V$ from $M_G$ \label{tab:coeffs}}
\tablehead{\colhead{Coefficient}  & \colhead{Value}}
\startdata
    $C_7$ & $+$0.000013832468  \\ 
    $C_6$ & $-$0.001085459656  \\ 
    $C_5$ & $+$0.035538756249  \\ 
    $C_4$ & $-$0.628958956779  \\
    $C_3$ & $+$6.498424582342  \\ 
    $C_2$ & $-$39.209614241954  \\ 
    $C_1$ & $+$129.200435014192 \\ 
    $C_0$ & $-$175.505733996809 \\ 
\enddata
\vspace{-0.8cm}
\tablecomments{...}
\end{deluxetable}

\subsubsection{Mass Function Result} \label{subsubsec:MFresult}

With $M_V$ values in-hand, we estimate masses using the new MLR in T.~Johns et al.~(submitted) and derive the MF for RMSTAR M dwarfs shown in the right panel of Figure \ref{fig:LFMF}. As with the LF, the histogram is stacked by subsample, and the same color coding is used. The widths and heights of each of the MF bins are given in Table \ref{tab:heights}. The same details relating to multiples apply for the MF as for the LF: (1) joint systems are those with at least one \textit{Gaia} parameter indicating a possible unresolved companion, as discussed in $\S$\ref{sec:unresolved}, (2) some systems flagged may not, in fact, be multiples, and (3) additional stellar components may be present for entries not flagged by \textit{Gaia} parameters in any of the four subsamples. For unresolved multiples, brighter $M_V$ values would generate larger masses, and when split, a single entry in the MF shown would convert to two (or more) entries at lower masses. Thus, as with the LF, the entire distribution will shift to the right, in this case to lower masses, via future work that characterizes the close stellar companions. Nonetheless, a key result is that by examining the red portion of the histogram alone --- the presumably single primaries, nearly all of which will turn out to be only single stars --- the MF rises towards the end of the main sequence. Fitting an exponential to the histogram for the full suite of objects with masses considered here, we find the MF to follow the functional form:

\begin{equation}
     \xi(\mathcal{M}) = Ce^{-\alpha\mathcal{M}}
\end{equation}

\noindent where $\xi(\mathcal{M})$ is the number of M dwarfs per pc$^3$ per mass bin, C$=$0.289\,pc$^{-3}$, $\alpha$$=$3.040, and $\mathcal{M}$ is the mass.

\begin{deluxetable}{cc|cc}[h]

%\tablecolumns{2}
\tablewidth{0pt}
\tablecaption{Luminosity Function (LF) and Mass Function (MF) bin widths and heights as seen in Figure \ref{fig:LFMF}.} 
\label{tab:heights}
\tablehead{\multicolumn{2}{c}{\textbf{Luminosity Function}} & \multicolumn{2}{c}{\textbf{Mass Function}} \\ \colhead{Bin Width}  & \colhead{Bin Height} & \colhead{Bin Width} & \colhead{Bin Height} \\ \colhead{$M_G$}  & \colhead{[N pc$^{-3}$ mag$^{-1}$]} & \colhead{[$\mathcal{M}/\mathcal{M}_{\odot}$]} & \colhead{[N pc$^{-3}$ $\mathcal{M}^{-1}$]}}
\startdata
7.9--8.8  & 0.00451&0.625--0.575&0.02902\\
8.8--9.7  & 0.00891&0.575--0.525&0.04033\\
9.7--10.6 & 0.01232&0.525--0.475&0.05561\\
10.6--11.5& 0.01304&0.475--0.425&0.06906\\
11.5--12.4& 0.01045&0.425--0.375&0.08770\\
12.4--13.3& 0.00719&0.375--0.325&0.11275\\
13.3--14.2& 0.00446&0.325--0.275&0.12803\\
14.2--15.1& 0.00229&0.275--0.225&0.14087\\
15.1--16.0& 0.00189&0.225--0.175&0.13781\\
16.0--16.9& 0.00146&0.175--0.125&0.15584\\
16.9--17.8& 0.00099&0.125--0.075&0.22521\\ 
\enddata
\vspace{-.8cm}
%\tablecomments{...}
\end{deluxetable}

Because these masses are estimated from the MLR rather than measured directly for each star, it is possible that the heights of some of the mass bins are shifted slightly from the true values. For example, as shown in the color-magnitude diagrams of Figure \ref{fig:HRDs}, the main sequence has significant width in luminosity, when either the observational $M_G$ values shown are considered or the $M_V$ values used for the MLR conversions to masses. Thus, direct one-to-one derivations of $M_V$ values to masses are estimates, although we believe that there are not likely to be large systematic shifts in sub-populations through the M dwarf regime.  In addition, there are relatively few very low mass stars with accurate dynamical masses, so Johns et al.~(submitted) interpolated masses down at the very end of the main sequence in the 0.075--0.080 $\mathcal{M}_{\odot}$ interval. This, combined with the lack of masses in the 0.080--0.100 $\mathcal{M}_{\odot}$ interval, makes the fit difficult to constrain. Therefore, the lowest mass bins in particular may actually contain more or fewer stars depending on how masses are estimated. Regardless, larger shifts are expected when all close companions have been resolved and accounted for in both the LF and MF, and we are working on improving the MLR at the very lowest masses, so future improvements will be forthcoming. Overall, small discrepancies in estimated masses that cause bumps and dips in the heights of mass function bins should not dictate the fit that is used to describe the distribution. There is no obvious physical explanation as to why the mass function for M dwarfs would change for specific mass bins, so we derive a single, unbroken, exponential equation that adequately describes the mass function of M dwarfs. 

%--------------------------------------------------------------------------------

\section{Discussion} \label{sec:Discussion}

This RMSTAR 2026 edition includes 3352 carefully vetted M dwarf systems and provides a major step forward in our understanding of the smallest stars. The lists of primaries and wide secondaries are not yet quite complete, although we anticipate that only a few percent at most of these two categories of members are missing, as discussed below. There are likely to be a few entries that are not M dwarfs at all because they are young brown dwarfs or brown dwarf multiples boosting sources to brighter than the $M_G$ = 17.8 cutoff. Again, only a few such entries are anticipated. Significant updates to future RMSTAR editions will be the result of adding many close companions as characterization efforts continue. 

In the next sections we discuss caveats related to RMSTAR membership, compare the current RMSTAR version to the latest 25 pc catalog in the {\it Catalog of Nearby Stars} series, and compare the multiplicity rate, LF, and MF to previous efforts. We wrap up with a brief discussion of a partner paper in this series for the sample of K dwarfs within 50 pc by T.~Johns et al.~(submitted), known as RKSTAR, and provide a final set of conclusions based upon RMSTAR 2026.

\subsection{RMSTAR 2026 Caveats}

\subsubsection{Missing Primaries}

There are at least four reasons why primaries may be missing from RMSTAR 2026, each outlined briefly here.

\textbf{No Astrometric Solution Available} --- There is a small number of M dwarf systems not captured in RMSTAR because there are no parallaxes yet available in GDR3, GDR2, \textit{Hipparcos}, or from ground-based programs. Of the 3352 systems in RMSTAR, there are currently 132 with parallaxes not from GDR3, corresponding to 3.9\% of the sample. These are typically close multiples for which the 34 months of data used to generate GDR3 results did not yield a solution, but for which longer-term astrometry like that of the RECONS program \citep{Henry_2018} does yield a reliable parallax. 

An additional statistical check on the number of missing systems can be made using the 10 pc sample of M dwarfs focused on by the RECONS team. As described by \cite{Henry_Jao_2024}, GDR3 is missing 22 of the 269 systems within 10 pc. Of these, 194 are RMSTAR M dwarf systems of which 8 (4.1\%) are close multiples that do not have \textit{Gaia} parallaxes. These multiples only have parallaxes because of targeted efforts, which is also true for many M dwarf systems between 10 and 25 pc, so the portion of missing systems is lower than 4\%. In addition, inspection of Figure \ref{fig:completeness} shows that the overall density of primaries is consistent throughout the 25 pc volume, so there is no significant distance-dependent effect on the sample due to missing systems. Thus, overall we predict that there will be (far) fewer than 100 RMSTAR systems remaining to be added to the current sample. Many of these systems will likely be included in GDR4.

{\bf Parallaxes that are Too Small} --- As with any large statistical sample, borderline cases must be considered, in this case for the defining distance measurement for RMSTAR membership. There are likely to be a few stars with parallaxes just less than the adopted 40 mas cutoff that will be added to the sample when additional data result in parallax revisions. All RMSTAR entries will be updated with GDR4 parallaxes (anticipated on 02 December 2026), at which point many of these ``just misses'' will join the sample.

{\bf Overluminous Multiple M Dwarfs} --- Unresolved M dwarf multiples may have sufficient extra flux in the \textit{Gaia} $G$ band that they are boosted to magnitudes brighter than the adopted K/M dwarf boundary of $M_G$ = 8.1. The overluminosity of these systems will place them in the K dwarf regime and they will not be in RMSTAR. Fortunately, the \textit{Gaia} multiplicity parameters can be used to identify likely unresolved systems among these ``K dwarfs'', and once they are resolved, revised $M_G$ values will be derived for the individual stars and systems with primaries having $M_G$ = 8.1 or fainter will be added to RMSTAR.

{\bf Overluminous Young M Dwarfs} --- As with unresolved multiples, there will be a small number of young M dwarfs within 25 pc that are brighter than the adopted K/M dwarf boundary of $M_G$ = 8.1. One example is the AU Mic triple system, in which the primary has $M_G$ = 7.906. We include this system in RMSTAR as a special case because there is no doubt that the primary is an M dwarf, and efforts are underway to identify additional young M dwarfs masquerading as K dwarfs. Determining accurate surface gravities would allow such stars to be identified, as stars young enough to emit sufficient extra flux to be lifted above the cutoff will be larger in size and have lower gravities. Based upon comprehensive survey work carried out on a volume-complete sample of K dwarfs \citep{Carrazco_2026}, it is estimated that $\sim$4\% of stellar systems in the solar neighborhood are younger than 1 Gyr, based on the presence of the Li~{\sc i} $\lambda$6708\,{\AA} doublet. Reasonably applying the same rate to the population of M dwarf systems within 25 pc implies the presence of $\sim$150 young M dwarfs in RMSTAR. However, only a small fraction of this number will fall in the upper range of M dwarf masses and thereby be sufficiently overluminous to be lifted above the cutoff, so we anticipate perhaps a few dozen at most will eventually be added to RMSTAR.

\subsubsection{Extra Primaries}

{\bf Parallaxes that are Too Large} --- As with the borderline cases that have parallaxes too small to make it into RMSTAR, there will be a modest number of stars that may not really be closer than 25 pc because their measured parallaxes are (slightly) too large. It is straightforward to assign entries to a suspect list by selecting those stars with parallaxes within 3$\sigma$ of 40 mas; the result is a set of 32 systems. Many of these will be swapped with M dwarf systems currently having parallaxes just shy of 40 mas, so the overall net effect of these swaps on the total number of RMSTAR systems, as well as the resulting LF and MF, is likely to be minimal.

{\bf Brown Dwarfs Masquerading as M Dwarfs} --- When young brown dwarfs land on the lower end of the main-sequence in the color-magnitude diagram, they are located at brighter magnitudes than the $M_G$ = 17.8 cutoff for the hydrogen-burning limit. This means that there are probably a few young brown dwarfs in RMSTAR that appear to be stars but are still cooling, and will ultimately slide to fainter magnitudes than the cutoff. In addition, there are likely a few brown dwarf-brown dwarf binary systems that are overluminous and masquerade as M dwarfs on the main sequence, such as RMS 0053-3631, RMS 1305-2541, and RMS 2135+7312 that are flagged with Gaia multiplicity parameters in GDR3. Additional clues that such objects are not stars can be gleaned using detections of lithium \citep{Rebolo_1992,Magazzu_1993}, space kinematics, near-infrared colors, and surface gravities inconsistent with very low mass stars.

\subsubsection{Missing Wide Companions}\label{subsec:widemissed}

{\bf No Astrometric Solutions Available} --- Several of the caveats for primaries apply to wide companions as well. The most likely of these is that RMSTAR is missing a few wide companions because the companions themselves are close multiples without astrometric solutions. This set of missing companions is somewhat ameliorated by our use of GDR2 and \textit{Hipparcos} to recover some secondaries. On the other hand, issues related to parallaxes that are too large or too small are minimized because if a companion has an erroneous parallax larger than 40 mas, it is not included unless the primary is in RMSTAR. If a companion has a parallax slightly less than 40 mas, as long as the primary makes all cuts and the companion's proper motions and parallax are nearly the same, it is included. Overluminous companions are similarly linked to their primaries, and each suspected wide companion that may appear to be too bright is verified to have astrometry consistent with the primary, and if so, is included in RMSTAR. As with the primaries, missing companions will be sorted out with GDR4 results. 

{\bf Missing Brown Dwarf Companions} --- RMSTAR reaches stars all the way to the end of the stellar main sequence at $M_G$ = 17.8 to the horizon at 25 pc, where the smallest star has $G$ = 19.8. However, \textit{Gaia's} faint limit of $G$ $\approx$ 21 does not reach many of the coolest brown dwarfs throughout much of the 25 pc volume of RMSTAR. For example, one of the nearest solitary brown dwarfs in GDR3, WISE 1506+7027, has $G$ = 20.2 and parallax 193.94 mas, yielding $M_G$ = 21.6. This object would have $G$ = 23.6 at 25 pc, a few magnitudes fainter than \textit{Gaia's} limit, and this is certainly not the intrinsically faintest brown dwarf known. Thus, RMSTAR 2026 should be considered to be a reliable source for wide stellar companions, but only touches upon the brown dwarf companion population, with a mere nine representatives in the current version.

\subsection{RMSTAR Comparison to CNS5}\label{subsec:CNS5}

The Fifth Catalog of Nearby Stars (CNS5) \citep{Golovin_2023} is a 25 pc volume-limited sample of stars and brown dwarfs compiled using \textit{Gaia} EDR3 and \textit{Hipparcos}, augmented with infrared parallax sources. While RMSTAR focuses on red dwarfs only, CNS5 includes all types of stars (N = 5230) as well as brown dwarfs (N = 701). RMSTAR is more complete than CNS5 for M dwarfs, containing 74 M dwarfs from GDR2, Hipparcos, and optical ground-based parallax work that are not listed in CNS5. RMSTAR also flags potential multiple star systems that need to be carefully considered for LF and MF analyses ($\S$\ref{subsubsec:LF} and $\S$\ref{subsubsec:MF}) because resolving these systems will have impacts on the population distributions.

\subsection{Multiplicity Rate Comparisons}

Here we compare our stellar multiplicity rate for wide companions to previous surveys of nearby M dwarfs. We do not reach further afield because once past 25 pc, many companions drop out as unresolved systems and the physical distances surveyed expand for given separations on the sky. Here we focus on the nearly complete sample of wide stellar companions with separations of at least 30 au (see Figure \ref{fig:logsepHist}), establishing a benchmark of 230 multiple M dwarfs systems among 3352 RMSTAR systems, yielding a multiplicity rate of 6.86$\pm$0.44\%\footnote{All errors given in this discussion represent binomial distribution statistical errors.}. We provide comparisons of this rate to multiplicity rates at all separations in others' work because that is what is typically reported, as well as assess companion rates with projected separations $\ge$30 au.

The first systematic assessment of nearby M dwarf multiplicity was the infrared speckle survey by \citet{Henry_1990}, who surveyed 27 M dwarf primaries within 5 pc and north of declination $-$30$^{\circ}$. Including visual pairs as well, they found 9 multiples, yielding a rate of 33$\pm$11\%. That effort was expanded to include 67 M dwarf primaries out to 8 pc by \citet{Henry_1991}, detecting 21 multiples and yielding a rate of 31$\pm$7\% for companions at all separations. Building upon that work, \citet{Fischer_1992} augmented the speckle work with radial velocity, infrared imaging, and visual binary efforts to derive a multiplicity rate of 42$\pm$9\% by synthesizing samples of various sizes from 28--62 M dwarf primaries depending on the different observing programs.

A few decades later, the multiplicity rate for M dwarfs garnered new attention, in large part because these small stars became key targets in searches for terrestrial planets. The most comprehensive survey for companions to nearby M dwarfs to date is that of \cite{Winters_2019}, who found a stellar multiplicity rate of 26.8$\pm$1.4\% over all separations for a volume-limited sample of 1120 M dwarf primaries within 25 pc. This rate relied upon corrections based on the distribution of companions to stars within 10 pc at small separations for stars at distances of 10--25 pc to capture unresolved companions with separations $\le$50 au. This was done because the companion population for M dwarfs within the 10 pc sample was well over 95\% complete at the time. \cite{Ward-Duong_2015} searched 245 late-K to mid-M (K7V--M6V) dwarfs within 15 pc and found a similar companion star fraction of 23.5$\pm$3.2\% for separations of 3--10,000 au. Finally, the POKEMON survey by \cite{Clark_2024} out to 15 pc targeted 455 stars of types M0V through M9V, and again found a similar multiplicity rate of 23.5$\pm$2.0\%.

To best compare multiplicity rates from the literature to this work, we have cross matched these three recent and similar M dwarf surveys with RMSTAR 2026\footnote{For more M dwarf samples and their multiplicity rates from the literature, see references in \citet{Winters_2019} Table 1 and \citet{Cifuentes_2025} Table 1.}. This allows us to calculate comparable multiplicity rates over the same separation regimes while also accounting for some stars being pushed out of the boundaries of their respective samples by updated parallaxes and capturing only M dwarfs as primaries. Considering only companions with projected separations of at least 30 au, we find multiplicity rates of 5.32--10.86\% for the three recent surveys described above; statistics are given in Table \ref{tab:MRcomparisons}. We find our rate of 6.86$\pm$0.44\% to be lower than rates in both \citet{Ward-Duong_2015} and \citet{Winters_2019}. This is because with the help of GDR3, RMSTAR contains many more faint stars than these two efforts, and as shown in the center panel of Figure \ref{fig:logsepHist}, intrinsically fainter M dwarfs have companions less often than their more luminous counterparts, thereby reducing the overall M dwarf multiplicity rate. Results from \citet{Clark_2024} agree within 2$\sigma$ our multiplicity rate, a consequence of the POKEMON survey also using GDR3 astrometry and photometry. Our statistical error is lower than the previous efforts because we are surveying more than 3000 M dwarfs and have 230 companions with separations of at least 30 au. Our current work to sweep up and characterize closer companions in our ongoing speckle imaging, radial velocity (J. G. Winters et al.,~in press), and long-term astrometry efforts \citep{Henry_2018} will increase the multiplicity rate significantly, likely to $\sim$30\%. Future RMSTAR editions will include close companions to provide a truly comprehensive version of the multiplicity rate of nearby M dwarfs at all separations.

\begin{deluxetable*}{ccccc} \label{tab:MRcomparisons}
%\tablecolumns{2}
\tablewidth{0pt}
\tablecaption{Multiplicity Rates for M Dwarfs at Separations $\geq$ 30 au}
\tablehead{\colhead{\textbf{Paper}}& \colhead{\textbf{Systems in}} & \colhead{\textbf{RMSTAR}} & \colhead{\textbf{Companions}} & \colhead{\textbf{Mult. Rate}} \\
\colhead{}&\colhead{\textbf{Paper}}&{\textbf{Systems}}& & \colhead{Proj. Sep.$\geq$ 30 au}}
\startdata
\citet{Ward-Duong_2015}  &  245 &  221 &   24 & 10.86$\pm$2.09\%  \\ 
\citet{Winters_2019}     & 1120 & 1062 &   90 &  8.47$\pm$0.85\%  \\ 
\citet{Clark_2024}       &  455 &  432 &   23 &  5.32$\pm$1.08\%  \\ 
\hline
This Work                &      & 3352 &  230 &  6.86$\pm$0.44\%  \\ 
\enddata
\vspace{-0.8cm}
\tablecomments{...}
\end{deluxetable*}

\subsection{Luminosity Function Comparisons}

Here we compare the RMSTAR luminosity function to that from the \textit{Gaia Catalogue of Nearby Stars} (GCNS) \citep{Smart_2021}, which is a good comparison sample because it was constructed with nearby stars from \textit{Gaia} Early Data Release 3 and reaches similar faintness limits as RMSTAR. The GCNS luminosity function peaks at $M_G=10.5$, which is slightly brighter than our peak at $M_G=10.8$ because the RMSTAR LF is more complete at the faint end at 25 pc than GCNS at 100 pc. LFs from earlier efforts were constructed using \textit{Hipparcos} datasets and/or ground-based observations (\citealp{Reid_2002}, \citealp{Reid_2003}, \citealp{Just_2015}) that are in different photometric bands and are incomplete at the faint end of the M dwarf regime. Therefore, they do not show a complete picture of the red dwarf LF. We reiterate that the LF will shift toward lower luminosity M dwarfs as the close companions are resolved and single points are broken into two or more intrinsically fainter components.

\subsection{Mass Function Comparisons}

\begin{figure}
    \centering
    \includegraphics[width=\linewidth]{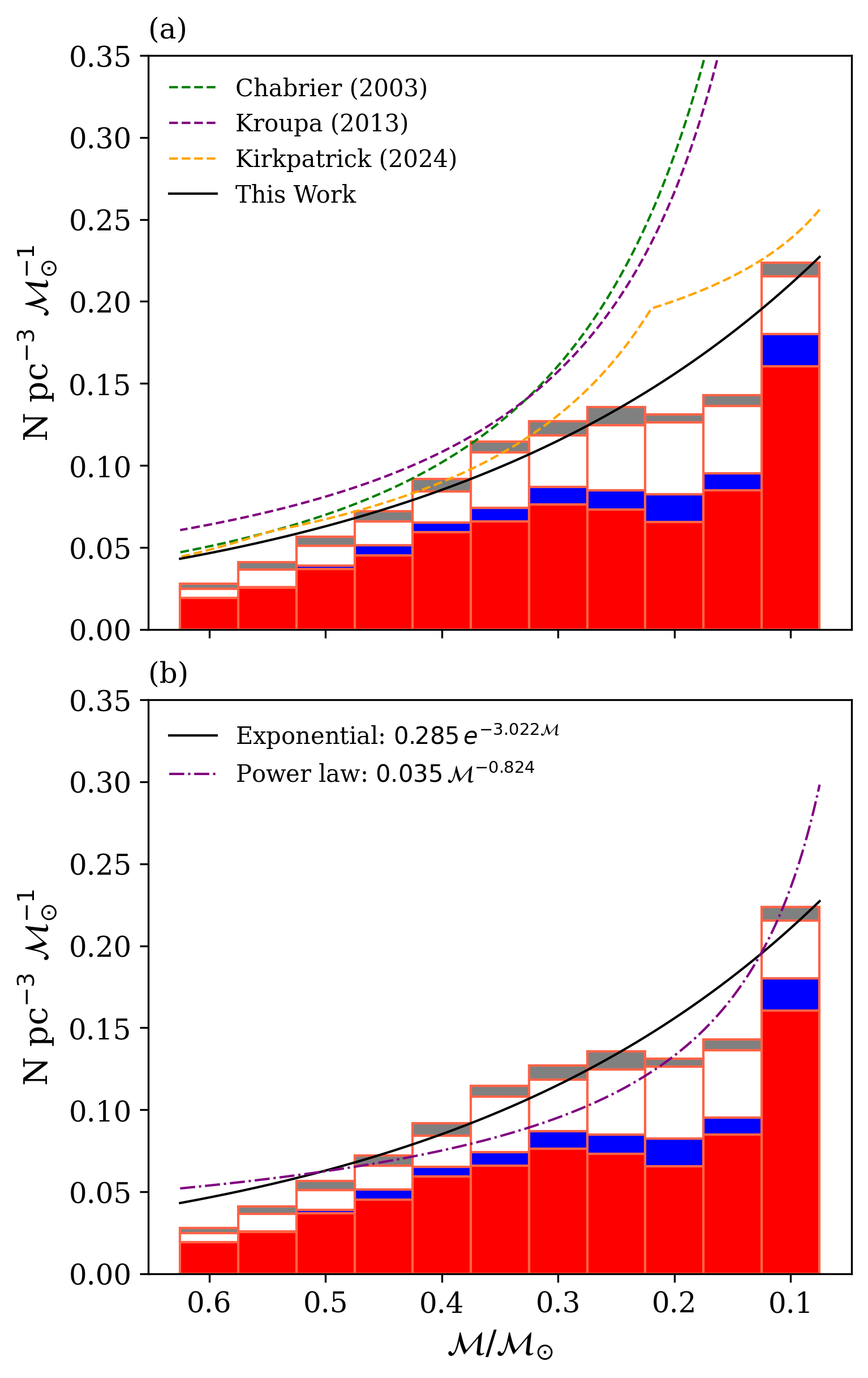}
    \caption{Comparison of the M dwarf mass function and fit from this work (black line) to (a) the fits from \citet{Chabrier_2003} (green dashed line), \citet{Kroupa_2002} (purple dashed line), and \citet{Kirkpatrick_2024} (orange dashed line) and to (b) a power law fit of the same data (purple dot-dashed line).}
    \label{fig:MFcomparison}
\end{figure}

We compare the RMSTAR MF from this work ($\S$\ref{subsubsec:MFresult}) to MFs from \citet{Chabrier_2003}, \citet{Kroupa_2013}, and \citet{Kirkpatrick_2024}, with the details provided in Table \ref{tab:imf}. \citet{Chabrier_2003} uses a log-normal function to describe the initial mass function of stars in the M dwarf regime, while \citet{Kroupa_2002} and \citet{Kirkpatrick_2024} use a series of broken power laws, each with breaks at different M dwarf masses. We have chosen to fit our mass function using a single exponential function with no breaks. As shown in Figure \ref{fig:MFcomparison}a, our single-equation, exponential fit traces the entire sequence of M dwarfs relatively well, whereas other equations fit only specific regions of M dwarf masses. 

We find that the bumps and dips in the the other MFs are a result of poor mass estimates, either from the blended photometry of unresolved multiple star systems or from a lack of dynamical masses in sections of the mass-luminosity relations utilized --- the features in the literature MF fits are not from intrinsic astrophysical reasons altering the MF for M dwarfs. We find that our adopted exponential function fits the data in Figures \ref{fig:LFMF} and \ref{fig:MFcomparison}a (the MF fit line is the same in both plots). Smooth, unbroken power laws were explored but yielded worse fits than the exponential based on the reduced $\chi^2$ values of the fits, as shown in Figure \ref{fig:MFcomparison}b, where the reduced $\chi^2 = 13.38$ for the exponential fit and $\chi^2 = 22.22$ for the power law.  We know of no physical reason why the MF should be described as a power law rather than our adopted exponential, and hence use the MF fit presented as a starting point to be improved upon when close companions have been added to the RMSTAR sample.

\begin{table*}[htbp]
\centering
\caption{Initial Mass Function Parameterizations} \label{tab:imf}
\begin{tabular}{lll}
\tablewidth{0pt}
\hline\hline
Source & Equation & Parameters \\
\hline
\\[6pt]
% ---- Kroupa 2013 ----
\citet{Kroupa_2002}
  & $\xi_{star}(\mathcal{M}) = k
       (\dfrac{\mathcal{M}}{0.08})^{-\alpha} $  
& $k = 0.877$ stars/(pc$^3$\,M$_{\odot}$), $\alpha = 1.3\pm0.5$\\ 
& & ($0.08 < \mathcal{M} \leq 0.5\mathcal{M}_\odot$)\\

  & & \\[6pt]
\hline
\\[6pt]
% ---- Chabrier 2003 ----
\multirow{2}{*}{\citet{Chabrier_2003}}
  & Log-normal ($\mathcal{M} < 1\,\mathcal{M}_\odot$): & \\
  & $\xi(\log \mathcal{M}) = A\exp\!\left[-\dfrac{(\log\mathcal{M} - \log m_c)^2}{2\sigma^2}\right]$
  &$A = 0.158$, $m_c = 0.079\,\mathcal{M}_\odot$, $\sigma = 0.69$ \\[6pt]
   \\
\hline 
\\[6pt]

% ---- Kirkpatrick 2024 ----
\multirow{3}{*}{\citet{Kirkpatrick_2024}}
  & \multirow{3}{*}{$\xi(\mathcal{M}) =C \mathcal{M}^{-\alpha}$}
  & $C_1 = 0.0150,\ \alpha_1 = 2.3\ \ (\mathcal{M} \geq 0.55\,\mathcal{M}_\odot)$ \\
  & &$C_2=0.0273,\ \alpha_2 = 1.3\ \ (0.22 \leq \mathcal{M} < 0.55\,\mathcal{M}_\odot)$ \\
  & &$C_3=0.1340,\ \alpha_3 = 0.25\ (0.05 \leq \mathcal{M} < 0.22\,\mathcal{M}_\odot)$ \\
%  & &$C_4=0.0469,\ \alpha_4 = 0.6\ \ (\mathcal{M} < 0.05\,\mathcal{M}_\odot)$ \\
  & & \\[6pt]
\hline
\\[6pt]
% ---- This work ----
This work
  & $\xi(\mathcal{M}) = Ce^{-\alpha\mathcal{M}}$
  & $C = 0.285$,\ $\alpha = 3.022 $ \\
 \\
\hline
\end{tabular}
\end{table*}

\subsection{The RKSTAR Catalog}\label{subsec:RKSTAR}

RMSTAR has been created in conjunction with the RKSTAR (RECONS K Star) effort by \citet{Johns_2026_AAS}, for which a companion paper has been submitted contemporaneously with this manuscript (T. Johns et al.~submitted). RKSTAR contains 4466 K dwarf systems over the entire sky within 50 pc, as well as lower mass stellar companions at wide separations, in the same spirit as RMSTAR. These two catalogs follow the same steps in their creation, with minor adjustments made when necessary. A more distant horizon was adopted for RKSTAR than RMSTAR to boost the sample to several thousand systems so that both lists would have comparable sizes for statistical evaluations. Because K dwarfs have $M_G$ = 5.5--8.1 it is possible to create a volume-limited sample out to 50 pc, with similar caveats to those for RMSTAR about sample completeness for both primaries and companions. The RMSTAR and RKSTAR combination represents the two most common types of stars in the solar neighborhood that account for 86\% of all stars \cite{Henry_Jao_2024}, and provides advantages for assessing both types of stars together, e.g., the smooth $M_G$ vs.~$M_V$ relation shown in Figure \ref{fig:MVvsMG}. Eventually, the key consideration for all stellar populations, the mass function, will be provided using the combination of RMSTAR and RKSTAR.

\section{Conclusions from RMSTAR 2026}\label{subsec:conclusions}

The RMSTAR catalog is a volume-limited sample of M dwarfs and their lower mass companions within a 25 pc radius of the Sun. Primary M dwarfs in RMSTAR come from the \textit{Gaia} and \textit{Hipparcos} surveys, with additional M dwarf systems from the literature. Key results from this work include:

$\bullet$ RMSTAR 2026 is anticipated to be nearly complete for M dwarf systems within 25 pc, of which 3352 are included in this edition. Fewer than 100 systems are anticipated to be missing from this edition --- these are primarily close multiples for which parallaxes are not yet available from \textit{Gaia}.

$\bullet$ There are 291 M dwarf systems with wide stellar companions (resolved by GDR2 or GDR3 with separations $\gtrsim 0.46\arcsec$) in RMSTAR, yielding a multiplicity rate of 8.68$\pm$0.49\%. Of the companions included here, 230 have projected separations $\geq$30 au, a separation regime which is anticipated to be nearly complete, yielding a multiplicity rate of 6.86$\pm$0.44\%.  The M dwarf multiplicity rate for stellar companions at projected separations of 100--1000 au is 3.22$\pm$0.30\% and the rate for 1000-30,000 au is only 1.01$\pm$0.17\%, showing an obvious drop in companions at wider separations.

$\bullet$ The luminosity function for M dwarfs peaks at $M_G$ = 10.8. This shifts to (slightly) fainter magnitudes when close companions are included.

$\bullet$ The mass function for M dwarfs rises all the way to the end of the stellar main sequence, and is represented by a smooth exponential function of the form $\xi(\mathcal{M}) = 0.289e^{-3.040\mathcal{M}}$. Similar to the LF, the rise in the numbers of lower mass stars becomes steeper when close companions are resolved and included.

Much work remains to be done to reach within the resolution limits of \textit{Gaia} and \textit{Hipparcos} to recover stellar companions within $\sim1\arcsec$, as well as brown dwarfs and planets at all separations. Multiplicity parameters in GDR3 point to a large population of unresolved companions, many of which have already been revealed via speckle, radial velocity, and long-term astrometric work. These close companions will be incorporated into future RMSTAR editions, resulting in a significant increase in the M dwarf multiplicity rate and shifts in the LF and MF toward lower luminosity and lower mass M dwarfs. Overall, this RMSTAR 2026 edition is a key step in providing a complete picture of red dwarfs near the Sun, and is provided to the community as a comprehensive resource for future astrophysical studies and exoplanet surveys. We look forward to continued exploration of the sample of the most common stars in the solar neighborhood and encourage others to join us in the journey.

\section{Acknowledgements}

This material is based upon work supported by the National Science Foundation Graduate Research Fellowship Program under grant number 2444106. Any opinions, findings, and conclusions or recommendations expressed in this material are those of the authors and do not necessarily reflect the views of the National Science Foundation. This work has been supported by the RECONS Institute, NSF grants AST-1910130 and AST-2510383, NASA grant 22-XRP22\_2-0187, Georgia State University, and via observations made at the SMARTS Observatory. M.R.L. thanks Sebasti\'{a}n Carrazco Gaxiola, Emma Galligan, and Melodie Sloneker for individual conversations in support of this work.

This research has made use of several astronomical catalogs and databases, including the Washington Double Star (WDS) Catalog maintained at the US Naval Observatory; the Set of Identifications, Measurements and Bibliography for Astronomical Data (SIMBAD) database, operated at the Centre de Données astronomiques de Strasbourg (CDS), France; and NASA’s Astrophysics Data System. This work has also made use of data from the European Space Agency (ESA) mission \textit{Gaia} (https://www.cosmos.esa.int/gaia), processed by the \textit{Gaia} Data Processing and Analysis Consortium (DPAC; https://www.cosmos.esa.int/web/gaia/dpac/consortium). Funding for the DPAC has been provided by national institutions, in particular the institutions participating in the \textit{Gaia} Multilateral Agreement.

\appendix

\section{Systems Worthy of Note} \label{sec:worthyofnote}

All systems listed in this Appendix are in the RMSTAR catalog.

\textbf{RMS 0939$+$3145B:} This system, also known as G 117-34, is comprised of two stars with angluar separation of 2\arcsec.00. The photometry of the B component of this system is reported in GDR3 as: $G = 16.45 \pm 0.00$, $BP = 16.20 \pm 0.01$, $RP = 14.52 \pm 0.18$ ($M_G=$14.81, $BP-RP=$1.68). The relatively large error on the $RP$ photometry causes the star to land in the white dwarf regime on the color-magnitude diagram, apparent in Figure \ref{fig:HRDs}c. However, there is no evidence to support that this object is a white dwarf and therefore it remains to be included as a M dwarf companion in RMSTAR.

\textbf{RMS 1006$-$1246:} The parallax for RMS 1006-1246, also known as 2MASS J10065210-1246543, had not been previously published and will therefore be published here. 

\textbf{RMS 1041$-$3653:} RMS 1041$-$3653A, also known as CD-36 6589A, has both photometric and astrometric solutions in the \textit{Gaia} and \textit{Hipparcos} catalogs, while RMS 1041$-$3653B has only astrometric solutions in the \textit{Gaia} catalog. The astrometric solutions of these two sources allowed us to determine that they are gravitationally bound and that RMS 1041$-$3653A is an M dwarf, however we were unable to confirm with Gaia and \textit{Hipparcos} data alone that RMS 1041$-$3653B is also an M dwarf. Further investigation led us to \citet{HIP_TYC_1997}, which contains deconvolved photometry, and thus individual \textit{Hipparcos} ($H_p$) magnitudes for both A ($H_p$$=$10.108) and B ($H_p$$=$12.75) in the system. With these individual $H_p$ magnitudes and the Gaia parallaxes, we confirm that both of these stars are M dwarfs belonging to the RMSTAR Catalog and calculate $M_G$ using Equation \ref{MG_MH} for A ($M_G$$=$8.33) and B ($M_G$$=$10.50).

\textbf{RMS 1103$+$1517:} RMS 1103$+$1517, also known as LTT 12957, was observed in both the north and south at the United States Naval Observatory Robotic Astrometric Telescope (URAT) and Cerro Tololo Interamerican Observatory (CTIO) and parallax measurements from these observations were reported by \citet{Finch_2018}. The northern observations yielded $\pi = 54.9 \pm 8.0$ mas and the southern observations yielded $\pi = 67.4 \pm 9.8$ mas. In RMSTAR, we report the weighted mean of these two measurements as $\pi = 59.9 \pm 8.8$ mas.

\textbf{RMS 1233$+$0901:} This system, also known as GJ 473 and Wolf 424, contains two stars: RMS 1233$+$0901A and RMS 1233$+$0901B. Though the convention of the catalog is that the brighter star in absolute $G$ magnitude be named ``A'' and the fainter star be named ``B'', this system is flipped so that the fainter star is named ``A'' and the brighter star named ``B''. This system's name is complicated by the fact that the stars are very similar in brightness -- $G$(A)$=11.23$ and $G$(B)$=11.24$ -- and the large errors in the GDR3 astrometry resulting in varying parallax measurements -- $\pi$(A)$=231.1\pm0.5$ mas and $\pi$(B)$=223.4\pm0.4$ mas. Therefore, we have named this system in agreement with how it has historically been named in the literature.

\textbf{RMS 1417$+$0851:} RMS 1417$+$0851, also known as PM J14171$+$0851, was observed in both the north and south at the United States Naval Observatory Robotic Astrometric Telescope (URAT) and Cerro Tololo Interamerican Observatory (CTIO) and parallax measurements from these observations were reported by \citet{Finch_2018}. The northern observations yielded $\pi = 54.3 \pm 6.3$ mas and the southern observations yielded $\pi = 64.7 \pm 6.9$ mas. In RMSTAR, we report the weighted mean of these two measurements as $\pi = 58.5 \pm 6.5$ mas.

\textbf{RMS 1510$-$5248:} RMS 1510$-$5248A, also known as L 262-74A,  has both photometric and astrometric solutions in the \textit{Gaia} and \textit{Hipparcos} catalogs, while RMS 1510$-$5248B has only astrometric solutions in the \textit{Gaia} catalog. The astrometric solutions of these two sources allowed us to determine that they are gravitationally bound and that RMS 1510$-$5248A is an M dwarf, however we were unable to confirm with Gaia and \textit{Hipparcos} data alone that RMS 1510$-$5248B is also an M dwarf. Further investigation led us to \citet{Mason_2001} which contains $K$ band photometry for both stars in the system -- $K$(A)$=7.17$ and $K$(B)$=8.34$. A $K$ band magnitude difference of 1.2 mag corresponds to a $V$ band magnitude difference of $\sim$2-3 mag, indicating that the fainter B component is in fact an M dwarf and remains in the RMSTAR catalog, however we do not list any photometric magnitude for this companion.

\textbf{RMS 1840$-$0959:} RMS 1840$-$0959, also known as 
2MASS J18405934$-$0959136, had no $G$ magnitude reported in GDR3, however it did have a $G$ magnitude reported in GDR2. Therefore, all data entered in RMSTAR for this star is from GDR3, except for photometry which is from GDR2. 

\textbf{RMS 2007$-$3145:} This system, also known as GJ 787.1, contains two stars: RMS 2007$-$3145A and RMS 2007$-$3145B. Similarly to RMS 1233$+$0901, this system is flipped so that the fainter star is named ``A'' and the brighter star named ``B''. Again, these stars have similar brightnesses --  $G$(A)$=11.13$ and $G$(B)$=11.07$ -- and we have named this system in agreement with how it has been historically named in the literature.

\textbf{RMS 2045$-$3120:} RMS 2045$-$3120 is a young, triple star system in which the brightest star -- RMS 2045$-$3120A -- is also known as AU Mic. AU Mic is technically brighter than the brightness limits of RMSTAR ($M_G = 7.90$), however, it is known to be an M dwarf via spectroscopic classification \citep{Keenan_1989}. 

\textbf{RMS 2227$+$5741:} It is apparent in Figure \ref{fig:onetoone}c that RMS 2227$+$5741AB, the well-known binary at 4 pc called Kruger 60 (also GJ 860), has a particularly large discrepancy in proper motion in declination of 462.783 mas\,yr$^{-1}$, which in fact reflects the relative orbital motions of the two components in the binary.

\section{Reasons for Removal} \label{sec:removals}

Throughout our search for M dwarfs within 25 pc, some candidate M dwarf primaries were removed from the RMSTAR catalog for various reasons. These are stars that were within the magnitude limits of the catalog but are white dwarfs, have white dwarf or higher mass companions that are not M dwarfs, or are faint with bad parallax measurements. The exact reason for each star's removal is listed in Table \ref{tab:removals}, along with their GDR3 ID and J2000 coordinates.

\begin{deluxetable}{c c c l}[h!]
\tablecaption{Examples of Candidate M Dwarf Primaries Removed from RMSTAR}\label{tab:removals}
\tablehead{
  \colhead{\textit{Gaia}  DR3 ID} & 
  \colhead{RA (J2000)} &
  \colhead{Dec (J2000)} &
  \colhead{Reason for removal}
}
\startdata
4706564427272810752 & 00 02 09.35 & $-$68 16 53.3 & primary with MG = 8.012 \\
4994877094997259264 & 00 02 10.72 & $-$43 09 55.3 & WD \\
386655019234959872	& 00 05 10.88 & $+$45 47 11.6 & triple with MG = 7.94 primary \\
418491408481398272	& 00 41 20.82 & $+$55 50 04.4 & companion to WD \\
4975284347548097152	& 00 45 41.56 & $-$47 33 30.8	& quadruple with MG = 4.75 primary \\
5601615988055197952	& 07 48 58.41 & $-$27 05 20.1	& bad parallax \\
3144837318276010624	& 07 50 14.58 & $+$07 11 48.9	& WD+WD \\
\enddata
\tablecomments{This table is available in its entirety in machine-readable form.}

\end{deluxetable}

\section{Photometry and Astrometry Reference Codes}\label{references}
All references for data used to build the RMSTAR Catalog are listed in Table \ref{tab:refs}.

\startlongtable
\begin{deluxetable*}{ll}
\tablecaption{Codes for the references cited in other Tables throughout this work.}\label{tab:refs}
\tablehead{
  \colhead{Code}&
  \colhead{References}
}
\startdata
Bar17 & \citet{Bartlett_2017} \\
Bar21 & \citet{Baroch_2021} \\
Ben16 & \citet{Benedict_2016} \\
Bes90 & \citet{Bessel_1990} \\
Bes91 & \citet{Bessel_1991} \\
Cos05 & \citet{Costa_2005} \\
Cos06 & \citet{Costa_2006} \\
Dah88 & \citet{Dahn88} \\
Dah02 & \citet{Dahn2002} \\
Dav15 & \citet{Dav15} \\
Del99 &\citet{Delfosse_1999} \\
Die14 & \citet{Dieterich_2014} \\
Dit14 & \citet{Dittmann_2014} \\
Dup17 & \citet{Dupuy_2017} \\
Fin18 & \citet{Finch_2018} \\
Gai18 & \citet{Gaia_2018} \\
Gai22 & \citet{Gaia_2023} \\
Gat09 & \citet{Gatewood_2009} \\
Giz98 & \citet{Gizis_1998} \\
Har80 & \citet{Harrington_1980} \\
Hei94 & \citet{Heintz_1994} \\
Hen04 & \citet{Henry_2004} \\
Hen06 & \citet{Henry_2006} \\
Hen18 & \citet{Henry_2018} \\
Hol18 & \citet{Hollands_2018} \\
Hon20 & \citet{Honaker_2020} \\
Hos15 & \citet{Hosey_2015} \\
Jao05 & \citet{Jao_2005} \\
Jao11 & \citet{Jao_2011} \\
Jao14 & \citet{Jao_2014}\\
Jao17 & \citet{Jao_2017} \\
Kar24 & \citet{Kar_2024} \\
Ker22 & \citet{Kervella_2022} \\
Khr10 & \citet{Khrutskaya_2010} \\
Kil98 & \citet{Kilkenny_1998} \\
Koe02 & \citet{Koen_2002} \\
Koe10 & \citet{Koen_2010} \\
Lan92 & \citet{Landolt_1992} \\
Lan09 & \citet{Landolt_2009} \\
Laz25 & \citet{Lazorenko_2025} \\
Lep09 & \citet{Lepine_2009} \\
Mac18& \citet{Mace_2018} \\
Mon92 & \citet{Monet_1992} \\
Rei02 & \citet{Reid_2002} \\
Rei03 & \citet{Reid_2003} \\
Rie10 & \citet{Riedel_2010} \\
Rie14 & \citet{Riedel_2014} \\
Rie18 & \citet{Riedel_2018} \\
Pat98 & \citet{Patterson_1998} \\
ScR12 & \citet{Schilbach_Roser_2012} \\
Sil02 & \citet{Silvestri_2002} \\
Sil19 & \citet{Silverstein_2019} \\
Sod99 & \citet{Soderhjelm_1999} \\
Tor10 & \citet{Torres_2010} \\
vAl95 & \citet{vanAltena_1995} \\
vLe07 & \citet{vanLee_2007} \\
Vri20 & \citet{Vrijmoet_2020} \\
Wei84 & \citet{Weis_1984} \\
Wei86 & \citet{Weis_1986} \\
Wei87 & \citet{Weis_1987} \\
Wei88 & \citet{Weis_1988} \\
Wei91a &\citet{Weis_1991a}  \\
Wei91b &\citet{Weis_1991b} \\
Wei93 & \citet{Weis_1993} \\
Wei94 & \citet{Weis_1994} \\
Wei96 & \citet{Weis_1996} \\
Wei99 & \citet{Weis_1999} \\
Win11 & \citet{Winters_2011} \\
Win15 & \citet{Winters_2015} \\
Win17 & \citet{Winters_2017} \\
Win19 & \citet{Winters_2019} \\
\enddata
\tablecomments{This table is available in its entirety in machine-readable form.}
\end{deluxetable*}

\bibliography{sample701}{}

@ARTICLE{Henry_2006,
       author = {{Henry}, Todd J. and {Jao}, Wei-Chun and {Subasavage}, John P. and {Beaulieu}, Thomas D. and {Ianna}, Philip A. and {Costa}, Edgardo and {M{\'e}ndez}, Ren{\'e} A.},
        title = "{The Solar Neighborhood. XVII. Parallax Results from the CTIOPI 0.9 m Program: 20 New Members of the RECONS 10 Parsec Sample}",
      journal = {\aj},
         year = 2006,
        month = dec,
       volume = {132},
       number = {6},
        pages = {2360-2371},
          doi = {10.1086/508233},
archivePrefix = {arXiv},
       eprint = {astro-ph/0608230},
 primaryClass = {astro-ph},
       adsurl = {https://ui.adsabs.harvard.edu/abs/2006AJ....132.2360H}
}

@ARTICLE{Benedict_2016,
       author = {{Benedict}, G.~F. and {Henry}, T.~J. and {Franz}, O.~G. and {McArthur}, B.~E. and {Wasserman}, L.~H. and {Jao}, Wei-Chun and {Cargile}, P.~A. and {Dieterich}, S.~B. and {Bradley}, A.~J. and {Nelan}, E.~P. and {Whipple}, A.~L.},
        title = "{The Solar Neighborhood. XXXVII: The Mass-Luminosity Relation for Main-sequence M Dwarfs}",
      journal = {\aj},
         year = 2016,
        month = nov,
       volume = {152},
       number = {5},
          eid = {141},
        pages = {141},
          doi = {10.3847/0004-6256/152/5/141},
archivePrefix = {arXiv},
       eprint = {1608.04775},
 primaryClass = {astro-ph.SR},
       adsurl = {https://ui.adsabs.harvard.edu/abs/2016AJ....152..141B}
}

@ARTICLE{Vrijmoet_2022,
       author = {{Vrijmoet}, Eliot Halley and {Tokovinin}, Andrei and {Henry}, Todd J. and {Winters}, Jennifer G. and {Horch}, Elliott and {Jao}, Wei-Chun},
        title = "{The Solar Neighborhood. XLIX. New Discoveries and Orbits of M-dwarf Multiples with Speckle Interferometry at SOAR}",
      journal = {\aj},
         year = 2022,
        month = apr,
       volume = {163},
       number = {4},
          eid = {178},
        pages = {178},
          doi = {10.3847/1538-3881/ac52f6},
archivePrefix = {arXiv},
       eprint = {2202.04688},
 primaryClass = {astro-ph.SR},
       adsurl = {https://ui.adsabs.harvard.edu/abs/2022AJ....163..178V}
}

@ARTICLE{Smart_2021,
       author = {{Gaia Collaboration} and {Smart}, R.~L. and {Sarro}, L.~M. and {Rybizki}, J. and {Reyl{\'e}}, C. and {Robin}, A.~C. and {Hambly}, N.~C. and {Abbas}, U. and {Barstow}, M.~A. and {de Bruijne}, J.~H.~J. and {Bucciarelli}, B. and {Carrasco}, J.~M. and {Cooper}, W.~J. and {Hodgkin}, S.~T. and {Masana}, E. and {Michalik}, D. and {Sahlmann}, J. and {Sozzetti}, A. and {Brown}, A.~G.~A. and {Vallenari}, A. and {Prusti}, T. and {Babusiaux}, C. and {Biermann}, M. and {Creevey}, O.~L. and {Evans}, D.~W. and {Eyer}, L. and {Hutton}, A. and {Jansen}, F. and {Jordi}, C. and {Klioner}, S.~A. and {Lammers}, U. and {Lindegren}, L. and {Luri}, X. and {Mignard}, F. and {Panem}, C. and {Pourbaix}, D. and {Randich}, S. and {Sartoretti}, P. and {Soubiran}, C. and {Walton}, N.~A. and {Arenou}, F. and {Bailer-Jones}, C.~A.~L. and {Bastian}, U. and {Cropper}, M. and {Drimmel}, R. and {Katz}, D. and {Lattanzi}, M.~G. and {van Leeuwen}, F. and {Bakker}, J. and {Casta{\~n}eda}, J. and {De Angeli}, F. and {Ducourant}, C. and {Fabricius}, C. and {Fouesneau}, M. and {Fr{\'e}mat}, Y. and {Guerra}, R. and {Guerrier}, A. and {Guiraud}, J. and {Jean-Antoine Piccolo}, A. and {Messineo}, R. and {Mowlavi}, N. and {Nicolas}, C. and {Nienartowicz}, K. and {Pailler}, F. and {Panuzzo}, P. and {Riclet}, F. and {Roux}, W. and {Seabroke}, G.~M. and {Sordo}, R. and {Tanga}, P. and {Th{\'e}venin}, F. and {Gracia-Abril}, G. and {Portell}, J. and {Teyssier}, D. and {Altmann}, M. and {Andrae}, R. and {Bellas-Velidis}, I. and {Benson}, K. and {Berthier}, J. and {Blomme}, R. and {Brugaletta}, E. and {Burgess}, P.~W. and {Busso}, G. and {Carry}, B. and {Cellino}, A. and {Cheek}, N. and {Clementini}, G. and {Damerdji}, Y. and {Davidson}, M. and {Delchambre}, L. and {Dell'Oro}, A. and {Fern{\'a}ndez-Hern{\'a}ndez}, J. and {Galluccio}, L. and {Garc{\'\i}a-Lario}, P. and {Garcia-Reinaldos}, M. and {Gonz{\'a}lez-N{\'u}{\~n}ez}, J. and {Gosset}, E. and {Haigron}, R. and {Halbwachs}, J. -L. and {Harrison}, D.~L. and {Hatzidimitriou}, D. and {Heiter}, U. and {Hern{\'a}ndez}, J. and {Hestroffer}, D. and {Holl}, B. and {Jan{\ss}en}, K. and {Jevardat de Fombelle}, G. and {Jordan}, S. and {Krone-Martins}, A. and {Lanzafame}, A.~C. and {L{\"o}ffler}, W. and {Lorca}, A. and {Manteiga}, M. and {Marchal}, O. and {Marrese}, P.~M. and {Moitinho}, A. and {Mora}, A. and {Muinonen}, K. and {Osborne}, P. and {Pancino}, E. and {Pauwels}, T. and {Recio-Blanco}, A. and {Richards}, P.~J. and {Riello}, M. and {Rimoldini}, L. and {Roegiers}, T. and {Siopis}, C. and {Smith}, M. and {Ulla}, A. and {Utrilla}, E. and {van Leeuwen}, M. and {van Reeven}, W. and {Abreu Aramburu}, A. and {Accart}, S. and {Aerts}, C. and {Aguado}, J.~J. and {Ajaj}, M. and {Altavilla}, G. and {{\'A}lvarez}, M.~A. and {{\'A}lvarez Cid-Fuentes}, J. and {Alves}, J. and {Anderson}, R.~I. and {Anglada Varela}, E. and {Antoja}, T. and {Audard}, M. and {Baines}, D. and {Baker}, S.~G. and {Balaguer-N{\'u}{\~n}ez}, L. and {Balbinot}, E. and {Balog}, Z. and {Barache}, C. and {Barbato}, D. and {Barros}, M. and {Bartolom{\'e}}, S. and {Bassilana}, J. -L. and {Bauchet}, N. and {Baudesson-Stella}, A. and {Becciani}, U. and {Bellazzini}, M. and {Bernet}, M. and {Bertone}, S. and {Bianchi}, L. and {Blanco-Cuaresma}, S. and {Boch}, T. and {Bombrun}, A. and {Bossini}, D. and {Bouquillon}, S. and {Bragaglia}, A. and {Bramante}, L. and {Breedt}, E. and {Bressan}, A. and {Brouillet}, N. and {Burlacu}, A. and {Busonero}, D. and {Butkevich}, A.~G. and {Buzzi}, R. and {Caffau}, E. and {Cancelliere}, R. and {C{\'a}novas}, H. and {Cantat-Gaudin}, T. and {Carballo}, R. and {Carlucci}, T. and {Carnerero}, M.~I. and {Casamiquela}, L. and {Castellani}, M. and {Castro-Ginard}, A. and {Castro Sampol}, P. and {Chaoul}, L. and {Charlot}, P. and {Chemin}, L. and {Chiavassa}, A. and {Cioni}, M. -R.~L. and {Comoretto}, G. and {Cornez}, T. and {Cowell}, S. and {Crifo}, F. and {Crosta}, M. and {Crowley}, C. and {Dafonte}, C. and {Dapergolas}, A. and {David}, M. and {David}, P. and {de Laverny}, P. and {De Luise}, F. and {De March}, R. and {De Ridder}, J. and {de Souza}, R. and {de Teodoro}, P. and {de Torres}, A. and {del Peloso}, E.~F. and {del Pozo}, E. and {Delgado}, A. and {Delgado}, H.~E. and {Delisle}, J. -B. and {Di Matteo}, P. and {Diakite}, S. and {Diener}, C. and {Distefano}, E. and {Dolding}, C. and {Eappachen}, D. and {Edvardsson}, B. and {Enke}, H. and {Esquej}, P. and {Fabre}, C. and {Fabrizio}, M. and {Faigler}, S. and {Fedorets}, G. and {Fernique}, P. and {Fienga}, A. and {Figueras}, F. and {Fouron}, C. and {Fragkoudi}, F. and {Fraile}, E. and {Franke}, F. and {Gai}, M. and {Garabato}, D. and {Garcia-Gutierrez}, A. and {Garc{\'\i}a-Torres}, M. and {Garofalo}, A. and {Gavras}, P. and {Gerlach}, E. and {Geyer}, R. and {Giacobbe}, P. and {Gilmore}, G. and {Girona}, S. and {Giuffrida}, G. and {Gomel}, R. and {Gomez}, A. and {Gonzalez-Santamaria}, I. and {Gonz{\'a}lez-Vidal}, J.~J. and {Granvik}, M. and {Guti{\'e}rrez-S{\'a}nchez}, R. and {Guy}, L.~P. and {Hauser}, M. and {Haywood}, M. and {Helmi}, A. and {Hidalgo}, S.~L. and {Hilger}, T. and {H{\l}adczuk}, N. and {Hobbs}, D. and {Holland}, G. and {Huckle}, H.~E. and {Jasniewicz}, G. and {Jonker}, P.~G. and {Juaristi Campillo}, J. and {Julbe}, F. and {Karbevska}, L. and {Kervella}, P. and {Khanna}, S. and {Kochoska}, A. and {Kontizas}, M. and {Kordopatis}, G. and {Korn}, A.~J. and {Kostrzewa-Rutkowska}, Z. and {Kruszy{\'n}ska}, K. and {Lambert}, S. and {Lanza}, A.~F. and {Lasne}, Y. and {Le Campion}, J. -F. and {Le Fustec}, Y. and {Lebreton}, Y. and {Lebzelter}, T. and {Leccia}, S. and {Leclerc}, N. and {Lecoeur-Taibi}, I. and {Liao}, S. and {Licata}, E. and {Lindstr{\o}m}, H.~E.~P. and {Lister}, T.~A. and {Livanou}, E. and {Lobel}, A. and {Madrero Pardo}, P. and {Managau}, S. and {Mann}, R.~G. and {Marchant}, J.~M. and {Marconi}, M. and {Marcos Santos}, M.~M.~S. and {Marinoni}, S. and {Marocco}, F. and {Marshall}, D.~J. and {Martin Polo}, L. and {Mart{\'\i}n-Fleitas}, J.~M. and {Masip}, A. and {Massari}, D. and {Mastrobuono-Battisti}, A. and {Mazeh}, T. and {McMillan}, P.~J. and {Messina}, S. and {Millar}, N.~R. and {Mints}, A. and {Molina}, D. and {Molinaro}, R. and {Moln{\'a}r}, L. and {Montegriffo}, P. and {Mor}, R. and {Morbidelli}, R. and {Morel}, T. and {Morris}, D. and {Mulone}, A.~F. and {Munoz}, D. and {Muraveva}, T. and {Murphy}, C.~P. and {Musella}, I. and {Noval}, L. and {Ord{\'e}novic}, C. and {Orr{\`u}}, G. and {Osinde}, J. and {Pagani}, C. and {Pagano}, I. and {Palaversa}, L. and {Palicio}, P.~A. and {Panahi}, A. and {Pawlak}, M. and {Pe{\~n}alosa Esteller}, X. and {Penttil{\"a}}, A. and {Piersimoni}, A.~M. and {Pineau}, F. -X. and {Plachy}, E. and {Plum}, G. and {Poggio}, E. and {Poretti}, E. and {Poujoulet}, E. and {Pr{\v{s}}a}, A. and {Pulone}, L. and {Racero}, E. and {Ragaini}, S. and {Rainer}, M. and {Raiteri}, C.~M. and {Rambaux}, N. and {Ramos}, P. and {Ramos-Lerate}, M. and {Re Fiorentin}, P. and {Regibo}, S. and {Ripepi}, V. and {Riva}, A. and {Rixon}, G. and {Robichon}, N. and {Robin}, C. and {Roelens}, M. and {Rohrbasser}, L. and {Romero-G{\'o}mez}, M. and {Rowell}, N. and {Royer}, F. and {Rybicki}, K.~A. and {Sadowski}, G. and {Sagrist{\`a} Sell{\'e}s}, A. and {Salgado}, J. and {Salguero}, E. and {Samaras}, N. and {Sanchez Gimenez}, V. and {Sanna}, N. and {Santove{\~n}a}, R. and {Sarasso}, M. and {Schultheis}, M. and {Sciacca}, E. and {Segol}, M. and {Segovia}, J.~C. and {S{\'e}gransan}, D. and {Semeux}, D. and {Shahaf}, S. and {Siddiqui}, H.~I. and {Siebert}, A. and {Siltala}, L. and {Slezak}, E. and {Solano}, E. and {Solitro}, F. and {Souami}, D. and {Souchay}, J. and {Spagna}, A. and {Spoto}, F. and {Steele}, I.~A. and {Steidelm{\"u}ller}, H. and {Stephenson}, C.~A. and {S{\"u}veges}, M. and {Szabados}, L. and {Szegedi-Elek}, E. and {Taris}, F. and {Tauran}, G. and {Taylor}, M.~B. and {Teixeira}, R. and {Thuillot}, W. and {Tonello}, N. and {Torra}, F. and {Torra}, J. and {Turon}, C. and {Unger}, N. and {Vaillant}, M. and {van Dillen}, E. and {Vanel}, O. and {Vecchiato}, A. and {Viala}, Y. and {Vicente}, D. and {Voutsinas}, S. and {Weiler}, M. and {Wevers}, T. and {Wyrzykowski}, {\L}. and {Yoldas}, A. and {Yvard}, P. and {Zhao}, H. and {Zorec}, J. and {Zucker}, S. and {Zurbach}, C. and {Zwitter}, T.},
        title = "{Gaia Early Data Release 3. The Gaia Catalogue of Nearby Stars}",
      journal = {\aap},
         year = 2021,
        month = may,
       volume = {649},
          eid = {A6},
        pages = {A6},
          doi = {10.1051/0004-6361/202039498},
archivePrefix = {arXiv},
       eprint = {2012.02061},
 primaryClass = {astro-ph.SR},
       adsurl = {https://ui.adsabs.harvard.edu/abs/2021A&A...649A...6G}
}

@ARTICLE{Dieterich_2014,
       author = {{Dieterich}, Sergio B. and {Henry}, Todd J. and {Jao}, Wei-Chun and {Winters}, Jennifer G. and {Hosey}, Altonio D. and {Riedel}, Adric R. and {Subasavage}, John P.},
        title = "{The Solar Neighborhood. XXXII. The Hydrogen Burning Limit}",
      journal = {\aj},
         year = 2014,
        month = may,
       volume = {147},
       number = {5},
          eid = {94},
        pages = {94},
          doi = {10.1088/0004-6256/147/5/94},
archivePrefix = {arXiv},
       eprint = {1312.1736},
 primaryClass = {astro-ph.SR},
       adsurl = {https://ui.adsabs.harvard.edu/abs/2014AJ....147...94D}
}

@article{Jao_2005,
doi = {10.1086/428489},
url = {https://dx.doi.org/10.1086/428489},
year = {2005},
month = {apr},
publisher = {},
volume = {129},
number = {4},
pages = {1954},
author = {Wei-Chun Jao and Todd J. Henry and John P. Subasavage and Misty A. Brown and Philip A. Ianna and Jennifer L. Bartlett and Edgardo Costa and René A. Méndez},
title = {The Solar Neighborhood. XIII. Parallax Results from the CTIOPI 0.9 Meter Program: Stars with μ ≥ 10 yr−1 (MOTION Sample)},
journal = {AJ}
}

@UNPUBLISHED{LL:LL-124,
   author = {L.~Lindegren},
   title={{R}e-normalising the astrometric chi-square in {G}aia {D}{R}2},
   institution={Lund Observatory},
   year={2018},
   month={August},
   url={http://www.rssd.esa.int/doc_fetch.php?id=3757412},
   note={GAIA-C3-TN-LU-LL-124},
   type={Technical Note}
}

@article{Ziegler_2021,
doi = {10.3847/1538-3881/ac17f6},
url = {https://dx.doi.org/10.3847/1538-3881/ac17f6},
year = {2021},
month = {oct},
publisher = {The American Astronomical Society},
volume = {162},
number = {5},
pages = {192},
author = {Carl Ziegler and Andrei Tokovinin and Madelyn Latiolais and César Briceño and Nicholas Law and Andrew W. Mann},
title = {SOAR TESS Survey. II. The Impact of Stellar Companions on Planetary Populations},
journal = {AJ}
}

@ARTICLE{Belokurov_2020,
       author = {{Belokurov}, Vasily and {Penoyre}, Zephyr and {Oh}, Semyeong and {Iorio}, Giuliano and {Hodgkin}, Simon and {Evans}, N. Wyn and {Everall}, Andrew and {Koposov}, Sergey E. and {Tout}, Christopher A. and {Izzard}, Robert and {Clarke}, Cathie J. and {Brown}, Anthony G.~A.},
        title = "{Unresolved stellar companions with Gaia DR2 astrometry}",
      journal = {\mnras},
         year = 2020,
        month = aug,
       volume = {496},
       number = {2},
        pages = {1922-1940},
          doi = {10.1093/mnras/staa1522},
archivePrefix = {arXiv},
       eprint = {2003.05467},
 primaryClass = {astro-ph.SR},
       adsurl = {https://ui.adsabs.harvard.edu/abs/2020MNRAS.496.1922B}
}

@article{GaiaEDR3,
	author ={{Lindegren}, L and {Klioner, S. A.} and {Hernández, J.} and {Bombrun, A.} and {Ramos-Lerate, M.} and {Steidelmüller, H.} and {Bastian, U.} and {Biermann, M.} and {de Torres, A.} and {Gerlach, E.} and {Geyer, R.} and {Hilger, T.} and {Hobbs, D.} and {Lammers, U.} and {McMillan, P. J.} and {Stephenson, C. A.} and {Castañeda, J.} and {Davidson, M.} and {Fabricius, C.} and {Gracia-Abril, G.} and {Portell, J.} and {Rowell, N.} and {Teyssier, D.} and {Torra, F.} and {Bartolomé, S.} and {Clotet, M.} and {Garralda, N.} and {González-Vidal, J. J.} and {Torra, J.} and {Abbas, U.} and {Altmann, M.} and {Anglada Varela, E.} and {Balaguer-Núñez, L.} and {Balog, Z.} and {Barache, C.} and {Becciani, U.} and {Bernet, M.} and {Bertone, S.} and {Bianchi, L.} and {Bouquillon, S.} and {Brown, A. G. A.} and {Bucciarelli, B.} and {Busonero, D.} and {Butkevich, A. G.} and {Buzzi, R.} and {Cancelliere, R.} and {Carlucci, T.} and {Charlot, P.} and {Cioni, M.-R. L.} and {Crosta, M.} and {Crowley, C.} and {del Peloso, E. F.} and {del Pozo, E.} and {Drimmel, R.} and {Esquej, P.} and {Fienga, A.} and {Fraile, E.} and {Gai, M.} and {Garcia-Reinaldos, M.} and {Guerra, R.} and {Hambly, N. C.} and {Hauser, M.} and {Janßen, K.} and {Jordan, S.} and {Kostrzewa-Rutkowska, Z.} and {Lattanzi, M. G.} and {Liao, S.} and {Licata, E.} and {Lister, T. A.} and {Löffler, W.} and {Marchant, J. M.} and {Masip, A.} and {Mignard, F.} and {Mints, A.} and {Molina, D.} and {Mora, A.} and {Morbidelli, R.} and {Murphy, C. P.} and {Pagani, C.} and {Panuzzo, P.} and {Peñalosa Esteller, X.} and {Poggio, E.} and {Re Fiorentin, P.} and {Riva, A.} and {Sagristà Sellés, A.} and {Sanchez Gimenez, V.} and {Sarasso, M.} and {Sciacca, E.} and {Siddiqui, H. I.} and {Smart, R. L.} and {Souami, D.} and {Spagna, A.} and {Steele, I. A.} and {Taris, F.} and {Utrilla, E.} and {van Reeven, W.} and {Vecchiato, A.}},
	title = {Gaia Early Data Release 3 - The astrometric solution},
	DOI= "10.1051/0004-6361/202039709",
	url= "https://doi.org/10.1051/0004-6361/202039709",
	journal = {A\&A},
	year = 2021,
	volume = 649,
	pages = "A2",
}

@article{Tokovinin_2023,
doi = {10.3847/1538-3881/acc464},
url = {https://dx.doi.org/10.3847/1538-3881/acc464},
year = {2023},
month = {mar},
publisher = {The American Astronomical Society},
volume = {165},
number = {4},
pages = {180},
author = {Andrei Tokovinin},
title = {Exploring Thousands of Nearby Hierarchical Systems with Gaia and Speckle Interferometry},
journal = {AJ}
}

@ARTICLE{Henry_2018,
       author = {{Henry}, Todd J. and {Jao}, Wei-Chun and {Winters}, Jennifer G. and {Dieterich}, Sergio B. and {Finch}, Charlie T. and {Ianna}, Philip A. and {Riedel}, Adric R. and {Silverstein}, Michele L. and {Subasavage}, John P. and {Vrijmoet}, Eliot Halley},
        title = "{The Solar Neighborhood XLIV: RECONS Discoveries within 10 parsecs}",
      journal = {\aj},
         year = 2018,
        month = jun,
       volume = {155},
       number = {6},
          eid = {265},
        pages = {265},
          doi = {10.3847/1538-3881/aac262},
archivePrefix = {arXiv},
       eprint = {1804.07377},
 primaryClass = {astro-ph.SR},
       adsurl = {https://ui.adsabs.harvard.edu/abs/2018AJ....155..265H}
}

@ARTICLE{Winters_2019,
       author = {{Winters}, Jennifer G. and {Henry}, Todd J. and {Jao}, Wei-Chun and {Subasavage}, John P. and {Chatelain}, Joseph P. and {Slatten}, Ken and {Riedel}, Adric R. and {Silverstein}, Michele L. and {Payne}, Matthew J.},
        title = "{The Solar Neighborhood. XLV. The Stellar Multiplicity Rate of M Dwarfs Within 25 pc}",
      journal = {\aj},
         year = 2019,
        month = jun,
       volume = {157},
       number = {6},
          eid = {216},
        pages = {216},
          doi = {10.3847/1538-3881/ab05dc},
archivePrefix = {arXiv},
       eprint = {1901.06364},
 primaryClass = {astro-ph.SR},
       adsurl = {https://ui.adsabs.harvard.edu/abs/2019AJ....157..216W}
}

@ARTICLE{Clark_2024,
       author = {{Clark}, Catherine A. and {van Belle}, Gerard T. and {Horch}, Elliott P. and {Ciardi}, David R. and {von Braun}, Kaspar and {Skiff}, Brian A. and {Winters}, Jennifer G. and {Lund}, Michael B. and {Everett}, Mark E. and {Hartman}, Zachary D. and {Llama}, Joe},
        title = "{The POKEMON Speckle Survey of Nearby M Dwarfs. III. The Stellar Multiplicity Rate of M Dwarfs within 15 pc}",
      journal = {\aj},
         year = 2024,
        month = apr,
       volume = {167},
       number = {4},
          eid = {174},
        pages = {174},
          doi = {10.3847/1538-3881/ad267d},
archivePrefix = {arXiv},
       eprint = {2401.14703},
 primaryClass = {astro-ph.SR},
       adsurl = {https://ui.adsabs.harvard.edu/abs/2024AJ....167..174C}
}

@ARTICLE{Vrijmoet_2020,
       author = {{Vrijmoet}, Eliot Halley and {Henry}, Todd J. and {Jao}, Wei-Chun and {Dieterich}, Serge B.},
        title = "{The Solar Neighborhood. XLVI. Revealing New M Dwarf Binaries and Their Orbital Architectures}",
      journal = {\aj},
         year = 2020,
        month = nov,
       volume = {160},
       number = {5},
          eid = {215},
        pages = {215},
          doi = {10.3847/1538-3881/abb4e9},
archivePrefix = {arXiv},
       eprint = {2009.00121},
 primaryClass = {astro-ph.SR},
       adsurl = {https://ui.adsabs.harvard.edu/abs/2020AJ....160..215V}
}

@ARTICLE{Babusiaux2023,
       author = {{Babusiaux}, C. and {Fabricius}, C. and {Khanna}, S. and {Muraveva}, T. and {Reyl{\'e}}, C. and {Spoto}, F. and {Vallenari}, A. and {Luri}, X. and {Arenou}, F. and {{\'A}lvarez}, M.~A. and {Anders}, F. and {Antoja}, T. and {Balbinot}, E. and {Barache}, C. and {Bauchet}, N. and {Bossini}, D. and {Busonero}, D. and {Cantat-Gaudin}, T. and {Carrasco}, J.~M. and {Dafonte}, C. and {Diakit{\'e}}, S. and {Figueras}, F. and {Garcia-Gutierrez}, A. and {Garofalo}, A. and {Helmi}, A. and {Jim{\'e}nez-Arranz}, {\'O}. and {Jordi}, C. and {Kervella}, P. and {Kostrzewa-Rutkowska}, Z. and {Leclerc}, N. and {Licata}, E. and {Manteiga}, M. and {Masip}, A. and {Mongui{\'o}}, M. and {Ramos}, P. and {Robichon}, N. and {Robin}, A.~C. and {Romero-G{\'o}mez}, M. and {S{\'a}ez}, A. and {Santove{\~n}a}, R. and {Spina}, L. and {Torralba Elipe}, G. and {Weiler}, M.},
        title = "{Gaia Data Release 3. Catalogue validation}",
      journal = {\aap},
         year = 2023,
        month = jun,
       volume = {674},
          eid = {A32},
        pages = {A32},
          doi = {10.1051/0004-6361/202243790},
archivePrefix = {arXiv},
       eprint = {2206.05989},
 primaryClass = {astro-ph.SR},
       adsurl = {https://ui.adsabs.harvard.edu/abs/2023A&A...674A..32B}
}

@ARTICLE{Gaia_2018,
       author = {{Gaia Collaboration} and {Brown}, A.~G.~A. and {Vallenari}, A. and {Prusti}, T. and {de Bruijne}, J.~H.~J. and {Babusiaux}, C. and {Bailer-Jones}, C.~A.~L. and {Biermann}, M. and {Evans}, D.~W. and {Eyer}, L. and {Jansen}, F. and {Jordi}, C. and {Klioner}, S.~A. and {Lammers}, U. and {Lindegren}, L. and {Luri}, X. and {Mignard}, F. and {Panem}, C. and {Pourbaix}, D. and {Randich}, S. and {Sartoretti}, P. and {Siddiqui}, H.~I. and {Soubiran}, C. and {van Leeuwen}, F. and {Walton}, N.~A. and {Arenou}, F. and {Bastian}, U. and {Cropper}, M. and {Drimmel}, R. and {Katz}, D. and {Lattanzi}, M.~G. and {Bakker}, J. and {Cacciari}, C. and {Casta{\~n}eda}, J. and {Chaoul}, L. and {Cheek}, N. and {De Angeli}, F. and {Fabricius}, C. and {Guerra}, R. and {Holl}, B. and {Masana}, E. and {Messineo}, R. and {Mowlavi}, N. and {Nienartowicz}, K. and {Panuzzo}, P. and {Portell}, J. and {Riello}, M. and {Seabroke}, G.~M. and {Tanga}, P. and {Th{\'e}venin}, F. and {Gracia-Abril}, G. and {Comoretto}, G. and {Garcia-Reinaldos}, M. and {Teyssier}, D. and {Altmann}, M. and {Andrae}, R. and {Audard}, M. and {Bellas-Velidis}, I. and {Benson}, K. and {Berthier}, J. and {Blomme}, R. and {Burgess}, P. and {Busso}, G. and {Carry}, B. and {Cellino}, A. and {Clementini}, G. and {Clotet}, M. and {Creevey}, O. and {Davidson}, M. and {De Ridder}, J. and {Delchambre}, L. and {Dell'Oro}, A. and {Ducourant}, C. and {Fern{\'a}ndez-Hern{\'a}ndez}, J. and {Fouesneau}, M. and {Fr{\'e}mat}, Y. and {Galluccio}, L. and {Garc{\'\i}a-Torres}, M. and {Gonz{\'a}lez-N{\'u}{\~n}ez}, J. and {Gonz{\'a}lez-Vidal}, J.~J. and {Gosset}, E. and {Guy}, L.~P. and {Halbwachs}, J. -L. and {Hambly}, N.~C. and {Harrison}, D.~L. and {Hern{\'a}ndez}, J. and {Hestroffer}, D. and {Hodgkin}, S.~T. and {Hutton}, A. and {Jasniewicz}, G. and {Jean-Antoine-Piccolo}, A. and {Jordan}, S. and {Korn}, A.~J. and {Krone-Martins}, A. and {Lanzafame}, A.~C. and {Lebzelter}, T. and {L{\"o}ffler}, W. and {Manteiga}, M. and {Marrese}, P.~M. and {Mart{\'\i}n-Fleitas}, J.~M. and {Moitinho}, A. and {Mora}, A. and {Muinonen}, K. and {Osinde}, J. and {Pancino}, E. and {Pauwels}, T. and {Petit}, J. -M. and {Recio-Blanco}, A. and {Richards}, P.~J. and {Rimoldini}, L. and {Robin}, A.~C. and {Sarro}, L.~M. and {Siopis}, C. and {Smith}, M. and {Sozzetti}, A. and {S{\"u}veges}, M. and {Torra}, J. and {van Reeven}, W. and {Abbas}, U. and {Abreu Aramburu}, A. and {Accart}, S. and {Aerts}, C. and {Altavilla}, G. and {{\'A}lvarez}, M.~A. and {Alvarez}, R. and {Alves}, J. and {Anderson}, R.~I. and {Andrei}, A.~H. and {Anglada Varela}, E. and {Antiche}, E. and {Antoja}, T. and {Arcay}, B. and {Astraatmadja}, T.~L. and {Bach}, N. and {Baker}, S.~G. and {Balaguer-N{\'u}{\~n}ez}, L. and {Balm}, P. and {Barache}, C. and {Barata}, C. and {Barbato}, D. and {Barblan}, F. and {Barklem}, P.~S. and {Barrado}, D. and {Barros}, M. and {Barstow}, M.~A. and {Bartholom{\'e} Mu{\~n}oz}, S. and {Bassilana}, J. -L. and {Becciani}, U. and {Bellazzini}, M. and {Berihuete}, A. and {Bertone}, S. and {Bianchi}, L. and {Bienaym{\'e}}, O. and {Blanco-Cuaresma}, S. and {Boch}, T. and {Boeche}, C. and {Bombrun}, A. and {Borrachero}, R. and {Bossini}, D. and {Bouquillon}, S. and {Bourda}, G. and {Bragaglia}, A. and {Bramante}, L. and {Breddels}, M.~A. and {Bressan}, A. and {Brouillet}, N. and {Br{\"u}semeister}, T. and {Brugaletta}, E. and {Bucciarelli}, B. and {Burlacu}, A. and {Busonero}, D. and {Butkevich}, A.~G. and {Buzzi}, R. and {Caffau}, E. and {Cancelliere}, R. and {Cannizzaro}, G. and {Cantat-Gaudin}, T. and {Carballo}, R. and {Carlucci}, T. and {Carrasco}, J.~M. and {Casamiquela}, L. and {Castellani}, M. and {Castro-Ginard}, A. and {Charlot}, P. and {Chemin}, L. and {Chiavassa}, A. and {Cocozza}, G. and {Costigan}, G. and {Cowell}, S. and {Crifo}, F. and {Crosta}, M. and {Crowley}, C. and {Cuypers}, J. and {Dafonte}, C. and {Damerdji}, Y. and {Dapergolas}, A. and {David}, P. and {David}, M. and {de Laverny}, P. and {De Luise}, F.},
        title = "{Gaia Data Release 2. Summary of the contents and survey properties}",
      journal = {\aap},
         year = 2018,
        month = aug,
       volume = {616},
          eid = {A1},
        pages = {A1},
          doi = {10.1051/0004-6361/201833051},
archivePrefix = {arXiv},
       eprint = {1804.09365},
 primaryClass = {astro-ph.GA},
       adsurl = {https://ui.adsabs.harvard.edu/abs/2018A&A...616A...1G}
}

@ARTICLE{ElBadry_2021,
       author = {{El-Badry}, Kareem and {Rix}, Hans-Walter and {Heintz}, Tyler M.},
        title = "{A million binaries from Gaia eDR3: sample selection and validation of Gaia parallax uncertainties}",
      journal = {\mnras},
         year = 2021,
        month = sep,
       volume = {506},
       number = {2},
        pages = {2269-2295},
          doi = {10.1093/mnras/stab323},
archivePrefix = {arXiv},
       eprint = {2101.05282},
 primaryClass = {astro-ph.SR},
       adsurl = {https://ui.adsabs.harvard.edu/abs/2021MNRAS.506.2269E}
}

@ARTICLE{SB9,
       author = {{Pourbaix}, D. and {Tokovinin}, A.~A. and {Batten}, A.~H. and {Fekel}, F.~C. and {Hartkopf}, W.~I. and {Levato}, H. and {Morrell}, N.~I. and {Torres}, G. and {Udry}, S.},
        title = "{S$_{B$^{9}$}$: The ninth catalogue of spectroscopic binary orbits}",
      journal = {\aap},
         year = 2004,
        month = sep,
       volume = {424},
        pages = {727-732},
          doi = {10.1051/0004-6361:20041213},
archivePrefix = {arXiv},
       eprint = {astro-ph/0406573},
 primaryClass = {astro-ph},
       adsurl = {https://ui.adsabs.harvard.edu/abs/2004A&A...424..727P}
}

@ARTICLE{Henry_Jao_2024,
       author = {{Henry}, Todd J. and {Jao}, Wei-Chun},
        title = "{The Character of M Dwarfs}",
      journal = {\araa},
         year = 2024,
        month = sep,
       volume = {62},
       number = {1},
        pages = {593-633},
          doi = {10.1146/annurev-astro-052722-102740},
       adsurl = {https://ui.adsabs.harvard.edu/abs/2024ARA&A..62..593H}
}

@ARTICLE{vanLee_2007,
       author = {{van Leeuwen}, F.},
        title = "{Validation of the new Hipparcos reduction}",
      journal = {\aap},
         year = 2007,
        month = nov,
       volume = {474},
       number = {2},
        pages = {653-664},
          doi = {10.1051/0004-6361:20078357},
archivePrefix = {arXiv},
       eprint = {0708.1752},
 primaryClass = {astro-ph},
       adsurl = {https://ui.adsabs.harvard.edu/abs/2007A&A...474..653V}
}

@ARTICLE{Perryman_1997,
       author = {{Perryman}, M.~A.~C. and {Lindegren}, L. and {Kovalevsky}, J. and {Hoeg}, E. and {Bastian}, U. and {Bernacca}, P.~L. and {Cr{\'e}z{\'e}}, M. and {Donati}, F. and {Grenon}, M. and {Grewing}, M. and {van Leeuwen}, F. and {van der Marel}, H. and {Mignard}, F. and {Murray}, C.~A. and {Le Poole}, R.~S. and {Schrijver}, H. and {Turon}, C. and {Arenou}, F. and {Froeschl{\'e}}, M. and {Petersen}, C.~S.},
        title = "{The HIPPARCOS Catalogue}",
      journal = {\aap},
         year = 1997,
        month = jul,
       volume = {323},
        pages = {L49-L52},
       adsurl = {https://ui.adsabs.harvard.edu/abs/1997A&A...323L..49P}
}

@ARTICLE{Kroupa_1991,
       author = {{Kroupa}, P. and {Tout}, C.~A. and {Gilmore}, G.},
        title = "{The effects of unresolved binary stars on the determination of the stellar mass function.}",
      journal = {\mnras},
         year = 1991,
        month = jul,
       volume = {251},
        pages = {293-302},
          doi = {10.1093/mnras/251.2.293},
       adsurl = {https://ui.adsabs.harvard.edu/abs/1991MNRAS.251..293K}
}

@ARTICLE{ElBadry_2019,
       author = {{El-Badry}, Kareem and {Rix}, Hans-Walter and {Ting}, Yuan-Sen and {Weisz}, Daniel R. and {Bergemann}, Maria and {Cargile}, Phillip and {Conroy}, Charlie and {Eilers}, Anna-Christina},
        title = "{Signatures of unresolved binaries in stellar spectra: implications for spectral fitting}",
      journal = {\mnras},
         year = 2018,
        month = feb,
       volume = {473},
       number = {4},
        pages = {5043-5049},
          doi = {10.1093/mnras/stx2758},
archivePrefix = {arXiv},
       eprint = {1709.03983},
 primaryClass = {astro-ph.SR},
       adsurl = {https://ui.adsabs.harvard.edu/abs/2018MNRAS.473.5043E}
}

@ARTICLE{Henry_1990,
       author = {{Henry}, Todd J. and {McCarthy}, Jr., D.~W.},
        title = "{A Systematic Search for Brown Dwarfs Orbiting Nearby Stars}",
      journal = {\apj},
         year = 1990,
        month = feb,
       volume = {350},
        pages = {334},
          doi = {10.1086/168387},
       adsurl = {https://ui.adsabs.harvard.edu/abs/1990ApJ...350..334H}
}

@ARTICLE{Simons_1996,
       author = {{Simons}, D.~A. and {Henry}, Todd J. and {Kirkpatrick}, J. Davy},
        title = "{The Solar Neighborhood.III.A Near-Infrared Search for Widely Separated Low-Mass Binaries}",
      journal = {\aj},
         year = 1996,
        month = nov,
       volume = {112},
        pages = {2238},
          doi = {10.1086/118176},
       adsurl = {https://ui.adsabs.harvard.edu/abs/1996AJ....112.2238S}
}

@ARTICLE{Tuchow_2025,
       author = {{Tuchow}, Noah W. and {Harada}, Caleb K. and {Mamajek}, Eric E. and {Tanner}, Angelle and {Hinkel}, Natalie R. and {Belikov}, Ruslan and {Sirbu}, Dan and {Ciardi}, David R. and {Stark}, Christopher C. and {Morgan}, Rhonda M. and {Savransky}, Dmitry and {Turmon}, Michael},
        title = "{HWO Target Stars and Systems: A Prioritized Community List of Potential Stellar Targets for the Habitable Worlds Observatory's ExoEarth Survey}",
      journal = {\pasp},
         year = 2025,
        month = oct,
       volume = {137},
       number = {10},
          eid = {104402},
        pages = {104402},
          doi = {10.1088/1538-3873/ae0a81},
archivePrefix = {arXiv},
       eprint = {2509.20544},
 primaryClass = {astro-ph.SR},
       adsurl = {https://ui.adsabs.harvard.edu/abs/2025PASP..137j4402T}
}

@ARTICLE{Raghavan_2010,
       author = {{Raghavan}, Deepak and {McAlister}, Harold A. and {Henry}, Todd J. and {Latham}, David W. and {Marcy}, Geoffrey W. and {Mason}, Brian D. and {Gies}, Douglas R. and {White}, Russel J. and {ten Brummelaar}, Theo A.},
        title = "{A Survey of Stellar Families: Multiplicity of Solar-type Stars}",
      journal = {\apjs},
         year = 2010,
        month = sep,
       volume = {190},
       number = {1},
        pages = {1-42},
          doi = {10.1088/0067-0049/190/1/1},
archivePrefix = {arXiv},
       eprint = {1007.0414},
 primaryClass = {astro-ph.SR},
       adsurl = {https://ui.adsabs.harvard.edu/abs/2010ApJS..190....1R}
}

@INPROCEEDINGS{Offner_2023,
       author = {{Offner}, S.~S.~R. and {Moe}, M. and {Kratter}, K.~M. and {Sadavoy}, S.~I. and {Jensen}, E.~L.~N. and {Tobin}, J.~J.},
        title = "{The Origin and Evolution of Multiple Star Systems}",
    booktitle = {Protostars and Planets VII},
         year = 2023,
       editor = {{Inutsuka}, S. and {Aikawa}, Y. and {Muto}, T. and {Tomida}, K. and {Tamura}, M.},
       series = {Astronomical Society of the Pacific Conference Series},
       volume = {534},
        month = jul,
        pages = {275},
          doi = {10.48550/arXiv.2203.10066},
archivePrefix = {arXiv},
       eprint = {2203.10066},
 primaryClass = {astro-ph.SR},
       adsurl = {https://ui.adsabs.harvard.edu/abs/2023ASPC..534..275O}
}

@ARTICLE{Chabrier_2003,
       author = {{Chabrier}, Gilles},
        title = "{Galactic Stellar and Substellar Initial Mass Function}",
      journal = {\pasp},
         year = 2003,
        month = jul,
       volume = {115},
       number = {809},
        pages = {763-795},
          doi = {10.1086/376392},
archivePrefix = {arXiv},
       eprint = {astro-ph/0304382},
 primaryClass = {astro-ph},
       adsurl = {https://ui.adsabs.harvard.edu/abs/2003PASP..115..763C}
}

@ARTICLE{Golovin_2023,
       author = {{Golovin}, Alex and {Reffert}, Sabine and {Just}, Andreas and {Jordan}, Stefan and {Vani}, Akash and {Jahrei{\ss}}, Hartmut},
        title = "{The Fifth Catalogue of Nearby Stars (CNS5)}",
      journal = {\aap},
         year = 2023,
        month = feb,
       volume = {670},
          eid = {A19},
        pages = {A19},
          doi = {10.1051/0004-6361/202244250},
archivePrefix = {arXiv},
       eprint = {2211.01449},
 primaryClass = {astro-ph.SR},
       adsurl = {https://ui.adsabs.harvard.edu/abs/2023A&A...670A..19G}
}

@ARTICLE{Finch_2018,
       author = {{Finch}, Charlie T. and {Zacharias}, Norbert and {Jao}, Wei-Chun},
        title = "{URAT South Parallax Results}",
      journal = {\aj},
         year = 2018,
        month = apr,
       volume = {155},
       number = {4},
          eid = {176},
        pages = {176},
          doi = {10.3847/1538-3881/aab2b1},
       adsurl = {https://ui.adsabs.harvard.edu/abs/2018AJ....155..176F}
}

@ARTICLE{Keenan_1989,
       author = {{Keenan}, Philip C. and {McNeil}, Raymond C.},
        title = "{The Perkins Catalog of Revised MK Types for the Cooler Stars}",
      journal = {\apjs},
         year = 1989,
        month = oct,
       volume = {71},
        pages = {245},
          doi = {10.1086/191373},
       adsurl = {https://ui.adsabs.harvard.edu/abs/1989ApJS...71..245K}
}

@ARTICLE{Bartlett_2017,
       author = {{Bartlett}, Jennifer L. and {Lurie}, John C. and {Riedel}, Adric and {Ianna}, Philip A. and {Jao}, Wei-Chun and {Henry}, Todd J. and {Winters}, Jennifer G. and {Finch}, Charlie T. and {Subasavage}, John P.},
        title = "{The Solar Neighborhood. XXXX. Parallax Results from the CTIOPI 0.9 m Program: New Young Stars Near the Sun}",
      journal = {\aj},
         year = 2017,
        month = oct,
       volume = {154},
       number = {4},
          eid = {151},
        pages = {151},
          doi = {10.3847/1538-3881/aa8457},
       adsurl = {https://ui.adsabs.harvard.edu/abs/2017AJ....154..151B}
}

@ARTICLE{Bessel_1990,
       author = {{Bessel}, M.~S.},
        title = "{BVRI photometry of the Gliese catalogue stars.}",
      journal = {\aaps},
         year = 1990,
        month = may,
       volume = {83},
        pages = {357-378},
       adsurl = {https://ui.adsabs.harvard.edu/abs/1990A&AS...83..357B}
}

@ARTICLE{Bessel_1991,
       author = {{Bessell}, M.~S.},
        title = "{The Late M Dwarfs}",
      journal = {\aj},
         year = 1991,
        month = feb,
       volume = {101},
        pages = {662},
          doi = {10.1086/115714},
       adsurl = {https://ui.adsabs.harvard.edu/abs/1991AJ....101..662B}
}

@ARTICLE{Costa_2005,
       author = {{Costa}, Edgardo and {M{\'e}ndez}, Ren{\'e} A. and {Jao}, W.-C. and {Henry}, Todd J. and {Subasavage}, John P. and {Brown}, Misty A. and {Ianna}, Philip A. and {Bartlett}, Jennifer},
        title = "{The Solar Neighborhood. XIV. Parallaxes from the Cerro Tololo Inter-American Observatory Parallax Investigation-First Results from the 1.5 m Telescope Program}",
      journal = {\aj},
         year = 2005,
        month = jul,
       volume = {130},
       number = {1},
        pages = {337-349},
          doi = {10.1086/430473},
       adsurl = {https://ui.adsabs.harvard.edu/abs/2005AJ....130..337C}
}

@ARTICLE{Costa_2006,
       author = {{Costa}, Edgardo and {M{\'e}ndez}, Ren{\'e} A. and {Jao}, W.-C. and {Henry}, Todd J. and {Subasavage}, John P. and {Ianna}, Philip A.},
        title = "{The Solar Neighborhood. XVI. Parallaxes from CTIOPI: Final Results from the 1.5 m Telescope Program}",
      journal = {\aj},
         year = 2006,
        month = sep,
       volume = {132},
       number = {3},
        pages = {1234-1247},
          doi = {10.1086/505706},
       adsurl = {https://ui.adsabs.harvard.edu/abs/2006AJ....132.1234C}
}

@ARTICLE{Dahn88,
       author = {{Dahn}, C.~C. and {Harrington}, R.~S. and {Kallarakal}, V.~V. and {Guetter}, H.~H. and {Luginbuhl}, C.~B. and {Riepe}, B.~Y. and {Walker}, R.~L. and {Pier}, J.~R. and {Vrba}, F.~J. and {Monet}, D.~G. and {Ables}, H.~D.},
        title = "{U.S. Naval Observatory Parallaxes of Faint Stars, List VIII}",
      journal = {\aj},
         year = 1988,
        month = jan,
       volume = {95},
        pages = {237},
          doi = {10.1086/114633},
       adsurl = {https://ui.adsabs.harvard.edu/abs/1988AJ.....95..237D}
}

@ARTICLE{Dahn2002,
       author = {{Dahn}, Conard C. and {Harris}, Hugh C. and {Vrba}, Frederick J. and {Guetter}, Harry H. and {Canzian}, Blaise and {Henden}, Arne A. and {Levine}, Stephen E. and {Luginbuhl}, Christian B. and {Monet}, Alice K.~B. and {Monet}, David G. and {Pier}, Jeffrey R. and {Stone}, Ronald C. and {Walker}, Richard L. and {Burgasser}, Adam J. and {Gizis}, John E. and {Kirkpatrick}, J. Davy and {Liebert}, James and {Reid}, I. Neill},
        title = "{Astrometry and Photometry for Cool Dwarfs and Brown Dwarfs}",
      journal = {\aj},
         year = 2002,
        month = aug,
       volume = {124},
       number = {2},
        pages = {1170-1189},
          doi = {10.1086/341646},
archivePrefix = {arXiv},
       eprint = {astro-ph/0205050},
 primaryClass = {astro-ph},
       adsurl = {https://ui.adsabs.harvard.edu/abs/2002AJ....124.1170D}
}

@ARTICLE{Dav15,
       author = {{Davison}, Cassy L. and {White}, R.~J. and {Henry}, T.~J. and {Riedel}, A.~R. and {Jao}, W.-C. and {Bailey}, III, J.~I. and {Quinn}, S.~N. and {Cantrell}, J.~R. and {Subasavage}, J.~P. and {Winters}, J.~G.},
        title = "{A 3D Search for Companions to 12 Nearby M Dwarfs}",
      journal = {\aj},
         year = 2015,
        month = mar,
       volume = {149},
       number = {3},
          eid = {106},
        pages = {106},
          doi = {10.1088/0004-6256/149/3/106},
archivePrefix = {arXiv},
       eprint = {1501.05012},
 primaryClass = {astro-ph.SR},
       adsurl = {https://ui.adsabs.harvard.edu/abs/2015AJ....149..106D}
}

@ARTICLE{Dittmann_2014,
       author = {{Dittmann}, Jason A. and {Irwin}, Jonathan M. and {Charbonneau}, David and {Berta-Thompson}, Zachory K.},
        title = "{Trigonometric Parallaxes for 1507 Nearby Mid-to-late M Dwarfs}",
      journal = {\apj},
         year = 2014,
        month = apr,
       volume = {784},
       number = {2},
          eid = {156},
        pages = {156},
          doi = {10.1088/0004-637X/784/2/156},
archivePrefix = {arXiv},
       eprint = {1312.3241},
 primaryClass = {astro-ph.SR},
       adsurl = {https://ui.adsabs.harvard.edu/abs/2014ApJ...784..156D}
}

@ARTICLE{Dupuy_2017,
       author = {{Dupuy}, Trent J. and {Liu}, Michael C.},
        title = "{Individual Dynamical Masses of Ultracool Dwarfs}",
      journal = {\apjs},
         year = 2017,
        month = aug,
       volume = {231},
       number = {2},
          eid = {15},
        pages = {15},
          doi = {10.3847/1538-4365/aa5e4c},
archivePrefix = {arXiv},
       eprint = {1703.05775},
 primaryClass = {astro-ph.SR},
       adsurl = {https://ui.adsabs.harvard.edu/abs/2017ApJS..231...15D}
}

@ARTICLE{Gatewood_2009,
       author = {{Gatewood}, George and {Coban}, Louis},
        title = "{Allegheny Observatory Parallaxes for Late M Dwarfs and White Dwarfs}",
      journal = {\aj},
         year = 2009,
        month = jan,
       volume = {137},
       number = {1},
        pages = {402-405},
          doi = {10.1088/0004-6256/137/1/402},
       adsurl = {https://ui.adsabs.harvard.edu/abs/2009AJ....137..402G}
}

@ARTICLE{Harrington_1980,
       author = {{Harrington}, R.~S. and {Dahn}, C.~C.},
        title = "{Summary of U.S. Naval Observatory parallaxes.}",
      journal = {\aj},
         year = 1980,
        month = apr,
       volume = {85},
        pages = {454-465},
          doi = {10.1086/112696},
       adsurl = {https://ui.adsabs.harvard.edu/abs/1980AJ.....85..454H}
}

@ARTICLE{Heintz_1994,
       author = {{Heintz}, W.~D.},
        title = "{Photographic Astrometry of Binary and Proper-Motion Stars. VII.}",
      journal = {\aj},
         year = 1994,
        month = dec,
       volume = {108},
        pages = {2338},
          doi = {10.1086/117247},
       adsurl = {https://ui.adsabs.harvard.edu/abs/1994AJ....108.2338H}
}

@ARTICLE{Henry_2004,
       author = {{Henry}, Todd J. and {Subasavage}, John P. and {Brown}, Misty A. and {Beaulieu}, Thomas D. and {Jao}, Wei-Chun and {Hambly}, Nigel C.},
        title = "{The Solar Neighborhood. X. New Nearby Stars in the Southern Sky and Accurate Photometric Distance Estimates for Red Dwarfs}",
      journal = {\aj},
         year = 2004,
        month = nov,
       volume = {128},
       number = {5},
        pages = {2460-2473},
          doi = {10.1086/425052},
archivePrefix = {arXiv},
       eprint = {astro-ph/0408240},
 primaryClass = {astro-ph},
       adsurl = {https://ui.adsabs.harvard.edu/abs/2004AJ....128.2460H}
}

@ARTICLE{Hosey_2015,
       author = {{Hosey}, Altonio D. and {Henry}, Todd J. and {Jao}, Wei-Chun and {Dieterich}, Sergio B. and {Winters}, Jennifer G. and {Lurie}, John C. and {Riedel}, Adric R. and {Subasavage}, John P.},
        title = "{The Solar Neighborhood. XXXVI. The Long-term Photometric Variability of Nearby Red Dwarfs in the VRI Optical Bands}",
      journal = {\aj},
         year = 2015,
        month = jul,
       volume = {150},
       number = {1},
          eid = {6},
        pages = {6},
          doi = {10.1088/0004-6256/150/1/6},
archivePrefix = {arXiv},
       eprint = {1503.02100},
 primaryClass = {astro-ph.SR},
       adsurl = {https://ui.adsabs.harvard.edu/abs/2015AJ....150....6H}
}

@ARTICLE{Jao_2011,
       author = {{Jao}, Wei-Chun and {Henry}, Todd J. and {Subasavage}, John P. and {Winters}, Jennifer G. and {Riedel}, Adric R. and {Ianna}, Philip A.},
        title = "{The Solar Neighborhood. XXIV. Parallax Results from the CTIOPI 0.9 m Program: Stars with {\ensuremath{\mu}} >= 1farcs0 yr$^{-1}$ (MOTION Sample) and Subdwarfs}",
      journal = {\aj},
         year = 2011,
        month = apr,
       volume = {141},
       number = {4},
          eid = {117},
        pages = {117},
          doi = {10.1088/0004-6256/141/4/117},
archivePrefix = {arXiv},
       eprint = {1102.0994},
 primaryClass = {astro-ph.SR},
       adsurl = {https://ui.adsabs.harvard.edu/abs/2011AJ....141..117J}
}

@ARTICLE{Jao_2017,
       author = {{Jao}, Wei-Chun and {Henry}, Todd J. and {Winters}, Jennifer G. and {Subasavage}, John P. and {Riedel}, Adric R. and {Silverstein}, Michele L. and {Ianna}, Philip A.},
        title = "{The Solar Neighborhood. XLII. Parallax Results from the CTIOPI 0.9 m Program{\textemdash}Identifying New Nearby Subdwarfs Using Tangential Velocities and Locations on the H-R Diagram}",
      journal = {\aj},
         year = 2017,
        month = nov,
       volume = {154},
       number = {5},
          eid = {191},
        pages = {191},
          doi = {10.3847/1538-3881/aa8b64},
archivePrefix = {arXiv},
       eprint = {1709.02713},
 primaryClass = {astro-ph.SR},
       adsurl = {https://ui.adsabs.harvard.edu/abs/2017AJ....154..191J}
}

@ARTICLE{Kar_2024,
       author = {{Kar}, Aman and {Henry}, Todd J. and {Couperus}, Andrew A. and {Vrijmoet}, Eliot Halley and {Jao}, Wei-Chun},
        title = "{The Solar Neighborhood LI: A Variability Survey of Nearby M Dwarfs with Planets from Months to Decades with TESS and the CTIO/SMARTS 0.9 m Telescope}",
      journal = {\aj},
         year = 2024,
        month = may,
       volume = {167},
       number = {5},
          eid = {196},
        pages = {196},
          doi = {10.3847/1538-3881/ad2ddc},
archivePrefix = {arXiv},
       eprint = {2402.14121},
 primaryClass = {astro-ph.SR},
       adsurl = {https://ui.adsabs.harvard.edu/abs/2024AJ....167..196K}
}

@ARTICLE{Kilkenny_1998,
       author = {{Kilkenny}, D. and {van Wyk}, F. and {Roberts}, G. and {Marang}, F. and {Cooper}, D.},
        title = "{Supplementary southern standards for UBV(RI)\_C photometry}",
      journal = {\mnras},
         year = 1998,
        month = feb,
       volume = {294},
       number = {1},
        pages = {93-104},
          doi = {10.1046/j.1365-8711.1998.01222.x},
       adsurl = {https://ui.adsabs.harvard.edu/abs/1998MNRAS.294...93K}
}

@ARTICLE{Koen_2002,
       author = {{Koen}, C. and {Kilkenny}, D. and {van Wyk}, F. and {Cooper}, D. and {Marang}, F.},
        title = "{UBV(RI)$_{C}$ photometry of Hipparcos red stars}",
      journal = {\mnras},
         year = 2002,
        month = jul,
       volume = {334},
       number = {1},
        pages = {20-38},
          doi = {10.1046/j.1365-8711.2002.05403.x},
       adsurl = {https://ui.adsabs.harvard.edu/abs/2002MNRAS.334...20K}
}

@ARTICLE{Koen_2010,
       author = {{Koen}, C. and {Kilkenny}, D. and {van Wyk}, F. and {Marang}, F.},
        title = "{UBV(RI)$_{C}$ JHK observations of Hipparcos-selected nearby stars}",
      journal = {\mnras},
         year = 2010,
        month = apr,
       volume = {403},
       number = {4},
        pages = {1949-1968},
          doi = {10.1111/j.1365-2966.2009.16182.x},
       adsurl = {https://ui.adsabs.harvard.edu/abs/2010MNRAS.403.1949K}
}

@ARTICLE{Landolt_2009,
       author = {{Landolt}, Arlo U.},
        title = "{UBVRI Photometric Standard Stars Around the Celestial Equator: Updates and Additions}",
      journal = {\aj},
         year = 2009,
        month = may,
       volume = {137},
       number = {5},
        pages = {4186-4269},
          doi = {10.1088/0004-6256/137/5/4186},
archivePrefix = {arXiv},
       eprint = {0904.0638},
 primaryClass = {astro-ph.SR},
       adsurl = {https://ui.adsabs.harvard.edu/abs/2009AJ....137.4186L}
}

@ARTICLE{Landolt_1992,
       author = {{Landolt}, Arlo U.},
        title = "{UBVRI Photometric Standard Stars in the Magnitude Range 11.5 < V < 16.0 Around the Celestial Equator}",
      journal = {\aj},
         year = 1992,
        month = jul,
       volume = {104},
        pages = {340},
          doi = {10.1086/116242},
       adsurl = {https://ui.adsabs.harvard.edu/abs/1992AJ....104..340L}
}

@ARTICLE{Lazorenko_2025,
       author = {{Lazorenko}, P.~F. and {Sahlmann}, J. and {Mayor}, M. and {Martin}, E.~L. and {Osorio}, M.-R. Zapatero and {Girard}, J.},
        title = "{Resolving the unresolved: Discovery and dynamical masses of the brown dwarf binary DE1756{\ensuremath{-}}45}",
      journal = {\aap},
         year = 2025,
        month = dec,
       volume = {704},
          eid = {A291},
        pages = {A291},
          doi = {10.1051/0004-6361/202557011},
       adsurl = {https://ui.adsabs.harvard.edu/abs/2025A&A...704A.291L}
}

@ARTICLE{Lepine_2009,
       author = {{L{\'e}pine}, S{\'e}bastien and {Thorstensen}, John R. and {Shara}, Michael M. and {Rich}, R. Michael},
        title = "{New Neighbors: Parallaxes of 18 Nearby Stars Selected from the LSPM-North Catalog}",
      journal = {\aj},
         year = 2009,
        month = may,
       volume = {137},
       number = {5},
        pages = {4109-4117},
          doi = {10.1088/0004-6256/137/5/4109},
archivePrefix = {arXiv},
       eprint = {0901.3756},
 primaryClass = {astro-ph.SR},
       adsurl = {https://ui.adsabs.harvard.edu/abs/2009AJ....137.4109L}
}

@ARTICLE{Monet_1992,
       author = {{Monet}, David G. and {Dahn}, Conard C. and {Vrba}, Frederick J. and {Harris}, Hugh C. and {Pier}, Jeffrey R. and {Luginbuhl}, Christian B. and {Ables}, Harold D.},
        title = "{U.S. Naval Observatory CCD Parallaxes of Faint Stars. I. Program Description and First Results}",
      journal = {\aj},
         year = 1992,
        month = feb,
       volume = {103},
        pages = {638},
          doi = {10.1086/116091},
       adsurl = {https://ui.adsabs.harvard.edu/abs/1992AJ....103..638M}
}

@ARTICLE{Reid_2002,
       author = {{Reid}, I. Neill and {Kilkenny}, D. and {Cruz}, K.~L.},
        title = "{Meeting the Cool Neighbors. II. Photometry of Southern NLTT Stars}",
      journal = {\aj},
         year = 2002,
        month = may,
       volume = {123},
       number = {5},
        pages = {2822-2827},
          doi = {10.1086/339700},
archivePrefix = {arXiv},
       eprint = {astro-ph/0202460},
 primaryClass = {astro-ph},
       adsurl = {https://ui.adsabs.harvard.edu/abs/2002AJ....123.2822R}
}

@ARTICLE{Reid_2003,
       author = {{Reid}, I. Neill and {Cruz}, Kelle L. and {Allen}, Peter and {Mungall}, F. and {Kilkenny}, D. and {Liebert}, James and {Hawley}, Suzanne L. and {Fraser}, Oliver J. and {Covey}, Kevin R. and {Lowrance}, Patrick},
        title = "{Meeting the Cool Neighbors. VII. Spectroscopy of Faint Red NLTT Dwarfs}",
      journal = {\aj},
         year = 2003,
        month = dec,
       volume = {126},
       number = {6},
        pages = {3007-3016},
          doi = {10.1086/379173},
archivePrefix = {arXiv},
       eprint = {astro-ph/0308380},
 primaryClass = {astro-ph},
       adsurl = {https://ui.adsabs.harvard.edu/abs/2003AJ....126.3007R}
}

@ARTICLE{Riedel_2010,
       author = {{Riedel}, Adric R. and {Subasavage}, John P. and {Finch}, Charlie T. and {Jao}, Wei-Chun and {Henry}, Todd J. and {Winters}, Jennifer G. and {Brown}, Misty A. and {Ianna}, Philip A. and {Costa}, Edgardo and {Mendez}, Rene A.},
        title = "{The Solar Neighborhood. XXII. Parallax Results from the CTIOPI 0.9 m Program: Trigonometric Parallaxes of 64 Nearby Systems with 0farcs5 <={\ensuremath{\mu}}<= 1farcs0 yr$^{-1}$ (SLOWMO Sample)}",
      journal = {\aj},
         year = 2010,
        month = sep,
       volume = {140},
       number = {3},
        pages = {897-911},
          doi = {10.1088/0004-6256/140/3/897},
archivePrefix = {arXiv},
       eprint = {1008.0648},
 primaryClass = {astro-ph.SR},
       adsurl = {https://ui.adsabs.harvard.edu/abs/2010AJ....140..897R}
}

@ARTICLE{Riedel_2014,
       author = {{Riedel}, Adric R. and {Finch}, Charlie T. and {Henry}, Todd J. and {Subasavage}, John P. and {Jao}, Wei-Chun and {Malo}, Lison and {Rodriguez}, David R. and {White}, Russel J. and {Gies}, Douglas R. and {Dieterich}, Sergio B. and {Winters}, Jennifer G. and {Davison}, Cassy L. and {Nelan}, Edmund P. and {Blunt}, Sarah C. and {Cruz}, Kelle L. and {Rice}, Emily L. and {Ianna}, Philip A.},
        title = "{The Solar Neighborhood. XXXIII. Parallax Results from the CTIOPI 0.9 m Program: Trigonometric Parallaxes of Nearby Low-mass Active and Young Systems}",
      journal = {\aj},
         year = 2014,
        month = apr,
       volume = {147},
       number = {4},
          eid = {85},
        pages = {85},
          doi = {10.1088/0004-6256/147/4/85},
archivePrefix = {arXiv},
       eprint = {1401.0722},
 primaryClass = {astro-ph.SR},
       adsurl = {https://ui.adsabs.harvard.edu/abs/2014AJ....147...85R}
}

@ARTICLE{Riedel_2018,
       author = {{Riedel}, Adric R. and {Silverstein}, Michele L. and {Henry}, Todd J. and {Jao}, Wei-Chun and {Winters}, Jennifer G. and {Subasavage}, John P. and {Malo}, Lison and {Hambly}, Nigel C.},
        title = "{The Solar Neighborhood. XLIII. Discovery of New Nearby Stars with {\ensuremath{\mu}} < 0.″18 yr$^{-1}$ (TINYMO Sample)}",
      journal = {\aj},
         year = 2018,
        month = aug,
       volume = {156},
       number = {2},
          eid = {49},
        pages = {49},
          doi = {10.3847/1538-3881/aaca33},
archivePrefix = {arXiv},
       eprint = {1804.08812},
 primaryClass = {astro-ph.SR},
       adsurl = {https://ui.adsabs.harvard.edu/abs/2018AJ....156...49R}
}

@ARTICLE{Soderhjelm_1999,
       author = {{S{\"o}derhjelm}, Staffan},
        title = "{Visual binary orbits and masses POST HIPPARCOS}",
      journal = {\aap},
         year = 1999,
        month = jan,
       volume = {341},
        pages = {121-140},
       adsurl = {https://ui.adsabs.harvard.edu/abs/1999A&A...341..121S}
}

@BOOK{vanAltena_1995,
       author = {{van Altena}, W.~F. and {Lee}, J.~T. and {Hoffleit}, E.~D.},
        title = "{The general catalogue of trigonometric [stellar] parallaxes}",
         year = 1995,
       adsurl = {https://ui.adsabs.harvard.edu/abs/1995gcts.book.....V}
}

@ARTICLE{Weis_1984,
       author = {{Weis}, E.~W.},
        title = "{Photometric parallaxes for selected stars of color class M from the NLTT catalog.}",
      journal = {\apjs},
         year = 1984,
        month = jun,
       volume = {55},
        pages = {289-299},
          doi = {10.1086/190956},
       adsurl = {https://ui.adsabs.harvard.edu/abs/1984ApJS...55..289W}
}

@ARTICLE{Weis_1986,
       author = {{Weis}, E.~W.},
        title = "{Photometric parallaxes for selected stars of color class M from the NLTT catalog. II. The declination zone 0 to +20.}",
      journal = {\aj},
         year = 1986,
        month = mar,
       volume = {91},
        pages = {626-639},
          doi = {10.1086/114045},
       adsurl = {https://ui.adsabs.harvard.edu/abs/1986AJ.....91..626W}
}

@ARTICLE{Weis_1987,
       author = {{Weis}, Edward W.},
        title = "{Photometric Parallaxes for Selected Stars of Color Class M from the NLTT Catalog. III. The Declination Zone + 20 Degrees to +45 Degrees}",
      journal = {\aj},
         year = 1987,
        month = feb,
       volume = {93},
        pages = {451},
          doi = {10.1086/114330},
       adsurl = {https://ui.adsabs.harvard.edu/abs/1987AJ.....93..451W}
}

@ARTICLE{Weis_1988,
       author = {{Weis}, Edward W.},
        title = "{Photometric Parallaxes for Selected Stars of Color Class M from the NLTT Catalog. IV. The Declination Zone +45 to +90}",
      journal = {\aj},
         year = 1988,
        month = nov,
       volume = {96},
        pages = {1710},
          doi = {10.1086/114923},
       adsurl = {https://ui.adsabs.harvard.edu/abs/1988AJ.....96.1710W}
}

@ARTICLE{Weis_1991a,
       author = {{Weis}, Edward W.},
        title = "{VRI Photometry of Late Dwarf Common Proper Motion Pairs}",
      journal = {\aj},
         year = 1991,
        month = may,
       volume = {101},
        pages = {1882},
          doi = {10.1086/115814},
       adsurl = {https://ui.adsabs.harvard.edu/abs/1991AJ....101.1882W}
}

@ARTICLE{Weis_1991b,
       author = {{Weis}, Edward W.},
        title = "{A Photometric Study of K and M Dwarf Stars Found by Stephenson}",
      journal = {\aj},
         year = 1991,
        month = nov,
       volume = {102},
        pages = {1795},
          doi = {10.1086/116003},
       adsurl = {https://ui.adsabs.harvard.edu/abs/1991AJ....102.1795W}
}

@ARTICLE{Weis_1993,
       author = {{Weis}, Edward W.},
        title = "{Photometry of Dwarf K and M Stars}",
      journal = {\aj},
         year = 1993,
        month = may,
       volume = {105},
        pages = {1962},
          doi = {10.1086/116571},
       adsurl = {https://ui.adsabs.harvard.edu/abs/1993AJ....105.1962W}
}

@ARTICLE{Weis_1994,
       author = {{Weis}, Edward W.},
        title = "{Long Term Variability in Dwarf M Stars}",
      journal = {\aj},
         year = 1994,
        month = mar,
       volume = {107},
        pages = {1135},
          doi = {10.1086/116925},
       adsurl = {https://ui.adsabs.harvard.edu/abs/1994AJ....107.1135W}
}

@ARTICLE{Weis_1996,
       author = {{Weis}, Edward W.},
        title = "{Photometry of Stars with Large Proper Motion}",
      journal = {\aj},
         year = 1996,
        month = nov,
       volume = {112},
        pages = {2300},
          doi = {10.1086/118183},
       adsurl = {https://ui.adsabs.harvard.edu/abs/1996AJ....112.2300W}
}

@ARTICLE{Weis_1999,
       author = {{Weis}, E.~W. and {Lee}, J.~T. and {Lee}, A.~H. and {Griese}, III, J.~W. and {Vincent}, J.~M. and {Upgren}, A.~R.},
        title = "{Parallaxes and Proper Motions. XX.}",
      journal = {\aj},
         year = 1999,
        month = feb,
       volume = {117},
       number = {2},
        pages = {1037-1041},
          doi = {10.1086/300747},
       adsurl = {https://ui.adsabs.harvard.edu/abs/1999AJ....117.1037W}
}

@ARTICLE{Winters_2011,
       author = {{Winters}, Jennifer G. and {Henry}, Todd J. and {Jao}, Wei-Chun and {Subasavage}, John P. and {Finch}, Charlie T. and {Hambly}, Nigel C.},
        title = "{The Solar Neighborhood. XXIII. CCD Photometric Distance Estimates of SCR Targets{\textemdash}77 M Dwarf Systems within 25 pc}",
      journal = {\aj},
         year = 2011,
        month = jan,
       volume = {141},
       number = {1},
          eid = {21},
        pages = {21},
          doi = {10.1088/0004-6256/141/1/21},
archivePrefix = {arXiv},
       eprint = {1012.2078},
 primaryClass = {astro-ph.SR},
       adsurl = {https://ui.adsabs.harvard.edu/abs/2011AJ....141...21W}
}

@ARTICLE{Winters_2015,
       author = {{Winters}, Jennifer G. and {Henry}, Todd J. and {Lurie}, John C. and {Hambly}, Nigel C. and {Jao}, Wei-Chun and {Bartlett}, Jennifer L. and {Boyd}, Mark R. and {Dieterich}, Sergio B. and {Finch}, Charlie T. and {Hosey}, Altonio D. and {Ianna}, Philip A. and {Riedel}, Adric R. and {Slatten}, Kenneth J. and {Subasavage}, John P.},
        title = "{The Solar Neighborhood. XXXV. Distances to 1404 m Dwarf Systems Within 25 pc in the Southern Sky}",
      journal = {\aj},
         year = 2015,
        month = jan,
       volume = {149},
       number = {1},
          eid = {5},
        pages = {5},
          doi = {10.1088/0004-6256/149/1/5},
archivePrefix = {arXiv},
       eprint = {1407.7837},
 primaryClass = {astro-ph.SR},
       adsurl = {https://ui.adsabs.harvard.edu/abs/2015AJ....149....5W}
}

@ARTICLE{Winters_2017,
       author = {{Winters}, Jennifer G. and {Sevrinsky}, R. Andrew and {Jao}, Wei-Chun and {Henry}, Todd J. and {Riedel}, Adric R. and {Subasavage}, John P. and {Lurie}, John C. and {Ianna}, Philip A. and {Finch}, Charlie T.},
        title = "{The Solar Neighborhood XXXVIII. Results from the CTIO/SMARTS 0.9m: Trigonometric Parallaxes for 151 Nearby M Dwarf Systems}",
      journal = {\aj},
         year = 2017,
        month = jan,
       volume = {153},
       number = {1},
          eid = {14},
        pages = {14},
          doi = {10.3847/1538-3881/153/1/14},
archivePrefix = {arXiv},
       eprint = {1610.07552},
 primaryClass = {astro-ph.SR},
       adsurl = {https://ui.adsabs.harvard.edu/abs/2017AJ....153...14W}
}

@ARTICLE{Patterson_1998,
       author = {{Patterson}, Richard J. and {Ianna}, Philip A. and {Begam}, Michael C.},
        title = "{The Solar Neighborhood. V. VRI Photometry of Southern Nearby Star Candidates}",
      journal = {\aj},
         year = 1998,
        month = apr,
       volume = {115},
       number = {4},
        pages = {1648-1652},
          doi = {10.1086/300311},
       adsurl = {https://ui.adsabs.harvard.edu/abs/1998AJ....115.1648P}
}

@PHDTHESIS{Silverstein_2019,
       author = {{Silverstein}, Michele Louise},
        title = "{Sizing up red dwarfs in the solar neighborhood}",
       school = {Georgia State University},
         year = 2019,
        month = jan,
       adsurl = {https://ui.adsabs.harvard.edu/abs/2019PhDT.......182S}
}

@ARTICLE{Torres_2010,
       author = {{Torres}, G. and {Andersen}, J. and {Gim{\'e}nez}, A.},
        title = "{Accurate masses and radii of normal stars: modern results and applications}",
      journal = {A\&AR},
         year = 2010,
        month = feb,
       volume = {18},
       number = {1-2},
        pages = {67-126},
          doi = {10.1007/s00159-009-0025-1},
archivePrefix = {arXiv},
       eprint = {0908.2624},
 primaryClass = {astro-ph.SR},
       adsurl = {https://ui.adsabs.harvard.edu/abs/2010A&ARv..18...67T}
}

@ARTICLE{Baroch_2021,
       author = {{Baroch}, D. and {Morales}, J.~C. and {Ribas}, I. and {B{\'e}jar}, V.~J.~S. and {Reffert}, S. and {Cardona Guill{\'e}n}, C. and {Reiners}, A. and {Caballero}, J.~A. and {Quirrenbach}, A. and {Amado}, P.~J. and {Anglada-Escud{\'e}}, G. and {Colom{\'e}}, J. and {Cort{\'e}s-Contreras}, M. and {Dreizler}, S. and {Galad{\'\i}-Enr{\'\i}quez}, D. and {Hatzes}, A.~P. and {Jeffers}, S.~V. and {Henning}, Th. and {Herrero}, E. and {Kaminski}, A. and {K{\"u}rster}, M. and {Lafarga}, M. and {Lodieu}, N. and {L{\'o}pez-Gonz{\'a}lez}, M.~J. and {Montes}, D. and {Pall{\'e}}, E. and {Perger}, M. and {Pollacco}, D. and {Rodr{\'\i}guez-L{\'o}pez}, C. and {Rodr{\'\i}guez}, E. and {Rosich}, A. and {Sch{\"o}fer}, P. and {Schweitzer}, A. and {Shan}, Y. and {Tal-Or}, L. and {Zechmeister}, M.},
        title = "{The CARMENES search for exoplanets around M dwarfs. Spectroscopic orbits of nine M-dwarf multiple systems, including two triples, two brown dwarf candidates, and one close M-dwarf-white dwarf binary}",
      journal = {\aap},
         year = 2021,
        month = sep,
       volume = {653},
          eid = {A49},
        pages = {A49},
          doi = {10.1051/0004-6361/202141031},
archivePrefix = {arXiv},
       eprint = {2105.14770},
 primaryClass = {astro-ph.SR},
       adsurl = {https://ui.adsabs.harvard.edu/abs/2021A&A...653A..49B}
}

@ARTICLE{Khrutskaya_2010,
       author = {{Khrutskaya}, E.~V. and {Izmailov}, I.~S. and {Khovrichev}, M. Yu.},
        title = "{Trigonometric parallaxes of 29 stars with large proper motions}",
      journal = {Astronomy Letters},
         year = 2010,
        month = aug,
       volume = {36},
       number = {8},
        pages = {576-583},
          doi = {10.1134/S1063773710080062},
       adsurl = {https://ui.adsabs.harvard.edu/abs/2010AstL...36..576K}
}

@ARTICLE{Jao_2014,
       author = {{Jao}, Wei-Chun and {Henry}, Todd J. and {Subasavage}, John P. and {Winters}, Jennifer G. and {Gies}, Douglas R. and {Riedel}, Adric R. and {Ianna}, Philip A.},
        title = "{The Solar Neighborhood. XXXI. Discovery of an Unusual Red+White Dwarf Binary at \raisebox{-0.5ex}\textasciitilde25 pc via Astrometry and UV Imaging}",
      journal = {\aj},
         year = 2014,
        month = jan,
       volume = {147},
       number = {1},
          eid = {21},
        pages = {21},
          doi = {10.1088/0004-6256/147/1/21},
archivePrefix = {arXiv},
       eprint = {1310.4746},
 primaryClass = {astro-ph.SR},
       adsurl = {https://ui.adsabs.harvard.edu/abs/2014AJ....147...21J}
}

@ARTICLE{Gizis_1998,
       author = {{Gizis}, John E.},
        title = "{High Chromospheric Activity in M Subdwarfs}",
      journal = {\aj},
         year = 1998,
        month = may,
       volume = {115},
       number = {5},
        pages = {2053-2058},
          doi = {10.1086/300325},
archivePrefix = {arXiv},
       eprint = {astro-ph/9801305},
 primaryClass = {astro-ph},
       adsurl = {https://ui.adsabs.harvard.edu/abs/1998AJ....115.2053G}
}

@ARTICLE{Mace_2018,
       author = {{Mace}, Gregory N. and {Mann}, Andrew W. and {Skiff}, Brian A. and {Sneden}, Christopher and {Kirkpatrick}, J. Davy and {Schneider}, Adam C. and {Kidder}, Benjamin and {Gosnell}, Natalie M. and {Kim}, Hwihyun and {Mulligan}, Brian W. and {Prato}, L. and {Jaffe}, Daniel},
        title = "{Wolf 1130: A Nearby Triple System Containing a Cool, Ultramassive White Dwarf}",
      journal = {\apj},
         year = 2018,
        month = feb,
       volume = {854},
       number = {2},
          eid = {145},
        pages = {145},
          doi = {10.3847/1538-4357/aaa8dd},
archivePrefix = {arXiv},
       eprint = {1802.04803},
 primaryClass = {astro-ph.SR},
       adsurl = {https://ui.adsabs.harvard.edu/abs/2018ApJ...854..145M}
}

@ARTICLE{Honaker_2020,
       author = {{Honaker}, Easton J. and {Mace}, Gregory N. and {Han}, Eunkyu and {Hussaini}, Maryam and {Lubar}, Emily},
        title = "{TESS Photometry of the Precataclysmic Variable Wolf 1130AB}",
      journal = {Research Notes of the American Astronomical Society},
         year = 2020,
        month = nov,
       volume = {4},
       number = {11},
          eid = {197},
        pages = {197},
          doi = {10.3847/2515-5172/abc6a4},
       adsurl = {https://ui.adsabs.harvard.edu/abs/2020RNAAS...4..197H}
}

@ARTICLE{Kervella_2022,
       author = {{Kervella}, Pierre and {Borgniet}, Simon and {Domiciano de Souza}, Armando and {M{\'e}rand}, Antoine and {Gallenne}, Alexandre and {Rivinius}, Thomas and {Lacour}, Sylvestre and {Carciofi}, Alex and {Faes}, Daniel Moser and {Le Bouquin}, Jean-Baptiste and {Taormina}, Monica and {Pilecki}, Bogumi{\l} and {Berger}, Jean-Philippe and {Bendjoya}, Philippe and {Klement}, Robert and {Millour}, Florentin and {Janot-Pacheco}, Eduardo and {Spang}, Alain and {Vakili}, Farrokh},
        title = "{The binary system of the spinning-top Be star Achernar}",
      journal = {\aap},
         year = 2022,
        month = nov,
       volume = {667},
          eid = {A111},
        pages = {A111},
          doi = {10.1051/0004-6361/202244009},
archivePrefix = {arXiv},
       eprint = {2209.07537},
 primaryClass = {astro-ph.SR},
       adsurl = {https://ui.adsabs.harvard.edu/abs/2022A&A...667A.111K}
}

@ARTICLE{Silvestri_2002,
       author = {{Silvestri}, Nicole M. and {Oswalt}, Terry D. and {Hawley}, Suzanne L.},
        title = "{Wide Binary Systems and the Nature of High-Velocity White Dwarfs}",
      journal = {\aj},
         year = 2002,
        month = aug,
       volume = {124},
       number = {2},
        pages = {1118-1126},
          doi = {10.1086/341382},
       adsurl = {https://ui.adsabs.harvard.edu/abs/2002AJ....124.1118S}
}

@ARTICLE{Schilbach_Roser_2012,
       author = {{Schilbach}, E. and {R{\"o}ser}, S.},
        title = "{New white dwarfs in the Hyades. Results from kinematic and photometric studies}",
      journal = {\aap},
         year = 2012,
        month = jan,
       volume = {537},
          eid = {A129},
        pages = {A129},
          doi = {10.1051/0004-6361/201117688},
archivePrefix = {arXiv},
       eprint = {1111.3959},
 primaryClass = {astro-ph.GA},
       adsurl = {https://ui.adsabs.harvard.edu/abs/2012A&A...537A.129S}
}

@ARTICLE{Mason_2001,
       author = {{Mason}, Brian D. and {Wycoff}, Gary L. and {Hartkopf}, William I. and {Douglass}, Geoffrey G. and {Worley}, Charles E.},
        title = "{The 2001 US Naval Observatory Double Star CD-ROM. I. The Washington Double Star Catalog}",
      journal = {\aj},
         year = 2001,
        month = dec,
       volume = {122},
       number = {6},
        pages = {3466-3471},
          doi = {10.1086/323920},
       adsurl = {https://ui.adsabs.harvard.edu/abs/2001AJ....122.3466M}
}

@ARTICLE{Delfosse_1999,
       author = {{Delfosse}, X. and {Forveille}, T. and {Beuzit}, J.-L. and {Udry}, S. and {Mayor}, M. and {Perrier}, C.},
        title = "{New neighbours. I. 13 new companions to nearby M dwarfs}",
      journal = {\aap},
         year = 1999,
        month = apr,
       volume = {344},
        pages = {897-910},
          doi = {10.48550/arXiv.astro-ph/9812008},
archivePrefix = {arXiv},
       eprint = {astro-ph/9812008},
 primaryClass = {astro-ph},
       adsurl = {https://ui.adsabs.harvard.edu/abs/1999A&A...344..897D}
}

@ARTICLE{Hollands_2018,
       author = {{Hollands}, M.~A. and {Tremblay}, P.-E. and {G{\"a}nsicke}, B.~T. and {Gentile-Fusillo}, N.~P. and {Toonen}, S.},
        title = "{The Gaia 20 pc white dwarf sample}",
      journal = {\mnras},
         year = 2018,
        month = nov,
       volume = {480},
       number = {3},
        pages = {3942-3961},
          doi = {10.1093/mnras/sty2057},
archivePrefix = {arXiv},
       eprint = {1805.12590},
 primaryClass = {astro-ph.SR},
       adsurl = {https://ui.adsabs.harvard.edu/abs/2018MNRAS.480.3942H}
}

@PHDTHESIS{Henry_1991,
       author = {{Henry}, Todd Jackson},
        title = "{A Systematic Search for Low Mass Companions Orbiting Nearby Stars and the Calibration of the End of the Stellar Main Sequence.}",
       school = {University of Arizona},
         year = 1991,
        month = jan,
       adsurl = {https://ui.adsabs.harvard.edu/abs/1991PhDT........11H}
}

@ARTICLE{Gaia_2023,
       author = {{Gaia Collaboration} and {Vallenari}, A. and {Brown}, A.~G.~A. and {Prusti}, T. and {de Bruijne}, J.~H.~J. and {Arenou}, F. and {Babusiaux}, C. and {Biermann}, M. and {Creevey}, O.~L. and {Ducourant}, C. and {Evans}, D.~W. and {Eyer}, L. and {Guerra}, R. and {Hutton}, A. and {Jordi}, C. and {Klioner}, S.~A. and {Lammers}, U.~L. and {Lindegren}, L. and {Luri}, X. and {Mignard}, F. and {Panem}, C. and {Pourbaix}, D. and {Randich}, S. and {Sartoretti}, P. and {Soubiran}, C. and {Tanga}, P. and {Walton}, N.~A. and {Bailer-Jones}, C.~A.~L. and {Bastian}, U. and {Drimmel}, R. and {Jansen}, F. and {Katz}, D. and {Lattanzi}, M.~G. and {van Leeuwen}, F. and {Bakker}, J. and {Cacciari}, C. and {Casta{\~n}eda}, J. and {De Angeli}, F. and {Fabricius}, C. and {Fouesneau}, M. and {Fr{\'e}mat}, Y. and {Galluccio}, L. and {Guerrier}, A. and {Heiter}, U. and {Masana}, E. and {Messineo}, R. and {Mowlavi}, N. and {Nicolas}, C. and {Nienartowicz}, K. and {Pailler}, F. and {Panuzzo}, P. and {Riclet}, F. and {Roux}, W. and {Seabroke}, G.~M. and {Sordo}, R. and {Th{\'e}venin}, F. and {Gracia-Abril}, G. and {Portell}, J. and {Teyssier}, D. and {Altmann}, M. and {Andrae}, R. and {Audard}, M. and {Bellas-Velidis}, I. and {Benson}, K. and {Berthier}, J. and {Blomme}, R. and {Burgess}, P.~W. and {Busonero}, D. and {Busso}, G. and {C{\'a}novas}, H. and {Carry}, B. and {Cellino}, A. and {Cheek}, N. and {Clementini}, G. and {Damerdji}, Y. and {Davidson}, M. and {de Teodoro}, P. and {Nu{\~n}ez Campos}, M. and {Delchambre}, L. and {Dell'Oro}, A. and {Esquej}, P. and {Fern{\'a}ndez-Hern{\'a}ndez}, J. and {Fraile}, E. and {Garabato}, D. and {Garc{\'\i}a-Lario}, P. and {Gosset}, E. and {Haigron}, R. and {Halbwachs}, J.-L. and {Hambly}, N.~C. and {Harrison}, D.~L. and {Hern{\'a}ndez}, J. and {Hestroffer}, D. and {Hodgkin}, S.~T. and {Holl}, B. and {Jan{\ss}en}, K. and {Jevardat de Fombelle}, G. and {Jordan}, S. and {Krone-Martins}, A. and {Lanzafame}, A.~C. and {L{\"o}ffler}, W. and {Marchal}, O. and {Marrese}, P.~M. and {Moitinho}, A. and {Muinonen}, K. and {Osborne}, P. and {Pancino}, E. and {Pauwels}, T. and {Recio-Blanco}, A. and {Reyl{\'e}}, C. and {Riello}, M. and {Rimoldini}, L. and {Roegiers}, T. and {Rybizki}, J. and {Sarro}, L.~M. and {Siopis}, C. and {Smith}, M. and {Sozzetti}, A. and {Utrilla}, E. and {van Leeuwen}, M. and {Abbas}, U. and {{\'A}brah{\'a}m}, P. and {Abreu Aramburu}, A. and {Aerts}, C. and {Aguado}, J.~J. and {Ajaj}, M. and {Aldea-Montero}, F. and {Altavilla}, G. and {{\'A}lvarez}, M.~A. and {Alves}, J. and {Anders}, F. and {Anderson}, R.~I. and {Anglada Varela}, E. and {Antoja}, T. and {Baines}, D. and {Baker}, S.~G. and {Balaguer-N{\'u}{\~n}ez}, L. and {Balbinot}, E. and {Balog}, Z. and {Barache}, C. and {Barbato}, D. and {Barros}, M. and {Barstow}, M.~A. and {Bartolom{\'e}}, S. and {Bassilana}, J.-L. and {Bauchet}, N. and {Becciani}, U. and {Bellazzini}, M. and {Berihuete}, A. and {Bernet}, M. and {Bertone}, S. and {Bianchi}, L. and {Binnenfeld}, A. and {Blanco-Cuaresma}, S. and {Blazere}, A. and {Boch}, T. and {Bombrun}, A. and {Bossini}, D. and {Bouquillon}, S. and {Bragaglia}, A. and {Bramante}, L. and {Breedt}, E. and {Bressan}, A. and {Brouillet}, N. and {Brugaletta}, E. and {Bucciarelli}, B. and {Burlacu}, A. and {Butkevich}, A.~G. and {Buzzi}, R. and {Caffau}, E. and {Cancelliere}, R. and {Cantat-Gaudin}, T. and {Carballo}, R. and {Carlucci}, T. and {Carnerero}, M.~I. and {Carrasco}, J.~M. and {Casamiquela}, L. and {Castellani}, M. and {Castro-Ginard}, A. and {Chaoul}, L. and {Charlot}, P. and {Chemin}, L. and {Chiaramida}, V. and {Chiavassa}, A. and {Chornay}, N. and {Comoretto}, G. and {Contursi}, G. and {Cooper}, W.~J. and {Cornez}, T. and {Cowell}, S. and {Crifo}, F. and {Cropper}, M. and {Crosta}, M. and {Crowley}, C. and {Dafonte}, C. and {Dapergolas}, A. and {David}, M. and {David}, P. and {de Laverny}, P. and {De Luise}, F. and {De March}, R.},
        title = "{Gaia Data Release 3. Summary of the content and survey properties}",
      journal = {\aap},
         year = 2023,
        month = jun,
       volume = {674},
          eid = {A1},
        pages = {A1},
          doi = {10.1051/0004-6361/202243940},
archivePrefix = {arXiv},
       eprint = {2208.00211},
 primaryClass = {astro-ph.GA},
       adsurl = {https://ui.adsabs.harvard.edu/abs/2023A&A...674A...1G}
}

@ARTICLE{Tokovinin_2017,
       author = {{Tokovinin}, Andrei},
        title = "{Orbit Alignment in Triple Stars}",
      journal = {\apj},
         year = 2017,
        month = aug,
       volume = {844},
       number = {2},
          eid = {103},
        pages = {103},
          doi = {10.3847/1538-4357/aa7746},
archivePrefix = {arXiv},
       eprint = {1706.00748},
 primaryClass = {astro-ph.SR},
       adsurl = {https://ui.adsabs.harvard.edu/abs/2017ApJ...844..103T}
}

@ARTICLE{Law_2010,
       author = {{Law}, N.~M. and {Dhital}, S. and {Kraus}, A. and {Stassun}, K.~G. and {West}, A.~A.},
        title = "{The High-order Multiplicity of Unusually Wide M Dwarf Binaries: Eleven New Triple and Quadruple Systems}",
      journal = {\apj},
         year = 2010,
        month = sep,
       volume = {720},
       number = {2},
        pages = {1727-1737},
          doi = {10.1088/0004-637X/720/2/1727},
archivePrefix = {arXiv},
       eprint = {1007.3735},
 primaryClass = {astro-ph.SR},
       adsurl = {https://ui.adsabs.harvard.edu/abs/2010ApJ...720.1727L}
}

@INCOLLECTION{Kroupa_2013,
       author = {{Kroupa}, Pavel and {Weidner}, Carsten and {Pflamm-Altenburg}, Jan and {Thies}, Ingo and {Dabringhausen}, J{\"o}rg and {Marks}, Michael and {Maschberger}, Thomas},
        title = "{The Stellar and Sub-Stellar Initial Mass Function of Simple and Composite Populations}",
    booktitle = {Planets, Stars and Stellar Systems. Volume 5: Galactic Structure and Stellar Populations},
         year = 2013,
       editor = {{Oswalt}, Terry D. and {Gilmore}, Gerard},
       volume = {5},
        pages = {115},
          doi = {10.1007/978-94-007-5612-0_4},
       adsurl = {https://ui.adsabs.harvard.edu/abs/2013pss5.book..115K}
}

@ARTICLE{Kirkpatrick_2024,
       author = {{Kirkpatrick}, J. Davy and {Marocco}, Federico and {Gelino}, Christopher R. and {Raghu}, Yadukrishna and {Faherty}, Jacqueline K. and {Bardalez Gagliuffi}, Daniella C. and {Schurr}, Steven D. and {Apps}, Kevin and {Schneider}, Adam C. and {Meisner}, Aaron M. and {Kuchner}, Marc J. and {Caselden}, Dan and {Smart}, R.~L. and {Casewell}, S.~L. and {Raddi}, Roberto and {Kesseli}, Aurora and {Stevnbak Andersen}, Nikolaj and {Antonini}, Edoardo and {Beaulieu}, Paul and {Bickle}, Thomas P. and {Bilsing}, Martin and {Chieng}, Raymond and {Colin}, Guillaume and {Deen}, Sam and {Dereveanco}, Alexandru and {Doll}, Katharina and {Durantini Luca}, Hugo A. and {Frazer}, Anya and {Gantier}, Jean Marc and {Gramaize}, L{\'e}opold and {Grant}, Kristin and {Hamlet}, Leslie K. and {Higashimura}, Hiro and {Hyogo}, Michiharu and {Ja{\l}owiczor}, Peter A. and {Jonkeren}, Alexander and {Kabatnik}, Martin and {Kiwy}, Frank and {Martin}, David W. and {Michaels}, Marianne N. and {Pendrill}, William and {Pessanha Machado}, Celso and {Pumphrey}, Benjamin and {Rothermich}, Austin and {Russwurm}, Rebekah and {Sainio}, Arttu and {Sanchez}, John and {Sapelkin-Tambling}, Fyodor Theo and {Sch{\"u}mann}, J{\"o}rg and {Selg-Mann}, Karl and {Singh}, Harshdeep and {Stenner}, Andres and {Sun}, Guoyou and {Tanner}, Christopher and {Th{\'e}venot}, Melina and {Ventura}, Maurizio and {Voloshin}, Nikita V. and {Walla}, Jim and {W{\k{e}}dracki}, Zbigniew and {Adorno}, Jose I. and {Aganze}, Christian and {Allers}, Katelyn N. and {Brooks}, Hunter and {Burgasser}, Adam J. and {Calamari}, Emily and {Connor}, Thomas and {Costa}, Edgardo and {Eisenhardt}, Peter R. and {Gagn{\'e}}, Jonathan and {Gerasimov}, Roman and {Gonzales}, Eileen C. and {Hsu}, Chih-Chun and {Kiman}, Rocio and {Li}, Guodong and {Low}, Ryan and {Mamajek}, Eric and {Pantoja}, Blake M. and {Popinchalk}, Mark and {Rees}, Jon M. and {Stern}, Daniel and {Su{\'a}rez}, Genaro and {Theissen}, Christopher and {Tsai}, Chao-Wei and {Vos}, Johanna M. and {Zurek}, David and {The Backyard Worlds: Planet 9 Collaboration}},
        title = "{The Initial Mass Function Based on the Full-sky 20 pc Census of {\ensuremath{\sim}}3600 Stars and Brown Dwarfs}",
      journal = {\apjs},
         year = 2024,
        month = apr,
       volume = {271},
       number = {2},
          eid = {55},
        pages = {55},
          doi = {10.3847/1538-4365/ad24e2},
archivePrefix = {arXiv},
       eprint = {2312.03639},
 primaryClass = {astro-ph.SR},
       adsurl = {https://ui.adsabs.harvard.edu/abs/2024ApJS..271...55K}
}

@INPROCEEDINGS{Carrazco_2026,
       author = {{Carrazco Gaxiola}, Sebastian and {Bieryla}, Allyson and {Henry}, Todd and {Latham}, David and {Jao}, Wei-Chun and {Johns}, Timothy and {Quinn}, Samuel and {White}, Russel},
        title = "{An All-Sky Spectroscopic Reconnaissance of more than 2100 K Dwarfs within 40 Parsecs using High-Resolution Spectra}",
    booktitle = {American Astronomical Society Meeting Abstracts},
         year = 2026,
       series = {American Astronomical Society Meeting Abstracts},
       volume = {247},
        month = feb,
          eid = {231.06},
        pages = {231.06},
       adsurl = {https://ui.adsabs.harvard.edu/abs/2026AAS...24723106C}
}

@ARTICLE{Fischer_1992,
       author = {{Fischer}, Debra A. and {Marcy}, Geoffrey W.},
        title = "{Multiplicity among M Dwarfs}",
      journal = {\apj},
         year = 1992,
        month = sep,
       volume = {396},
        pages = {178},
          doi = {10.1086/171708},
       adsurl = {https://ui.adsabs.harvard.edu/abs/1992ApJ...396..178F}
}

@ARTICLE{Ward-Duong_2015,
       author = {{Ward-Duong}, K. and {Patience}, J. and {De Rosa}, R.~J. and {Bulger}, J. and {Rajan}, A. and {Goodwin}, S.~P. and {Parker}, Richard J. and {McCarthy}, D.~W. and {Kulesa}, C.},
        title = "{The M-dwarfs in Multiples (MINMS) survey - I. Stellar multiplicity among low-mass stars within 15 pc}",
      journal = {\mnras},
         year = 2015,
        month = may,
       volume = {449},
       number = {3},
        pages = {2618-2637},
          doi = {10.1093/mnras/stv384},
archivePrefix = {arXiv},
       eprint = {1503.00724},
 primaryClass = {astro-ph.SR},
       adsurl = {https://ui.adsabs.harvard.edu/abs/2015MNRAS.449.2618W}
}

@INPROCEEDINGS{Johns_2026_AAS,
       author = {{Johns}, Timothy and {LeBlanc}, Madison and {Henry}, Todd and {Carrazco Gaxiola}, Jose and {Hubbard-James}, Hodari-Sadiki and {Jao}, Wei-Chun and {Paredes}, Leonardo},
        title = "{The RKSTAR Catalog: Getting to Know *ALL* of the 4466 K Dwarf Systems Within 50 Parsecs}",
    booktitle = {American Astronomical Society Meeting Abstracts},
         year = 2026,
       series = {American Astronomical Society Meeting Abstracts},
       volume = {247},
        month = feb,
          eid = {211.02},
        pages = {211.02},
       adsurl = {https://ui.adsabs.harvard.edu/abs/2026AAS...24721102J}
}

@ARTICLE{Just_2015,
       author = {{Just}, A. and {Fuchs}, B. and {Jahrei{\ss}}, H. and {Flynn}, C. and {Dettbarn}, C. and {Rybizki}, J.},
        title = "{The local stellar luminosity function and mass-to-light ratio in the near-infrared}",
      journal = {\mnras},
         year = 2015,
        month = jul,
       volume = {451},
       number = {1},
        pages = {149-158},
          doi = {10.1093/mnras/stv858},
archivePrefix = {arXiv},
       eprint = {1504.05808},
 primaryClass = {astro-ph.GA},
       adsurl = {https://ui.adsabs.harvard.edu/abs/2015MNRAS.451..149J}
}

@dataset{HIP_TYC_1997,
       author = {{Esa}, 1997},
        title = "{VizieR Online Data Catalog: The Hipparcos and Tycho Catalogues (ESA 1997)}",
 howpublished = {VizieR On-line Data Catalog: I/239.  Originally published in: 1997HIP...C......0E},
         year = 1997,
        month = feb,
          eid = {I/239},
       adsurl = {https://ui.adsabs.harvard.edu/abs/1997yCat.1239....0E}
}

@ARTICLE{Kroupa_2002,
       author = {{Kroupa}, Pavel},
        title = "{The Initial Mass Function of Stars: Evidence for Uniformity in Variable Systems}",
      journal = {Science},
         year = 2002,
        month = jan,
       volume = {295},
       number = {5552},
        pages = {82-91},
          doi = {10.1126/science.1067524},
archivePrefix = {arXiv},
       eprint = {astro-ph/0201098},
 primaryClass = {astro-ph},
       adsurl = {https://ui.adsabs.harvard.edu/abs/2002Sci...295...82K}
}

@ARTICLE{GaiaDR2_2016,
       author = {{Gaia Collaboration} and {Prusti}, T. and {de Bruijne}, J.~H.~J. and {Brown}, A.~G.~A. and {Vallenari}, A. and {Babusiaux}, C. and {Bailer-Jones}, C.~A.~L. and {Bastian}, U. and {Biermann}, M. and {Evans}, D.~W. and {Eyer}, L. and {Jansen}, F. and {Jordi}, C. and {Klioner}, S.~A. and {Lammers}, U. and {Lindegren}, L. and {Luri}, X. and {Mignard}, F. and {Milligan}, D.~J. and {Panem}, C. and {Poinsignon}, V. and {Pourbaix}, D. and {Randich}, S. and {Sarri}, G. and {Sartoretti}, P. and {Siddiqui}, H.~I. and {Soubiran}, C. and {Valette}, V. and {van Leeuwen}, F. and {Walton}, N.~A. and {Aerts}, C. and {Arenou}, F. and {Cropper}, M. and {Drimmel}, R. and {H{\o}g}, E. and {Katz}, D. and {Lattanzi}, M.~G. and {O'Mullane}, W. and {Grebel}, E.~K. and {Holland}, A.~D. and {Huc}, C. and {Passot}, X. and {Bramante}, L. and {Cacciari}, C. and {Casta{\~n}eda}, J. and {Chaoul}, L. and {Cheek}, N. and {De Angeli}, F. and {Fabricius}, C. and {Guerra}, R. and {Hern{\'a}ndez}, J. and {Jean-Antoine-Piccolo}, A. and {Masana}, E. and {Messineo}, R. and {Mowlavi}, N. and {Nienartowicz}, K. and {Ord{\'o}{\~n}ez-Blanco}, D. and {Panuzzo}, P. and {Portell}, J. and {Richards}, P.~J. and {Riello}, M. and {Seabroke}, G.~M. and {Tanga}, P. and {Th{\'e}venin}, F. and {Torra}, J. and {Els}, S.~G. and {Gracia-Abril}, G. and {Comoretto}, G. and {Garcia-Reinaldos}, M. and {Lock}, T. and {Mercier}, E. and {Altmann}, M. and {Andrae}, R. and {Astraatmadja}, T.~L. and {Bellas-Velidis}, I. and {Benson}, K. and {Berthier}, J. and {Blomme}, R. and {Busso}, G. and {Carry}, B. and {Cellino}, A. and {Clementini}, G. and {Cowell}, S. and {Creevey}, O. and {Cuypers}, J. and {Davidson}, M. and {De Ridder}, J. and {de Torres}, A. and {Delchambre}, L. and {Dell'Oro}, A. and {Ducourant}, C. and {Fr{\'e}mat}, Y. and {Garc{\'\i}a-Torres}, M. and {Gosset}, E. and {Halbwachs}, J.-L. and {Hambly}, N.~C. and {Harrison}, D.~L. and {Hauser}, M. and {Hestroffer}, D. and {Hodgkin}, S.~T. and {Huckle}, H.~E. and {Hutton}, A. and {Jasniewicz}, G. and {Jordan}, S. and {Kontizas}, M. and {Korn}, A.~J. and {Lanzafame}, A.~C. and {Manteiga}, M. and {Moitinho}, A. and {Muinonen}, K. and {Osinde}, J. and {Pancino}, E. and {Pauwels}, T. and {Petit}, J.-M. and {Recio-Blanco}, A. and {Robin}, A.~C. and {Sarro}, L.~M. and {Siopis}, C. and {Smith}, M. and {Smith}, K.~W. and {Sozzetti}, A. and {Thuillot}, W. and {van Reeven}, W. and {Viala}, Y. and {Abbas}, U. and {Abreu Aramburu}, A. and {Accart}, S. and {Aguado}, J.~J. and {Allan}, P.~M. and {Allasia}, W. and {Altavilla}, G. and {{\'A}lvarez}, M.~A. and {Alves}, J. and {Anderson}, R.~I. and {Andrei}, A.~H. and {Anglada Varela}, E. and {Antiche}, E. and {Antoja}, T. and {Ant{\'o}n}, S. and {Arcay}, B. and {Atzei}, A. and {Ayache}, L. and {Bach}, N. and {Baker}, S.~G. and {Balaguer-N{\'u}{\~n}ez}, L. and {Barache}, C. and {Barata}, C. and {Barbier}, A. and {Barblan}, F. and {Baroni}, M. and {Barrado y Navascu{\'e}s}, D. and {Barros}, M. and {Barstow}, M.~A. and {Becciani}, U. and {Bellazzini}, M. and {Bellei}, G. and {Bello Garc{\'\i}a}, A. and {Belokurov}, V. and {Bendjoya}, P. and {Berihuete}, A. and {Bianchi}, L. and {Bienaym{\'e}}, O. and {Billebaud}, F. and {Blagorodnova}, N. and {Blanco-Cuaresma}, S. and {Boch}, T. and {Bombrun}, A. and {Borrachero}, R. and {Bouquillon}, S. and {Bourda}, G. and {Bouy}, H. and {Bragaglia}, A. and {Breddels}, M.~A. and {Brouillet}, N. and {Br{\"u}semeister}, T. and {Bucciarelli}, B. and {Budnik}, F. and {Burgess}, P. and {Burgon}, R. and {Burlacu}, A. and {Busonero}, D. and {Buzzi}, R. and {Caffau}, E. and {Cambras}, J. and {Campbell}, H. and {Cancelliere}, R. and {Cantat-Gaudin}, T. and {Carlucci}, T. and {Carrasco}, J.~M. and {Castellani}, M. and {Charlot}, P. and {Charnas}, J. and {Charvet}, P. and {Chassat}, F. and {Chiavassa}, A. and {Clotet}, M. and {Cocozza}, G. and {Collins}, R.~S. and {Collins}, P. and {Costigan}, G.},
        title = "{The Gaia mission}",
      journal = {\aap},
         year = 2016,
        month = nov,
       volume = {595},
          eid = {A1},
        pages = {A1},
          doi = {10.1051/0004-6361/201629272},
archivePrefix = {arXiv},
       eprint = {1609.04153},
 primaryClass = {astro-ph.IM},
       adsurl = {https://ui.adsabs.harvard.edu/abs/2016A&A...595A...1G}
}

@ARTICLE{GaiaDR2_2018b,
       author = {{Gaia Collaboration} and {Brown}, A.~G.~A. and {Vallenari}, A. and {Prusti}, T. and {de Bruijne}, J.~H.~J. and {Babusiaux}, C. and {Bailer-Jones}, C.~A.~L. and {Biermann}, M. and {Evans}, D.~W. and {Eyer}, L. and {Jansen}, F. and {Jordi}, C. and {Klioner}, S.~A. and {Lammers}, U. and {Lindegren}, L. and {Luri}, X. and {Mignard}, F. and {Panem}, C. and {Pourbaix}, D. and {Randich}, S. and {Sartoretti}, P. and {Siddiqui}, H.~I. and {Soubiran}, C. and {van Leeuwen}, F. and {Walton}, N.~A. and {Arenou}, F. and {Bastian}, U. and {Cropper}, M. and {Drimmel}, R. and {Katz}, D. and {Lattanzi}, M.~G. and {Bakker}, J. and {Cacciari}, C. and {Casta{\~n}eda}, J. and {Chaoul}, L. and {Cheek}, N. and {De Angeli}, F. and {Fabricius}, C. and {Guerra}, R. and {Holl}, B. and {Masana}, E. and {Messineo}, R. and {Mowlavi}, N. and {Nienartowicz}, K. and {Panuzzo}, P. and {Portell}, J. and {Riello}, M. and {Seabroke}, G.~M. and {Tanga}, P. and {Th{\'e}venin}, F. and {Gracia-Abril}, G. and {Comoretto}, G. and {Garcia-Reinaldos}, M. and {Teyssier}, D. and {Altmann}, M. and {Andrae}, R. and {Audard}, M. and {Bellas-Velidis}, I. and {Benson}, K. and {Berthier}, J. and {Blomme}, R. and {Burgess}, P. and {Busso}, G. and {Carry}, B. and {Cellino}, A. and {Clementini}, G. and {Clotet}, M. and {Creevey}, O. and {Davidson}, M. and {De Ridder}, J. and {Delchambre}, L. and {Dell'Oro}, A. and {Ducourant}, C. and {Fern{\'a}ndez-Hern{\'a}ndez}, J. and {Fouesneau}, M. and {Fr{\'e}mat}, Y. and {Galluccio}, L. and {Garc{\'\i}a-Torres}, M. and {Gonz{\'a}lez-N{\'u}{\~n}ez}, J. and {Gonz{\'a}lez-Vidal}, J.~J. and {Gosset}, E. and {Guy}, L.~P. and {Halbwachs}, J.-L. and {Hambly}, N.~C. and {Harrison}, D.~L. and {Hern{\'a}ndez}, J. and {Hestroffer}, D. and {Hodgkin}, S.~T. and {Hutton}, A. and {Jasniewicz}, G. and {Jean-Antoine-Piccolo}, A. and {Jordan}, S. and {Korn}, A.~J. and {Krone-Martins}, A. and {Lanzafame}, A.~C. and {Lebzelter}, T. and {L{\"o}ffler}, W. and {Manteiga}, M. and {Marrese}, P.~M. and {Mart{\'\i}n-Fleitas}, J.~M. and {Moitinho}, A. and {Mora}, A. and {Muinonen}, K. and {Osinde}, J. and {Pancino}, E. and {Pauwels}, T. and {Petit}, J.-M. and {Recio-Blanco}, A. and {Richards}, P.~J. and {Rimoldini}, L. and {Robin}, A.~C. and {Sarro}, L.~M. and {Siopis}, C. and {Smith}, M. and {Sozzetti}, A. and {S{\"u}veges}, M. and {Torra}, J. and {van Reeven}, W. and {Abbas}, U. and {Abreu Aramburu}, A. and {Accart}, S. and {Aerts}, C. and {Altavilla}, G. and {{\'A}lvarez}, M.~A. and {Alvarez}, R. and {Alves}, J. and {Anderson}, R.~I. and {Andrei}, A.~H. and {Anglada Varela}, E. and {Antiche}, E. and {Antoja}, T. and {Arcay}, B. and {Astraatmadja}, T.~L. and {Bach}, N. and {Baker}, S.~G. and {Balaguer-N{\'u}{\~n}ez}, L. and {Balm}, P. and {Barache}, C. and {Barata}, C. and {Barbato}, D. and {Barblan}, F. and {Barklem}, P.~S. and {Barrado}, D. and {Barros}, M. and {Barstow}, M.~A. and {Bartholom{\'e} Mu{\~n}oz}, S. and {Bassilana}, J.-L. and {Becciani}, U. and {Bellazzini}, M. and {Berihuete}, A. and {Bertone}, S. and {Bianchi}, L. and {Bienaym{\'e}}, O. and {Blanco-Cuaresma}, S. and {Boch}, T. and {Boeche}, C. and {Bombrun}, A. and {Borrachero}, R. and {Bossini}, D. and {Bouquillon}, S. and {Bourda}, G. and {Bragaglia}, A. and {Bramante}, L. and {Breddels}, M.~A. and {Bressan}, A. and {Brouillet}, N. and {Br{\"u}semeister}, T. and {Brugaletta}, E. and {Bucciarelli}, B. and {Burlacu}, A. and {Busonero}, D. and {Butkevich}, A.~G. and {Buzzi}, R. and {Caffau}, E. and {Cancelliere}, R. and {Cannizzaro}, G. and {Cantat-Gaudin}, T. and {Carballo}, R. and {Carlucci}, T. and {Carrasco}, J.~M. and {Casamiquela}, L. and {Castellani}, M. and {Castro-Ginard}, A. and {Charlot}, P. and {Chemin}, L. and {Chiavassa}, A. and {Cocozza}, G. and {Costigan}, G. and {Cowell}, S. and {Crifo}, F. and {Crosta}, M. and {Crowley}, C. and {Cuypers}, J. and {Dafonte}, C. and {Damerdji}, Y. and {Dapergolas}, A. and {David}, P. and {David}, M. and {de Laverny}, P. and {De Luise}, F.},
        title = "{Gaia Data Release 2. Summary of the contents and survey properties}",
      journal = {\aap},
         year = 2018,
        month = aug,
       volume = {616},
          eid = {A1},
        pages = {A1},
          doi = {10.1051/0004-6361/201833051},
archivePrefix = {arXiv},
       eprint = {1804.09365},
 primaryClass = {astro-ph.GA},
       adsurl = {https://ui.adsabs.harvard.edu/abs/2018A&A...616A...1G}
}

@ARTICLE{Lepine_2005,
       author = {{L{\'e}pine}, S{\'e}bastien},
        title = "{Nearby Stars from the LSPM-North Proper-Motion Catalog. I. Main-Sequence Dwarfs and Giants within 33 Parsecs of the Sun}",
      journal = {\aj},
         year = 2005,
        month = oct,
       volume = {130},
       number = {4},
        pages = {1680-1692},
          doi = {10.1086/432792},
archivePrefix = {arXiv},
       eprint = {astro-ph/0506152},
 primaryClass = {astro-ph},
       adsurl = {https://ui.adsabs.harvard.edu/abs/2005AJ....130.1680L}
}

@ARTICLE{Scholz_2005,
       author = {{Scholz}, R.-D. and {Meusinger}, H. and {Jahrei{\ss}}, H.},
        title = "{Search for nearby stars among proper motion stars selected by optical-to-infrared photometry. III. Spectroscopic distances of 322 NLTT stars}",
      journal = {\aap},
         year = 2005,
        month = oct,
       volume = {442},
       number = {1},
        pages = {211-227},
          doi = {10.1051/0004-6361:20053004},
archivePrefix = {arXiv},
       eprint = {astro-ph/0507284},
 primaryClass = {astro-ph},
       adsurl = {https://ui.adsabs.harvard.edu/abs/2005A&A...442..211S}
}

@ARTICLE{Kaminski_2025,
       author = {{Kaminski}, A. and {Sabotta}, S. and {Kemmer}, J. and {Chaturvedi}, P. and {Burn}, R. and {Morales}, J.~C. and {Caballero}, J.~A. and {Ribas}, I. and {Reiners}, A. and {Quirrenbach}, A. and {Amado}, P.~J. and {B{\'e}jar}, V.~J.~S. and {Dreizler}, S. and {Guenther}, E.~W. and {Hatzes}, A.~P. and {Henning}, Th. and {K{\"u}rster}, M. and {Montes}, D. and {Nagel}, E. and {Pall{\'e}}, E. and {Pinter}, V. and {Reffert}, S. and {Schlecker}, M. and {Shan}, Y. and {Trifonov}, T. and {Osorio}, M.~R. Zapatero and {Zechmeister}, M.},
        title = "{The CARMENES search for exoplanets around M dwarfs: Occurrence rates of Earth-like planets around very low-mass stars}",
      journal = {\aap},
         year = 2025,
        month = apr,
       volume = {696},
          eid = {A101},
        pages = {A101},
          doi = {10.1051/0004-6361/202453381},
archivePrefix = {arXiv},
       eprint = {2504.03364},
 primaryClass = {astro-ph.EP},
       adsurl = {https://ui.adsabs.harvard.edu/abs/2025A&A...696A.101K}
}

@ARTICLE{Ribas_2023,
       author = {{Ribas}, I. and {Reiners}, A. and {Zechmeister}, M. and {Caballero}, J.~A. and {Morales}, J.~C. and {Sabotta}, S. and {Baroch}, D. and {Amado}, P.~J. and {Quirrenbach}, A. and {Abril}, M. and {Aceituno}, J. and {Anglada-Escud{\'e}}, G. and {Azzaro}, M. and {Barrado}, D. and {B{\'e}jar}, V.~J.~S. and {Ben{\'\i}tez de Haro}, D. and {Bergond}, G. and {Bluhm}, P. and {Calvo Ortega}, R. and {Cardona Guill{\'e}n}, C. and {Chaturvedi}, P. and {Cifuentes}, C. and {Colom{\'e}}, J. and {Cont}, D. and {Cort{\'e}s-Contreras}, M. and {Czesla}, S. and {D{\'\i}ez-Alonso}, E. and {Dreizler}, S. and {Duque-Arribas}, C. and {Espinoza}, N. and {Fern{\'a}ndez}, M. and {Fuhrmeister}, B. and {Galad{\'\i}-Enr{\'\i}quez}, D. and {Garc{\'\i}a-L{\'o}pez}, A. and {Gonz{\'a}lez-{\'A}lvarez}, E. and {Gonz{\'a}lez Hern{\'a}ndez}, J.~I. and {Guenther}, E.~W. and {de Guindos}, E. and {Hatzes}, A.~P. and {Henning}, Th. and {Herrero}, E. and {Hintz}, D. and {Huelmo}, {\'A}. L. and {Jeffers}, S.~V. and {Johnson}, E.~N. and {de Juan}, E. and {Kaminski}, A. and {Kemmer}, J. and {Khaimova}, J. and {Khalafinejad}, S. and {Kossakowski}, D. and {K{\"u}rster}, M. and {Labarga}, F. and {Lafarga}, M. and {Lalitha}, S. and {Lamp{\'o}n}, M. and {Lillo-Box}, J. and {Lodieu}, N. and {L{\'o}pez Gonz{\'a}lez}, M.~J. and {L{\'o}pez-Puertas}, M. and {Luque}, R. and {Mag{\'a}n}, H. and {Mancini}, L. and {Marfil}, E. and {Mart{\'\i}n}, E.~L. and {Mart{\'\i}n-Ruiz}, S. and {Molaverdikhani}, K. and {Montes}, D. and {Nagel}, E. and {Nortmann}, L. and {Nowak}, G. and {Pall{\'e}}, E. and {Passegger}, V.~M. and {Pavlov}, A. and {Pedraz}, S. and {Perdelwitz}, V. and {Perger}, M. and {Ram{\'o}n-Ballesta}, A. and {Reffert}, S. and {Revilla}, D. and {Rodr{\'\i}guez}, E. and {Rodr{\'\i}guez-L{\'o}pez}, C. and {Sadegi}, S. and {S{\'a}nchez Carrasco}, M. {\'A}. and {S{\'a}nchez-L{\'o}pez}, A. and {Sanz-Forcada}, J. and {Sch{\"a}fer}, S. and {Schlecker}, M. and {Schmitt}, J.~H.~M.~M. and {Sch{\"o}fer}, P. and {Schweitzer}, A. and {Seifert}, W. and {Shan}, Y. and {Skrzypinski}, S.~L. and {Solano}, E. and {Stahl}, O. and {Stangret}, M. and {Stock}, S. and {St{\"u}rmer}, J. and {Tabernero}, H.~M. and {Tal-Or}, L. and {Trifonov}, T. and {Vanaverbeke}, S. and {Yan}, F. and {Zapatero Osorio}, M.~R.},
        title = "{The CARMENES search for exoplanets around M dwarfs. Guaranteed time observations Data Release 1 (2016-2020)}",
      journal = {\aap},
         year = 2023,
        month = feb,
       volume = {670},
          eid = {A139},
        pages = {A139},
          doi = {10.1051/0004-6361/202244879},
archivePrefix = {arXiv},
       eprint = {2302.10528},
 primaryClass = {astro-ph.EP},
       adsurl = {https://ui.adsabs.harvard.edu/abs/2023A&A...670A.139R}
}

@ARTICLE{Cifuentes_2025,
       author = {{Cifuentes}, C. and {Caballero}, J.~A. and {Gonz{\'a}lez-Payo}, J. and {Amado}, P.~J. and {B{\'e}jar}, V.~J.~S. and {Burgasser}, A.~J. and {Cort{\'e}s-Contreras}, M. and {Lodieu}, N. and {Montes}, D. and {Quirrenbach}, A. and {Reiners}, A. and {Ribas}, I. and {Sanz-Forcada}, J. and {Seifert}, W. and {Zapatero Osorio}, M.~R.},
        title = "{CARMENES input catalogue of M dwarfs: IX. Multiplicity from close spectroscopic binaries to ultra-wide systems}",
      journal = {\aap},
         year = 2025,
        month = jan,
       volume = {693},
          eid = {A228},
        pages = {A228},
          doi = {10.1051/0004-6361/202452527},
archivePrefix = {arXiv},
       eprint = {2412.12264},
 primaryClass = {astro-ph.SR},
       adsurl = {https://ui.adsabs.harvard.edu/abs/2025A&A...693A.228C}
}

@ARTICLE{Janson_2012,
       author = {{Janson}, Markus and {Hormuth}, Felix and {Bergfors}, Carolina and {Brandner}, Wolfgang and {Hippler}, Stefan and {Daemgen}, Sebastian and {Kudryavtseva}, Natalia and {Schmalzl}, Eva and {Schnupp}, Carolin and {Henning}, Thomas},
        title = "{The AstraLux Large M-dwarf Multiplicity Survey}",
      journal = {\apj},
         year = 2012,
        month = jul,
       volume = {754},
       number = {1},
          eid = {44},
        pages = {44},
          doi = {10.1088/0004-637X/754/1/44},
archivePrefix = {arXiv},
       eprint = {1205.4718},
 primaryClass = {astro-ph.SR},
       adsurl = {https://ui.adsabs.harvard.edu/abs/2012ApJ...754...44J}
}

@ARTICLE{Bonfils_2013,
       author = {{Bonfils}, X. and {Lo Curto}, G. and {Correia}, A.~C.~M. and {Laskar}, J. and {Udry}, S. and {Delfosse}, X. and {Forveille}, T. and {Astudillo-Defru}, N. and {Benz}, W. and {Bouchy}, F. and {Gillon}, M. and {H{\'e}brard}, G. and {Lovis}, C. and {Mayor}, M. and {Moutou}, C. and {Naef}, D. and {Neves}, V. and {Pepe}, F. and {Perrier}, C. and {Queloz}, D. and {Santos}, N.~C. and {S{\'e}gransan}, D.},
        title = "{The HARPS search for southern extra-solar planets. XXXIV. A planetary system around the nearby M dwarf <ASTROBJ>GJ 163</ASTROBJ>, with a super-Earth possibly in the habitable zone}",
      journal = {\aap},
         year = 2013,
        month = aug,
       volume = {556},
          eid = {A110},
        pages = {A110},
          doi = {10.1051/0004-6361/201220237},
archivePrefix = {arXiv},
       eprint = {1306.0904},
 primaryClass = {astro-ph.EP},
       adsurl = {https://ui.adsabs.harvard.edu/abs/2013A&A...556A.110B}
}

@BOOK{Kippenhahn_1990,
       author = {{Kippenhahn}, Rudolf and {Weigert}, Alfred},
        title = "{Stellar Structure and Evolution}",
         year = 1990,
       adsurl = {https://ui.adsabs.harvard.edu/abs/1990sse..book.....K}
}

@ARTICLE{Weinberg_1987,
       author = {{Weinberg}, Martin D. and {Shapiro}, Stuart L. and {Wasserman}, Ira},
        title = "{The Dynamical Fate of Wide Binaries in the Solar Neighborhood}",
      journal = {\apj},
         year = 1987,
        month = jan,
       volume = {312},
        pages = {367},
          doi = {10.1086/164883},
       adsurl = {https://ui.adsabs.harvard.edu/abs/1987ApJ...312..367W}
}

@ARTICLE{Jiang_2010,
       author = {{Jiang}, Yan-Fei and {Tremaine}, Scott},
        title = "{The evolution of wide binary stars}",
      journal = {\mnras},
         year = 2010,
        month = jan,
       volume = {401},
       number = {2},
        pages = {977-994},
          doi = {10.1111/j.1365-2966.2009.15744.x},
archivePrefix = {arXiv},
       eprint = {0907.2952},
 primaryClass = {astro-ph.GA},
       adsurl = {https://ui.adsabs.harvard.edu/abs/2010MNRAS.401..977J}
}

@BOOK{Binney_2008,
       author = {{Binney}, James and {Tremaine}, Scott},
        title = "{Galactic Dynamics: Second Edition}",
         year = 2008,
       adsurl = {https://ui.adsabs.harvard.edu/abs/2008gady.book.....B}
}

@ARTICLE{Reipurth_2012,
       author = {{Reipurth}, Bo and {Mikkola}, Seppo},
        title = "{Formation of the widest binary stars from dynamical unfolding of triple systems}",
      journal = {\nat},
         year = 2012,
        month = dec,
       volume = {492},
       number = {7428},
        pages = {221-224},
          doi = {10.1038/nature11662},
archivePrefix = {arXiv},
       eprint = {1212.1246},
 primaryClass = {astro-ph.GA},
       adsurl = {https://ui.adsabs.harvard.edu/abs/2012Natur.492..221R}
}

@ARTICLE{Magazzu_1993,
       author = {{Magazzu}, Antonio and {Martin}, Eduardo L. and {Rebolo}, Rafael},
        title = "{A Spectroscopic Test for Substellar Objects}",
      journal = {\apjl},
         year = 1993,
        month = feb,
       volume = {404},
        pages = {L17},
          doi = {10.1086/186733},
       adsurl = {https://ui.adsabs.harvard.edu/abs/1993ApJ...404L..17M}
}

@ARTICLE{Rebolo_1992,
       author = {{Rebolo}, Rafael and {Martin}, Eduardo L. and {Magazzu}, Antonio},
        title = "{Spectroscopy of a Brown Dwarf Candidate in the alpha Persei Open Cluster}",
      journal = {\apjl},
         year = 1992,
        month = apr,
       volume = {389},
        pages = {L83},
          doi = {10.1086/186354},
       adsurl = {https://ui.adsabs.harvard.edu/abs/1992ApJ...389L..83R}
}

@ARTICLE{Gonzalez-Payo_2026,
       author = {{Gonz{\'a}lez-Payo}, J. and {Caballero}, J.~A. and {Cifuentes}, C. and {Cort{\'e}s-Contreras}, M. and {Rica}, F.},
        title = "{Characterization of all known multiple stellar systems within 10 pc}",
      journal = {\mnras},
         year = 2026,
        month = jun,
       volume = {549},
       number = {1},
          eid = {stag838},
        pages = {stag838},
          doi = {10.1093/mnras/stag838},
archivePrefix = {arXiv},
       eprint = {2605.04094},
 primaryClass = {astro-ph.SR},
       adsurl = {https://ui.adsabs.harvard.edu/abs/2026MNRAS.549ag838G}
}
\bibliographystyle{aasjournalv7}

\end{document}